\documentclass{article}

\usepackage[fleqn]{amsmath}
\usepackage{cite,latexsym,amssymb,tabularx,supertabular,bm,url}
\usepackage{graphicx}
\usepackage{color}

\usepackage[letterpaper]{geometry}
\usepackage{caption}
\newcommand{\tr}{{\rm tr}}

\def\lsi{\raise0.3ex\hbox{$<$\kern-0.75em\raise-1.1ex\hbox{$\sim$}}}
\def\gsi{\raise0.3ex\hbox{$>$\kern-0.75em\raise-1.1ex\hbox{$\sim$}}}
\newcommand{\lsim}{\mathop{\lsi}}
\newcommand{\gsim}{\mathop{\gsi}}

\newcommand{\as}{\alpha_s}

\begin{document}

\noindent{\Large\bf Subleading-$N_c$ corrections in non-linear small-$x$ evolution}

\vspace{0.5cm} { { Yuri V.\ Kovchegov$^1$, Janne Kuokkanen$^2$,
    Kari Rummukainen$^2$, and Heribert Weigert$^2$} }
\\

{\small 
 {$^1$ Department of Physics, The Ohio State University, Columbus, 
   OH 43210, USA}\\
  {$^2$ Department of Physical Sciences, University of Oulu, P.O. Box 3000,
    FI-90014 Oulu, Finland}\\[5mm]
}

\noindent\begin{center}
\begin{minipage}{.92\textwidth}
  {\sf 
    We explore the subleading-$N_c$ corrections to the large-$N_c$
    Balitsky--Kovchegov (BK) evolution equation by comparing its
    solution to that of the all-$N_c$
    Jalilian-Marian--Iancu--McLerran--Weigert--Leonidov--Kovner
    (JIMWLK) equation. In earlier simulations it was observed that the
    difference between the solutions of JIMWLK and BK is unusually
    small for a quark dipole scattering amplitude, of the order of
    $0.1 \%$, which is two orders of magnitude smaller than the
    naively expected $1/N_c^2 \approx 11 \%$. In this paper we argue
    that this smallness is not accidental.  We provide analytical
    arguments showing that saturation effects and correlator
    coincidence limits fixed by group theory constraints conspire with
    the particular structure of the dipole kernel to suppress
    subleading-$N_c$ corrections reducing the difference between the
    solutions of JIMWLK and BK to $0.1\%$. We solve the JIMWLK
    equation with improved numerical accuracy and verify that the
    remaining $1/N_c$ corrections, while small, still manage to slow
    down the rapidity-dependence of JIMWLK evolution compared to that
    of BK. We demonstrate that a truncation of JIMWLK evolution in the
    form of a minimal Gaussian generalization of the BK equation
    captures some of the remaining $1/N_c$ contributions leading to an
    even better agreement with JIMWLK evolution.  As the $1/N_c$
    corrections to BK include multi--reggeon exchanges one may
    conclude that the net effect of multi--reggeon exchanges on the
    dipole amplitude is rather small.
}
\end{minipage}
\end{center}
\vspace{1cm}      

\section{Introduction}
\label{sec:introduction}

Little is known about the features of small-$x$ evolution in the Color
Glass Condensate (CGC) picture \cite{Gribov:1984tu, Mueller:1986wy,
  Mueller:1994rr, Mueller:1994jq, Mueller:1995gb, McLerran:1993ka,
  McLerran:1994ni, McLerran:1994ka, McLerran:1994vd, Kovchegov:1996ty,
  Kovchegov:1997pc, Jalilian-Marian:1997xn, Jalilian-Marian:1997jx,
  Jalilian-Marian:1997gr, Jalilian-Marian:1997dw,
  Jalilian-Marian:1998cb, Kovner:2000pt, Weigert:2000gi, Iancu:2000hn,
  Ferreiro:2001qy, Kovchegov:1999yj, Kovchegov:1999ua,
  Balitsky:1996ub, Balitsky:1997mk, Balitsky:1998ya, Iancu:2003xm,
  Weigert:2005us, Jalilian-Marian:2005jf} beyond the
Balitsky--Kovchegov (BK) truncation \cite{Balitsky:1996ub,
  Balitsky:1997mk,Balitsky:1998ya,Kovchegov:1999yj, Kovchegov:1999ua}
of the Balitsky hierarchy of evolution equations
\cite{Balitsky:1996ub, Balitsky:1997mk,Balitsky:1998ya}. Besides the
theoretical work deriving the
Jalilian-Marian--Iancu--McLerran--Weigert--Leonidov--Kovner (JIMWLK)
equations that summarize the Balitsky hierarchies in a compact form,
only a single numerical study of generic properties of the full
evolution equations is available, carried out by Rummukainen and
Weigert~\cite{Rummukainen:2003ns}. All other studies employ some
additional approximation, typically in form of the BK truncation or
even more schematic approximation. The BK truncation, as the Mueller
dipole model \cite{Mueller:1994rr,Mueller:1994jq,Mueller:1995gb} it is
based on, explicitly neglects $1/N_c$ corrections to the full
$\ln(1/x)$ evolution of QCD observables at high energy. Nevertheless,
both JIMWLK evolution and its BK truncation correctly reproduce the
$N_c$-dependence of the linear Balitsky--Fadin--Kuraev--Lipatov (BFKL)
\cite{Kuraev:1977fs,Bal-Lip} evolution equation in their respective
low density limits. This implies that in the linear, low density
(BFKL) domain subleading $1/N_c$ corrections are manifestly absent
from JIMWLK evolution. The influence of $1/N_c$ corrections on the
non-linear part of the full, untruncated evolution equations is much
harder to estimate.

The only study of the full leading-$\ln(1/x)$ JIMWLK equation
available~\cite{Rummukainen:2003ns} has established, albeit only summarily,
that the $1/N_c$ corrections appear to be much smaller than the $1/N_c^2$
naively expected for the gluon-dominated evolution. Instead of expected
$1/N_c^2 \approx 10 \%$ corrections, the JIMWLK solution for the scattering
amplitude of a quark dipole on a target nucleus found
in~\cite{Rummukainen:2003ns} differs from the solution of the BK equation for
the same quantity by only $0.1 \%$. This has established the BK equation as a reasonable tool to predict the energy dependence of CGC
cross sections, at least after running coupling and some DGLAP corrections are
included~\cite{Gardi:2006rp,
  Kovchegov:2006vj,Balitsky:2006wa,Albacete:2007yr}.\footnote{Kuokkanen,
  Rummukainen, Weigert, in preparation.}  However, the question remains
whether the unexpectedly small difference found in~\cite{Rummukainen:2003ns}
is accidental, being perhaps due to either some intrinsic properties of the
calculated dipole amplitude or to some features of the numerical setup used
in~\cite{Rummukainen:2003ns}. In this paper we argue that the smallness of the
$1/N_c$ corrections found in~\cite{Rummukainen:2003ns} is not accidental. In
fact it is imposed by an interplay of group theoretical properties and
saturation effects of the CGC. As a result non-linear small-$x$ evolution
turns out to be an example of a system in which the $1/N_c$ corrections are
much smaller than naively expected.

We should emphasize that our discussion remains strictly within the context of
JIMWLK evolution and within that only explores the contextual neighborhood of
the BK truncation. The JIMWLK evolution equation is valid for scattering on a
large target that provides a strong gluon field, e.g. for a nucleus with a
large atomic number $A$. It does not include contributions of diagrams which
are not enhanced by the strong target field or, for a large nucleus, are
subleading in powers of $A$. This excludes, right from the start any
discussion of pomeron loop contributions, as they are not included in the
JIMWLK framework. Indeed for small targets with a weaker gluon field like a
proton, which has $A=1$, pomeron loops are not parametrically suppressed
anymore. While pomeron loop induced fluctuations have also recently been
identified in~\cite{Avsar:2008ph} as a source of possible large factorization
violations for such small targets with some parametric
uncertainty,~\cite{Dumitru:2007ew} had found that running coupling corrections
tend to strongly numerically suppress such fluctuations, so that we feel that
our exclusion of pomeron loops from the analysis should not lead to a very
severe restriction for the applicability of our results.

In addition to the large gluon field in the target required for JIMWLK
evolution, the BK evolution equation induces a correlator factorization
assumption that is valid only in the large-$N_c$ limit. [See
Eq.~(\ref{eq:BKfact}) below.] Hence the BK factorization (\ref{eq:BKfact}) has
two types of corrections: those suppressed by the powers of $A$ and those
suppressed by powers of $N_c$. In this paper we are interested in the second
kind of corrections only, in $1/N_c$ corrections, which are resummed to all
orders in the JIMWLK equation but are excluded in the BK equation.  

Even within the purview of JIMWLK evolution we restrict ourselves to a subset
of phenomena: We only discuss how $1/N_c$ suppressed contributions affect
{\em dipole} evolution.  Quantities that have no good approximation in terms
of (multi-) dipole projectiles scattering on dense targets are beyond the
scope of our discussion.  An example not addressed here would be $p A$
scattering at high energies: any realistic description of a proton projectile
lies far outside the standard dipole large $N_c$ approximation, despite the
fact that JIMWLK evolution does cover this example faithfully.  Obviously, in
situations like this, where even a leading order large $N_c$ dipole
description is unavailable any discussion of the size of $1/N_c$-corrections
is moot.

One should also note that JIMWLK and BK equations were both first derived at
leading order $\alpha_s\ln(1/x)$, but have a whole tower of
$\alpha_s$-suppressed corrections, of which only the next to leading order
(NLO) terms are partially available. Running coupling
corrections~\cite{Gardi:2006rp, Kovchegov:2006vj, Balitsky:2006wa} have been
calculated and partial results for the remaining contributions (new physics
channels) are available~\cite{Balitsky:2006wa,Balitsky:2008zz}.
These corrections will change quantitative features to some extent, but should
not completely distort the qualitative structures found at leading order. Our
discussion and simulations will therefore focus on the leading order situation
and only comment on NLO corrections where possible.

In Sect.~\ref{sect:dip} we prepare the ground for our arguments, reminding the
reader about the differences between the BK equation and the JIMWLK equation
for a $q \bar q$ dipole scattering amplitude. The removal of subleading
$1/N_c$ corrections in the BK equation is operationally achieved by
factorizing the expectation value of the product of a pair of dipole operators
into a product of their expectation values. For this reason we refer to the BK
equation as a factorized truncation of the JIMWLK equation. The difference of
the unfactorized and the factorized expectation values measures the size of
the factorization violations. The factorization violation $\Delta$ is defined
in Sect.~\ref{sect:dip} [see \eqref{eq:Delta-variants}], where we also present
its main features. At one loop accuracy, a vanishing $\Delta$ would imply a
complete decoupling of all $1/N_c$ corrections from dipole correlators and is
thus the crucial quantity to explore.\footnote{At NLO, running coupling
  corrections primarily modify the evolution kernel and thus mainly modify how
  strongly a non-vanishing $\Delta$ affects evolution speed [see also
  Sect.~\ref{sec:quant-cons-fact}]. Other NLO corrections generically
  introduce new $1/N_c$ suppressed contributions but are accompanied by an
  additional power of $\alpha_s$.}

In Sect.~\ref{sec:orig-fact} we will clarify the reason for the smallness of
the factorization violations observed in~\cite{Rummukainen:2003ns}, for
``typical'' factorization violations $\Delta$. A
more in depth discussion of this issue than that offered
in~\cite{Rummukainen:2003ns} must first note that the correlator $\Delta$
measuring the factorization violation itself has in fact contributions that do
reach all the way up to their natural size of $1/N_c^2$ in certain regions of
configuration space. (Note that $\Delta$ depends on three transverse
coordinates: the positions of the original quark and anti-quark, and of the
emitted gluon. Varying those coordinates gives different values of $\Delta$.)
However, as observed in~\cite{Rummukainen:2003ns}, the typical contributions
to $\Delta$ in the majority of configuration space are in fact tiny
compared to $1/N_c^2$. In Sect.~\ref{sec:orig-fact} we will systematically map
out configuration space to identify all regions with factorization
violations. We will argue on general grounds that the factorization violation
$\Delta$ is indeed much smaller than $1/N_c^2$ in the majority of its
configuration space, in agreement with the result of numerical simulations
presented in~\cite{Rummukainen:2003ns}. We will also demonstrate analytically
that the evolution kernel wipes out all contributions from the only region
where the factorization violation $\Delta$ is of the naively expected order
$1/N_c^2$. We thus will complete the proof of the statement that $1/N_c$
corrections to BK evolution, which are consistently included into JIMWLK
evolution, are indeed much smaller than $1/N_c^2$. This constitutes our first
main result.

The basis of our mapping out of configuration space in a systematic way is the
insight that the origin of the factorization violations is to be found in a
set of group theoretical identities that apply to coincidence points of
(s-channel) $n$-point functions involved in the Balitsky hierarchies (i.e.,
the limits in which any pair of the transverse coordinates overlaps). The
identities are shown in~\eqref{eq:3-point-generic-coincidence}. They are, by
construction, respected in JIMWLK evolution, but automatically broken at the
$1/N_c^2$ level by the correlator factorization assumption underlying the BK
truncation.

In Sect.~\ref{sect:gauss} we note that it is possible to extend the BK
equations in a minimal manner that reinstates these group theoretical
constraints for {\em all} eikonal correlators in high energy
scattering. The inspiration comes from calculating all involved Wilson
line correlators in the quasi-classical approximation known as the
McLerran-Venugopalan (MV) model \cite{McLerran:1993ka,
  McLerran:1994ni, McLerran:1994ka}. One can sum up all
Glauber-Mueller (GM) multiple rescatterings \cite{Mueller:1989st} to
calculate various 2- and 3-point functions (see e.g.
\cite{Jalilian-Marian:1997xn, Kopeliovich:1998nw, Kopeliovich:1999am,
  Kovchegov:2001ni, Kovner:2001vi}). Using the resulting correlation
functions one can construct the factorization violation $\Delta$ and
study its properties. This allows us to revisit our earlier general
observation on the structure of $\Delta$ in configuration space and
amend it with explicit expressions for the correlators, albeit within
a model. However, as was noted in \cite{Kovner:2001vi} and as we will
explain in Sect.~\ref{sec:rest-coinc-limits}, one can also insert the
2- and 3-point correlators obtained in the GM/MV approximation into
the JIMWLK evolution equation for the 2-point correlator (the lowest
order equation in the corresponding Balitsky hierarchy). One can then
suggest treating the resulting equation as an evolution equation in
its own right \cite{Kovner:2001vi}, though no parametric
justification/proof of this statement exists. This equation (see
\eqref{eq:tilde-G-evo-short} below) is thus only a guess for the
evolution equation beyond the leading-$N_c$ BK equation.  The result
will be referred to as a Gaussian truncation (GT) of the Balitsky
hierarchies or equivalently the JIMWLK equation.  This Gaussian
truncation had been introduced originally in~\cite{Kovner:2001vi} as
an ``exponential parametrization'' for $q\Bar q$ dipoles and a certain
set of other correlators, with an evolution equation derived
explicitly for the $q\Bar q$ dipole operator. On this level it was
also explicitly used in~\cite{Weigert:2005us} to unify a diversity of
``McLerran-Venugopalan models.''\footnote{This is different in content
  as well as in spirit from the Gaussian approximation discussed
  in~\cite{Iancu:2002aq}.}
  
The relationship of the Gaussian truncation to the BK equation turns
out to be unexpectedly subtle: On the one hand it extends the BK
truncation in the sense that it includes a set of subleading $1/N_c$
corrections, those ``minimally'' required to reinstate the coincidence
limits violated in the BK factorization. Consistently, the Gaussian
truncation reduces to BK in the large $N_c$ limit.  On the other hand,
Eq.~\eqref{eq:tilde-G-evo-short}, the evolution equation in the
Gaussian truncation turns out to be equivalent to the BK evolution
equation with respect to dynamical content. The only changes occur in
the way this content is mapped onto the expressions for correlators.

In Sect.~\ref{sec:qual-struct-fact} we compare the factorization violation
given by GT and by JIMWLK, and find them similar qualitatively, but still
quite different quantitatively.  Since we view GT as a truncation of JIMWLK
evolution and hence the Balitsky hierarchies, we will also clarify where GT
breaks consistency with JIMWLK: GT remains only an approximation to the full
JIMWLK evolution.

Our analytical arguments are complemented in Sect.~\ref{sec:quant-cons-fact}
by a new numerical study of the JIMWLK evolution equation that goes beyond
that of~\cite{Rummukainen:2003ns} with simulations on much larger lattices in
the transverse space, extending the $48^2-512^2$-range covered earlier with
simulations on $512^2-4096^2$ lattices.  We emphasize that the simulations
presented here are done for fixed coupling only: at present the numerical
simulation of the exact JIMWLK kernel with the running coupling corrections
found in~\cite{Gardi:2006rp, Kovchegov:2006vj,Balitsky:2006wa} would render
the numerical cost prohibitive. To efficiently include them would require us
to find an alternative representation that allows a factorized form of the
JIMWLK Hamiltonian akin to that used at leading order. This remains beyond the
scope of this paper. Nevertheless, the additional numerical effort allows us
to reduce extrapolation errors considerably (they arise mostly from the
infinite volume limit as it turns out) and establish reliably that JIMWLK
evolution is in fact slightly slower than factorized BK evolution: Subleading
$1/N_c$ corrections indeed slow down evolution, just as was observed earlier
with running coupling corrections. Evolution speed turns out to be
particularly sensitive to factorization violations, which is in keeping with
the integral expressions of Eq.~(\ref{eq:lambda-int-def}) below. At one loop
order, we observe numerically a 3-5\% slowdown induced by factorization
violations where our simulations approach the scaling region. We argue that
running coupling corrections should suppress the UV part of phase space
leading to a strong reduction of this difference of evolution speeds between
JIMWLK and BK. We cannot estimate the influence of other NLO corrections
which may well have their own offsetting effects, but they should not
completely distort the leading order picture. We conclude that, while the net
effect remains small, $1/N_c$ corrections pull our predictions towards
evolution speeds compatible with experiment, not in the opposite direction.
This qualitative slowdown effect is our second main conclusion.

In Sect.~\ref{sec:jimwlk-beyond-gauss} we concentrate on the physical origin
of $1/N_c$ corrections to the BK evolution equation. As noted above, the BK
truncation reproduces the $N_c$-dependence of the linear BFKL evolution
equation, corresponding to a two-reggeon state in the $t$-channel. However,
among other $1/N_c$ corrections, the BK truncation neglects contributions of
multiple reggeon exchanges
\cite{Bartels:1978fc,Bartels:1980pe,Kwiecinski:1980wb,Jaroszewicz:1980mq}.
Some of those omitted higher reggeon exchanges, like the odderon contribution
corresponding to a $C$-odd three-reggeon exchange \cite{Lukaszuk:1973nt,
  Nicolescu:2007ji, Janik:1998xj, Korchemsky:2001nx, Bartels:1999yt}, have
been included in the BK-truncated CGC formalism by a minimal modification of
the truncated evolution equations \cite{Kovchegov:2003dm, Hatta:2005as,
  Kovner:2005qj}. Higher-order reggeon exchanges usually require substantial
modification of Mueller's dipole model to a more generic $s$-channel picture
as one generically is required to include $1/N_c$ suppressed multipole
correlators on top of simple dipoles: see \cite{Chen:1995pa} for an analogue
of the Bartels--Jaroszewicz--Kwiecinski--Praszalowicz (BJKP) evolution
equation~\cite{Bartels:1978fc, Bartels:1980pe, Kwiecinski:1980wb,
  Jaroszewicz:1980mq} in the $s$-channel formalism. Generically not much has
been done to identify the contributions of higher $n$-reggeon exchange
contributions~\cite{Derkachov:2002wz} to nonlinear JIMWLK evolution in any
systematic way.

Nevertheless, the fact that the odderon \cite{Kovchegov:2003dm, Hatta:2005as,
  Kovner:2005qj} and 4-reggeon \cite{Chen:1995pa} exchanges are included in
the $s$-channel evolution picture allows us to conjecture that all
multi--reggeon exchanges are included in the JIMWLK evolution equation. The
JIMWLK equation also probably includes some multi--reggeon vertices containing
more legs on the target side of the evolution than on the projectile side. If
this conjecture is true, one concludes that the difference between the dipole
amplitude given by BK and by JIMWLK is at least partially due to an aggregate
of multiple--reggeon effects. The smallness of this difference then, in turn,
would indicate the smallness of multiple-reggeon exchange effects.

The link of multi--reggeon exchanges with subleading $1/N_c$ corrections gives
a natural explanation for the slowdown of JIMWLK evolution compared to BK
observed in Sect.~\ref{sec:quant-cons-fact}.  Generically one would argue that
nonlinear effects will work to temper any influence of multi--reggeon
contributions, which would complement the power suppression of $1/N_c$
contributions via the kernel observed earlier in our line of argument. If
true, this is testable numerically, but it is not a priori clear how to test
this.  Identifying the Gaussian truncation with iterated two reggeon exchange
gives a handle on this as well: we may filter out the multi--reggeon exchanges
by comparing the Gaussian truncation with full JIMWLK evolution. It turns out
that the Gaussian truncation has a distinctive feature that is naturally
violated by multi--reggeon exchanges: the Gaussian truncation would predict
strict Casimir scaling of dipole correlators in different
representations. (Casimir scaling is defined in
\eqref{eq:casimir-scaling-app}.) In Sect.~\ref{sec:jimwlk-beyond-gauss} we
illustrate this statement by extending GT to include the simplest
multi--reggeon contribution in the form of an odderon exchange: we then show
that it indeed violates the Casimir scaling. Therefore we argue that the size
of Casimir scaling violations can quantify the net contribution of all
multi--reggeon exchanges. We thus can numerically explore the effect of
multi--reggeon exchanges by measuring the violations of Casimir scaling of the
dipole correlators. Casimir scaling violation in the numerical solution of
JIMWLK that we performed is studied in Sect.~\ref{sec:jimwlk-beyond-gauss}. It
turns out that the Casimir scaling violations (which summarize the collective
effect of all multi--reggeon exchanges included in JIMWLK evolution) are
generically small and do not grow with energy (see e.g.
Fig.~\ref{fig:casimir-scaling-viol}). This is our third main result.

We review our results and methods in Sect.~\ref{sec:conclusions}.

\section{Dipole evolution in JIMWLK and BK frameworks}
\label{sect:dip}

JIMWLK evolution is equivalent to sets of coupled infinite hierarchies
of evolution equations, the simplest of which is based on the equation
for the $q\Bar q$-dipole correlators $\langle \Hat S_{\bm{x
    y}}^{q\Bar q}\rangle(Y)$ for the scattering on a target at high energies in
which the scattering of the $q$ and $\Bar q$ is expressed via
light-like Wilson lines in the fundamental representation $U_{\bm x}$
or $U_{\bm y}^\dagger$ respectively (at fixed transverse positions
$\bm x, \bm y$),
\begin{equation}
\label{eq:hatSqqbardef}
  \Hat S_{\bm{x y}}^{q\Bar q} := \frac{\tr\left(U_{\bm x} U_{\bm
        y}^\dagger\right)}{N_c}
\ .
\end{equation}
This operator is gauge invariant in the sense that the contributions that
close the trace at $x^+=\pm \infty$ are unity to leading order in
$\ln(1/x)$.\footnote{This is true strictly speaking only if one does not force
  the $U$-factors shown to be one by gauge choice. In this case one would need
  to display the contributions that connect the trace at $x^+=\pm\infty$ to a
  closed Wilson loop. They would carry the full contribution, but would remain
  independent of the paths used to connect $\bm x$ to $\bm y$.} Averaging the
operator in Eq.~(\ref{eq:hatSqqbardef}) over all states in the target wave
function yields the $Y$-dependent S-matrix for the scattering of a dipole on
that specific target.  The evolution equation for this average, $\langle \Hat
S_{\bm{x y}}\rangle(Y)$, involves a gluon Wilson line operator $\Tilde U$ in
the adjoint representation on its right-hand side. At fixed coupling can be
written either as \cite{Balitsky:1997mk,Balitsky:1996ub,Weigert:2000gi}
  \begin{align}
    \label{eq:preBKU}
    \frac{d}{dY} 
 \langle 
 \tr(
  U_{\bm{x}}
  U^\dagger_{\bm{y}}) 
\rangle(Y)    
=\frac{\alpha_s}{\pi^2}\int d^2z\ {\cal K}_{\bm{x z y}} 
  \left(
  \langle
\big[\Tilde U_{\bm{z}}\big]^{a b}\
    \tr(
  t^a 
  U_{\bm{x}}
  t^b 
  U^\dagger_{\bm{y}}) 
  \rangle(Y)
  -C_f  \langle 
 \tr(
  U_{\bm{x}}
  U^\dagger_{\bm{y}})  \rangle(Y)    
  \right)
  \end{align}
or, using~(\ref{eq:hatSqqbardef}) and the Fierz identity
\begin{equation}
  \label{eq:Fierz}
  \big[\Tilde U_{\bm{z}}\big]^{a b}
  2 \tr(t^a U_{\bm{x}}t^b U^\dagger_{\bm{y}}) =
  \tr( U_{\bm{x}}U^\dagger_{\bm{z}}) \
           \tr( U_{\bm{z}}U^\dagger_{\bm{y}})
           -\frac{1}{N_c}\tr( U_{\bm{x}}U^\dagger_{\bm{y}})
\end{equation}
as
\begin{align}
  \label{eq:preBKS}
 \frac{d}{d Y} \langle \Hat S_{\bm{x y}} \rangle(Y)
  =\frac{\alpha_s N_c}{2\pi^2}\int d^2z\,
  \ {\cal K}_{\bm{x z y}} \
 \langle \Hat S_{\bm{x z}} 
\Hat S_{\bm{z y}} 
 -
\Hat S_{\bm{x y}} \rangle(Y)
\ .
\end{align}
The integral kernel in both~(\ref{eq:preBKU}) and~(\ref{eq:preBKS})
is given by~\cite{Mueller:1994rr,Kovchegov:1999yj}
\begin{align}\label{K}
  {\cal K}_{\bm{x z y}} \, := \, \frac{({\bm x} - {\bm y})^2}{({\bm x}
    - {\bm z})^2 \ ({\bm z} - {\bm y})^2}.
\end{align}
Eqs.~(\ref{eq:preBKU}) and~(\ref{eq:preBKS}) are completely
equivalent versions of the first equation in the Balitsky hierarchy of
the quark dipole operator~(\ref{eq:hatSqqbardef}).
Eqs.~(\ref{eq:preBKU}) and~(\ref{eq:preBKS}) obviously do not
represent closed equations since the evolution of $\langle \tr(
U_{\bm{x}} U^\dagger_{\bm{y}}) \rangle(Y)$ depends on an operator with
an additional gluon operator $\tilde U$ insertion. The evolution
equation of that new operator, $\langle \big[\Tilde U_{\bm{z}}\big]^{a
  b}\ \tr( t^a U_{\bm{x}} t^b U^\dagger_{\bm{y}}) \rangle(Y)$, in turn
will involve yet one more insertion of a gluon operator $\tilde U$,
iteratively creating an infinite coupled hierarchy of evolution
equations, the Balitsky hierarchy of the quark dipole
operator~(\ref{eq:hatSqqbardef})
\cite{Balitsky:1997mk, Balitsky:1996ub}. JIMWLK evolution summarizes
the totality of all such hierarchies, based on any (gauge invariant)
combination of multipole operators but can only be solved
numerically~\cite{Rummukainen:2003ns} at considerable numerical cost.
The situation can be simplified for the price of introducing an additional
approximation that truncates the hierarchy. The most widely used truncation is
known as the BK approximation. It assumes the factorization
\begin{align}
  \label{eq:BKfact}
  \langle \Hat S_{\bm{x z}} 
\Hat S_{\bm{z y}} \rangle(Y) \to 
 \langle \Hat S_{\bm{x z}} \rangle(Y)\ \langle
\Hat S_{\bm{z y}} \rangle(Y)
\ ,
\end{align}
which turns Eq.~\eqref{eq:preBKS} into a closed equation in terms of
$\langle\Hat S_{\bm{x y}} \rangle(Y)$ only and thus decouples the rest of the
Balitsky hierarchy. The BK truncation is valid and is parametrically justified
in the large-$N_c$ limit for scattering on a large dilute nuclear
target. Using \eqref{eq:BKfact} in \eqref{eq:preBKS} we obtain the BK
evolution equation
\begin{align}
  \label{eq:BK}
  \frac{d}{d Y} \langle \Hat S_{\bm{x y}} \rangle(Y) =\frac{\alpha_s
    N_c}{2\pi^2}\int d^2z\, \ {\cal K}_{\bm{x z y}} \ \left[ \langle
    \Hat S_{\bm{x z}} \rangle (Y) \ \langle \Hat S_{\bm{z y}} \rangle
    (Y) - \langle \Hat S_{\bm{x y}} \rangle(Y) \right].
\end{align}

Provided that the dipole correlator shapes $\langle \Hat S_{\bm{x y}}
\rangle(Y)$ are not too different (this notion with be refined in
Sec.~\ref{sec:quant-cons-fact}) in JIMWLK (without
factorization~(\ref{eq:BKfact})) and BK (with factorization, as shown
in \eqref{eq:BK}) the factorization violation that creates the difference
between the two
\begin{subequations}
  \label{eq:Delta-variants}
\begin{align}
  \label{eq:Delta-old}
  \Delta_{\bm{x z y}}(Y) :=   \langle \Hat S_{\bm{x z}} 
\Hat S_{\bm{z y}} \rangle(Y) - 
 \langle \Hat S_{\bm{x z}} \rangle(Y)\ \langle
\Hat S_{\bm{z y}} \rangle(Y)
\end{align}
can be simply interpreted as the difference of the correlators on the
right-hand side of Eq.~(\ref{eq:preBKU}) and its BK counterpart
(\ref{eq:BK}), i.e.
\begin{align}
  \label{eq:fact-rhsdiff}
 \Delta_{\bm{x z y}}(Y) =   \bigl[\langle \Hat S_{\bm{x z}} 
\Hat S_{\bm{z y}} 
 -
\Hat S_{\bm{x y}} \rangle(Y)
\bigr]
-\bigl[\langle \Hat S_{\bm{x z}} \rangle(Y)\ \langle
\Hat S_{\bm{z y}} \rangle(Y)
-
\langle \Hat S_{\bm{x y}} \rangle(Y)
\bigr]
\end{align}
(the $\langle \Hat S_{\bm{x y}} \rangle(Y)$ term is the same under these
conditions and cancels trivially) or as a fluctuation away from a mean field
value
\begin{align}
  \label{eq:Delta-mean-field}
  \Delta_{\bm{x z y}}(Y) = \bigl\langle
    \bigl( 
    \Hat S_{\bm{x z}} -\langle \Hat S_{\bm{x z}} \rangle(Y)
    \bigr)
    \bigl(
      \Hat S_{\bm{z y}} -\langle \Hat S_{\bm{z y}} \rangle(Y)
    \bigr)
    \bigr\rangle(Y)
\ .
\end{align}
\end{subequations}
The first interpretation directly leads us to consider factorization
violations as a source for a difference in evolution speed in JIMWLK
and BK, the second interpretation will lead us to the question of what
kind of degrees of freedom (which are absent in BK but included in
JIMWLK) would be associated with these fluctuations. We meet the
latter question repeatedly in all remaining sections, here we will
first look at the individual terms in Eq.~(\ref{eq:fact-rhsdiff}) to
get a generic idea of the structure of configuration space and how it
affects evolution and then give a first glimpse at how JIMWLK
evolution via~(\ref{eq:preBKS}) might differ from BK evolution
(\ref{eq:BK}).

\begin{figure}[!thb]
  \centering
  \includegraphics[width=.49\linewidth]{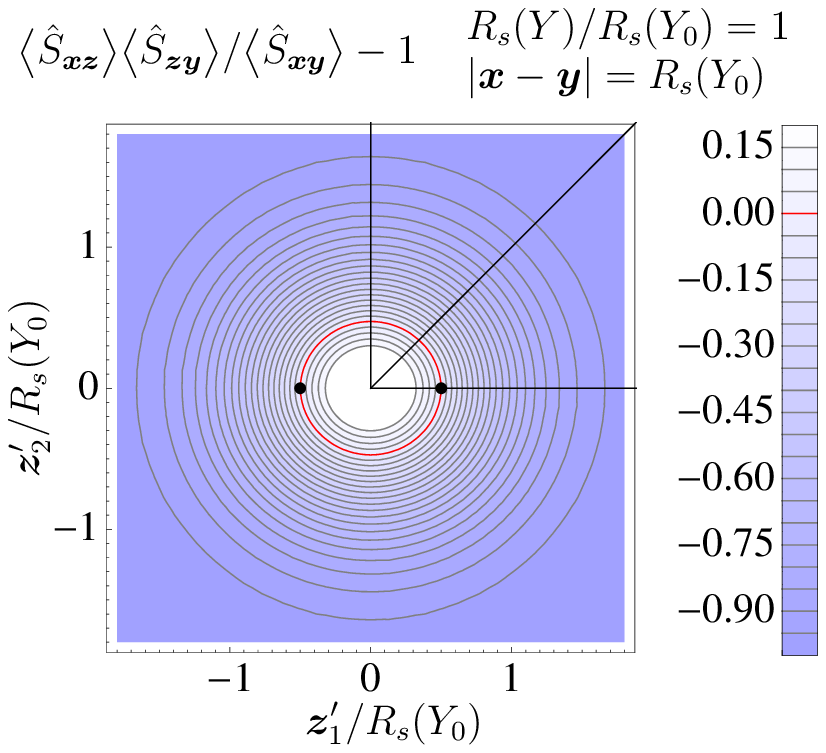}\hfill
  \includegraphics[width=.49\linewidth]{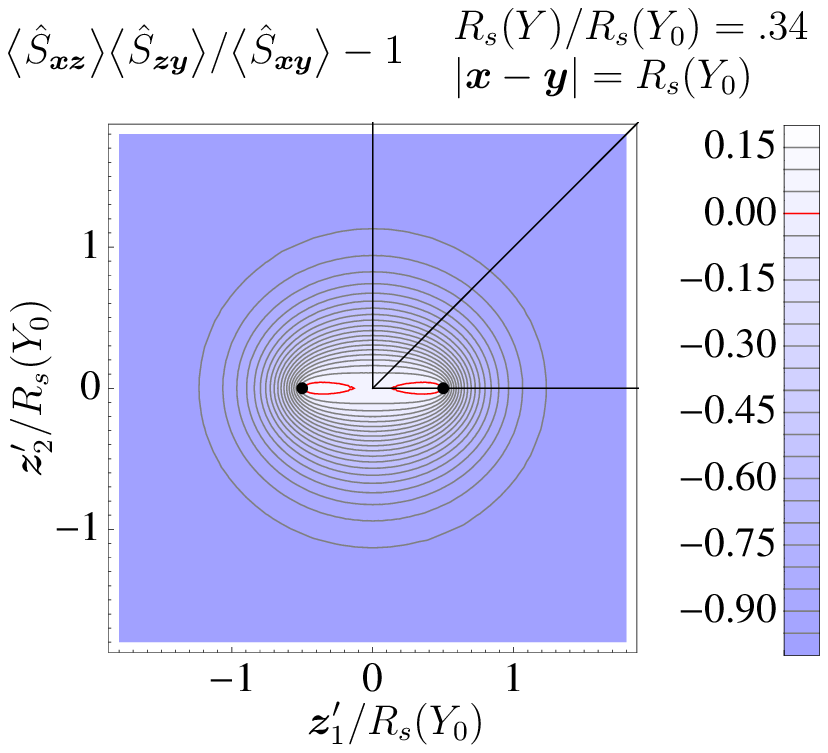}
  \caption{Contour plot of the correlators on the right-hand side of the BK
    equation (\ref{eq:BK}) (divided by $\bigl\langle \Hat S_{\bm{x
        y}}\bigr\rangle$ to normalize the large $\bm z$ asymptotics to $-1$)
    in the $\bm z'$-plane at different stages in the evolution, for fixed
    $|\bm x-\bm y|= R_s (Y_0)$. Here ${\bm z}' = (z'_1, z'_2)$. The left panel
    shows the initial conditions for the evolution, while the right panel
    displays the evolved distribution (see text). The dots mark $\bm x$ and
    $\bm y$, the locations of the parent $q$ and $\Bar q$. These points always
    fall on the boundaries between positive and negative contributions to the
    right-hand side of \eqref{eq:BK} (marked by contour lines going through
    the dots). The half-rays denote the angles at which the factorization
    violations will be plotted in Fig.~\ref{fig:fact-viols-fixed-xmy}.}
  \label{fig:z-plane}
\end{figure}

In both cases in an otherwise translationally invariant system with a
given parent dipole the integrands (correlators and kernel separately
-- that is why we will leave the latter aside) have a twofold mirror
symmetry in the $\bm z$-plane: one with respect to the $(\bm x-\bm
y)$-axis, the other with respect to an axis perpendicular to $\bm
x-\bm y$, through the midpoint $(\bm x+\bm y)/2$. The latter only
holds if $ \langle \tr( U_{\bm{x}} U^\dagger_{\bm{y}}) \rangle(Y) =
\langle \tr( U_{\bm{x}} U^\dagger_{\bm{y}})^\dagger \rangle(Y)
=\langle \tr( U_{\bm{y}} U^\dagger_{\bm{x}}) \rangle(Y) $, i.e., it is
real (and thus symmetric in $\bm x\leftrightarrow\bm y$) as is the
case if we study its contribution to the total DIS cross section at
high energy \cite{Balitsky:1996ub,Kovchegov:1999yj}.  In this context
it is useful to introduce a $\bm z$ coordinate with respect to $(\bm
x+\bm y)/2$ as the origin
\begin{equation}
  \label{eq:z'def}
  \bm z':=\bm z-(\bm x+\bm y)/2
\ .
\end{equation}
There are strong zeroes in the correlators on the right-hand side of the
evolution equations as well as in $\Delta$ when $\bm z\to \bm x$ or $\bm
y$. They are needed to cancel the kernel singularities at these points and
have their origin in real virtual cancellations.  Generically, these zeroes
are not isolated but lie on lines that separate the positive from the negative
contributions to the integrand of the evolution equations. Picking out the
positive sign regions in the integrand, the mirror symmetries allow two
situations: one in which there are two separate such regions adjacent to the
$q$ and $\Bar q$ respectively, and another where the regions are joined
together [generically in situations with dipole correlators not too dissimilar
from the Golec-Biernat--W\"usthoff (GB-W) case \cite{GolecBiernat:1998js}
(also known as Glauber-Mueller multiple rescatterings) which serves as our
initial conditions; the initial conditions are in fact radially
symmetric]. The generic patterns are shown in Fig.~\ref{fig:z-plane}, which
presents contour plots of the right-hand side of the BK equation (\ref{eq:BK})
(divided by $\bigl\langle \Hat S_{\bm{x y}}\bigr\rangle (Y)$) obtained by
performing a numerical solution of that equation. The horizontal and vertical
axis on each panel show $z'_1$ and $z'_2$, which are the two components of the
two-dimensional vector $\bm z'$. The coordinates are plotted in the units of
the initial correlation length $R_s (Y_0)$ of the system (formally defined as
the distance at which the dipole correlator falls to $1/2$, i.e. via
$\langle\Hat S_{|\bm r|=R_s(Y)}\rangle(Y)=1/2$).  $R_s(Y)$ can be thought of
as the inverse of the saturation scale $Q_s (Y)$: $R_s (Y) = 1/Q_s (Y)$. $Y_0$
is the rapidity of the initial conditions for \eqref{eq:BK}. The dots in
Fig.~\ref{fig:z-plane} denote the positions $\bm x$ and $\bm y$ of the quark
and the anti-quark, with $|\bm x-\bm y|$ taken in Fig.~\ref{fig:z-plane} to be
equal to $R_s (Y_0)$. They lie on the contour lines that separate positive
from negative regions.  The left panel of Fig.~\ref{fig:z-plane} corresponds
to the initial conditions for the BK evolution ($R_s (Y) / R_s (Y_0) =1$),
while the right panel corresponds to a higher rapidity $Y>Y_0$ where $R_s (Y)
/ R_s (Y_0) =0.34$, i.e., after running the evolution for some time.

For fixed parent dipole size $|\bm x -\bm y|$ the most extreme
situations arise when we vary $\bm z'$ along the ${\bm x} - {\bm y}$
axis and the axis perpendicular to it, all other directions in the
$\bm z'$-plane interpolate smoothly. The two axes, along with one
intermediate $45^\circ$ axis, are also shown in both panels of
Fig.~\ref{fig:z-plane}.
Evolution speed, in either JIMWLK or BK, is then a consequence of a
numerically delicate balance of the negative and positive regions of the
quantity plotted in Fig.~\ref{fig:z-plane}. Since we are talking of evolution
for $S$ in which generically $S$ is driven to smaller values at fixed dipole
sizes as rapidity $Y$ increases, the negative regions in
Fig.~\ref{fig:z-plane} push evolution forward (these contributions are
generically those at large $|\bm z'|$), while the positive regions in
Fig.~\ref{fig:z-plane} (generically near $\bm x$ and $\bm y$) slow it down.
At fixed coupling, {\em any} change of evolution speed can be mapped onto a
change of relative weight of these two contributions.  Starting from a
non-scaling initial condition like the GB-W model, evolution typically speeds
up until scaling is reached and evolution speed is maximal. (Scaling here is
defined as the situation in which all rapidity dependence is carried by the
saturation scale so that observables like $\langle\Hat S_{\bm r}\rangle(Y)$
become functions of scaling ratios like $\bm r/R_s(Y)$ only
\cite{Stasto:2000er}.)  This is mirrored perfectly in a shrinking of the
positive regions from the radially symmetric situation of the GB-W initial
condition (Fig.~\ref{fig:z-plane}, left) to a situation in which there are two
separate positive regions near the $q$ and $\Bar q$ positions
(Fig.~\ref{fig:z-plane}, right).

The $\bm z'$ plane symmetries of the dipole evolution equations
translate directly into the factorization violations $\Delta_{\bm{x z
    y}}(Y)$ from Eqs. (\ref{eq:Delta-variants}), and also the zeroes
at the $q$ and $\Bar q$ positions carry over.
In~\cite{Rummukainen:2003ns} two of us (Rummukainen and Weigert) had
observed numerically that all factorization violations tested were
positive (i.e., qualitatively acted to slow down evolution compared to
BK), and unexpectedly small, at least in the regions that contribute
to evolution: instead of $\Delta \sim 1/N_c^2 \sim 10\%$ at $N_c=3$
one found contributions roughly another magnitude smaller.
\begin{figure}[!thb]
  \centering
  \includegraphics[width=.32\textwidth]{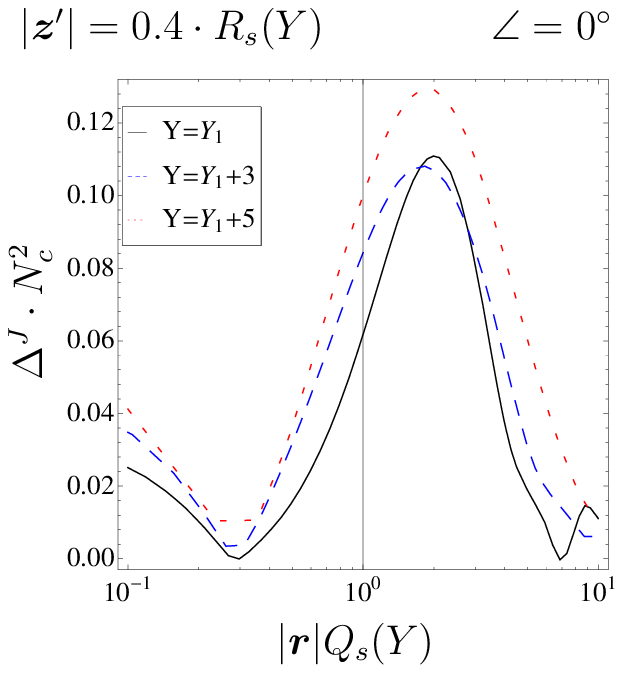}\hfill
  \includegraphics[width=.32\textwidth]{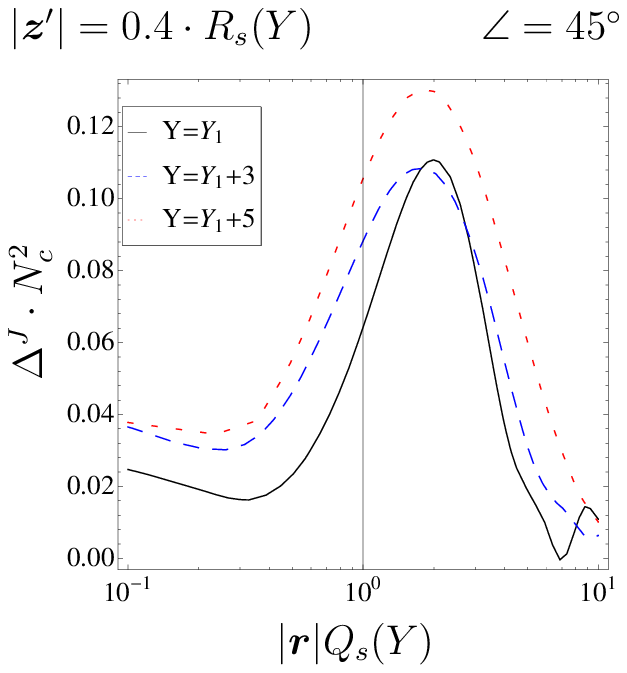}\hfill
  \includegraphics[width=.32\textwidth]{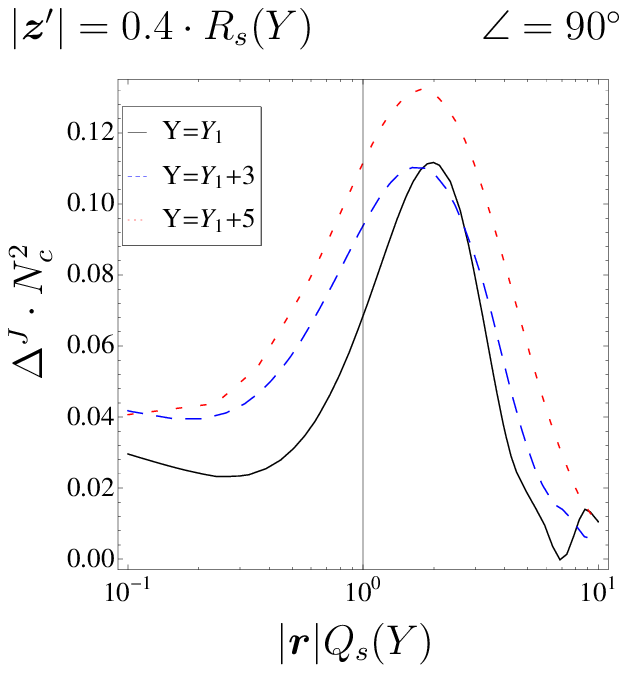}
  \caption{Factorization violations from JIMWLK evolution (scaled up by
    $N_c^2$) plotted against varying parent dipole size at fixed $|\bm
    z'| = 0.4\cdot R_s(Y)$. The angles $0^\circ$ (left), $45^\circ$
    (middle), and $90^\circ$ (right) are the angles between $\bm z'$
    and $\bm r$ and refer to rays in the $\bm z'$-plane as indicated
    in Fig~\ref{fig:z-plane} for one fixed $|\bm r| = |{\bm x} - {\bm
      y}|$. Shown are three different rapidities each. One obtains a
    reduction by a factor of 10 compared to the natural size $\Delta^J
    \sim 1/N_c^2$ (or $\Delta^J \, N_c^2 \sim 1$). This was observed
    earlier in~\cite{Rummukainen:2003ns}, with any differences being
    due to the slightly different correlator geometries chosen here
    for ease of comparison with the discussion below. Shown are
    ``typical'' regions that contribute to evolution, see
    Sect.~\ref{sec:orig-fact} for details. Note also that only the
    $0^\circ$ ray shows special structure since it contains strict
    coincidence limits (i.e. the limits where $\bm x$, $\bm y$ or $\bm
    z$ overlap), all other angles are qualitatively well represented
    by the $90^\circ$ case.}
  \label{fig:fact-viols-fixed-xmy}
\end{figure}
Fig.~\ref{fig:fact-viols-fixed-xmy} re-illustrates the observation
of\cite{Rummukainen:2003ns}, namely that the factorization violation $\Delta$
is about an order of magnitude smaller than the naively expected $10\%$ in
regions relevant for evolution.  Fig.~\ref{fig:fact-viols-fixed-xmy} plots
$\Delta_{\bm{x z y}}(Y)$ (henceforth referred to as simply $\Delta$ without
the arguments) from the numerical solution of JIMWLK evolution equation which
we will describe below. $\Delta_{\bm{x z y}}(Y)$ is plotted in
Fig.~\ref{fig:fact-viols-fixed-xmy} as a function of the parent dipole size
\begin{align}
  {\bm r} = {\bm x} - {\bm y}
\end{align}
for fixed $|{\bm z}'|$ with the angle between $\bm z'$ and $\bm r$
being $0^\circ$ (left panel), $45^\circ$ (middle panel) and $90^\circ$
(right panel). These directions were also shown in Fig.~\ref{fig:z-plane}.
In each panel of Fig.~\ref{fig:fact-viols-fixed-xmy} the factorization
violation $\Delta$ is plotted for three different rapidities: $Y_1$,
$Y_1 + 3$, and $Y_1 + 5$, with the exact numerical value of $Y_1$
being irrelevant here (along with the value of the fixed coupling
constant $\as$), as our goal in this Section is only to demonstrate
the size of the typical factorization violations.

While the correlator geometries in Fig.~\ref{fig:fact-viols-fixed-xmy}
differ slightly from those shown in~\cite{Rummukainen:2003ns}, the
magnitudes are comparable. We see again that instead of naively
expected $\Delta \, N_c^2 \sim 1$ one gets $\Delta \, N_c^2 \sim 0.1$,
which is an order of magnitude smaller. As rapidity increases beyond
the values shown in Fig.~\ref{fig:fact-viols-fixed-xmy}, the factorization
violation $\Delta$ does not grow significantly beyond the values
achieved in the figure.  In \cite{Rummukainen:2003ns} no attempt was
made to clarify in which regions of configuration space the
factorization violation $\Delta$ is small, and no generic discussion
of relative importance of configuration space regions was given. To
fully understand the statement of the smallness of corrections one
must expand on the discussion given there and first gain a better
understanding of where to expect sizable contributions, since mapping
out {\em all} of the configuration space in $\bm z'$, $\bm x-\bm y$
and $Y$ is otherwise not feasible. This will also provide the
underlying reason for the observed smallness.

\section{Origin and smallness of the factorization violation: an interplay of
  saturation and coincidence limits}
\label{sec:orig-fact}

Smallness of the specific factorization violations~(\ref{eq:Delta-variants})
are only one facet of a more generic question: what kind of deviations from
full JIMWLK evolution are caused by the factorization
assumption~(\ref{eq:BKfact}) with its associated truncation of the Balitsky
hierarchy of the quark dipole operator?

Full JIMWLK evolution does not only couple in a full hierarchy of evolution
equations for the quark dipole operator, it has an even wider scope: It
consistently incorporates hierarchies based on any $n$-point
correlator. Examples for such {\em distinct} hierarchies are obtained by
considering the infinite set of dipole correlators labeled by all finite
dimensional unitary representations ${\cal R}$.  Each of them has its own
distinct evolution equation, that can be summarily written as
  \begin{align}
    \label{eq:prefactUR}
    \frac{d}{dY} 
 \bigl\langle 
\overset{{\scriptscriptstyle\cal R}}\tr(
\overset{{\scriptscriptstyle\cal R}} U_{\bm{x}}
\overset{{\scriptscriptstyle\cal R}}
  U^\dagger_{\bm{y}}) 
\bigr\rangle(Y)    
=\frac{\alpha_s}{\pi^2}\int\! d^2z\ {\cal K}_{\bm{x z y}} \
  \biggl(
  \bigl\langle
\big[\Tilde U_{\bm{z}}\big]^{a b}\
   \overset{{\scriptscriptstyle\cal R}}\tr(
\overset{{\scriptscriptstyle\cal R}} t^a 
\overset{{\scriptscriptstyle\cal R}} U_{\bm{x}}
\overset{{\scriptscriptstyle\cal R}} t^b 
\overset{{\scriptscriptstyle\cal R}}
  U^\dagger_{\bm{y}})
\bigr\rangle(Y) 
  -C_{\cal R}  
\bigl\langle
\overset{{\scriptscriptstyle\cal R}}\tr(
\overset{{\scriptscriptstyle\cal R}} U_{\bm{x}}
\overset{{\scriptscriptstyle\cal R}}
  U^\dagger_{\bm{y}})  \bigr\rangle(Y)    
\biggr)
\ .
\end{align}
Here $\overset{{\scriptscriptstyle\cal R}} U$ refers to the group
element in the representation ${\cal R}$, with analogous notations for
the trace, generators and conjugate representation. $C_{\cal R}$
denotes the second Casimir of the representation of the dipole, i.e.,
for the $q\Bar q$ correlator of the BK case it equals
$C_f=\frac{N_c^2-1}{2N_c}$ or for a $gg$ dipole it would be $C_A=N_c$.
The gluon produced in the evolution step is denoted by $\Tilde U_{\bm
  z}$ and is of course always in the adjoint representation.

The hierarchies based on the dipole equations~(\ref{eq:prefactUR}) are
by no means all independent (group constraints and coincidence limits
may reveal that the $Y$ dependence of the same multi--$U$ correlator
does appear in several hierarchies), nor do they exhaust all the
information contained in JIMWLK evolution (for instance operators with
non-vanishing triality are absent from the family of dipole
hierarchies). What is important here is that JIMWLK evolution treats
this multitude of correlator equations consistently -- as long as no
truncation assumptions are made.

The BK approximation greatly simplifies this intricately interlinked
set of hierarchies and may not capture all of its features in the
process: since there is no simple generalization of the Fierz
identity~(\ref{eq:Fierz}) for $\bigl\langle \big[ \Tilde
U_{\bm{z}}\big]^{a b}\ \overset{{\scriptscriptstyle\cal R}}\tr(
\overset{{\scriptscriptstyle\cal R}} t^a
\overset{{\scriptscriptstyle\cal R}} U_{\bm{x}}
\overset{{\scriptscriptstyle\cal R}} t^b
\overset{{\scriptscriptstyle\cal R}}
U^\dagger_{\bm{y}})\bigr\rangle(Y)$ in an arbitrary representation
${\cal R}$, it may become impossible to consistently generalize the BK
approximation to even this class of equations, despite the fact that
one can write expressions for the BK (large $N_c$) limit of generic
dipole operators. (For the cases of ${\cal R}$ being fundamental or
adjoint representations the BK approximation is indeed possible and is
done routinely.) To consistently include all equations
(\ref{eq:prefactUR}) is to go at least one step beyond the BK
approximation and below, in Sect. \ref{sect:gauss}, we shall see that
this can indeed be achieved quite elegantly.

The key feature satisfied by JIMWLK evolution that is violated by BK
factorization beyond the leading-$N_c$ limit is a set of group identities for
the three point correlators $\bigl\langle \big[\Tilde U_{\bm{z}}\big]^{a b}\
\overset{{\scriptscriptstyle\cal R}}\tr( \overset{{\scriptscriptstyle\cal R}}
t^a \overset{{\scriptscriptstyle\cal R}} U_{\bm{x}}
\overset{{\scriptscriptstyle\cal R}} t^b \overset{{\scriptscriptstyle\cal R}}
U^\dagger_{\bm{y}})\bigr\rangle(Y)$ on the right-hand side of the evolution
equations (\ref{eq:prefactUR}). In what follows we will generically use the
term {\sl coincidence limits} to refer to the limits were any pair of points
($\bm x$ and $\bm y$, $\bm x$ and $\bm z$, or $\bm z$ and $\bm y$) or all
three of them ($\bm x$, $\bm y$, and $\bm z$) coincide with each other. At the
coincidence limits the correlator $\bigl\langle \big[\Tilde U_{\bm{z}}\big]^{a
  b}\ \overset{{\scriptscriptstyle\cal R}}\tr(
\overset{{\scriptscriptstyle\cal R}} t^a \overset{{\scriptscriptstyle\cal R}}
U_{\bm{x}} \overset{{\scriptscriptstyle\cal R}} t^b
\overset{{\scriptscriptstyle\cal R}} U^\dagger_{\bm{y}})\bigr\rangle(Y)$
should inherit relationships that JIMWLK evolution respects on the operator
level (see Appendix~\ref{sec:gener-coinc-limits} for their derivation):
\begin{subequations}
  \label{eq:3-point-generic-coincidence}
\begin{align}
  \label{eq:generic-x=y-limit}
\lim\limits_{y\to x}     \big[\Tilde U_{\bm{z}}\big]^{a b}\
   \overset{{\scriptscriptstyle\cal R}}\tr(
\overset{{\scriptscriptstyle\cal R}} t^a 
\overset{{\scriptscriptstyle\cal R}} U_{\bm{x}}
\overset{{\scriptscriptstyle\cal R}} t^b 
\overset{{\scriptscriptstyle\cal R}}
  U^\dagger_{\bm{y}})
& =  C_{\cal R} \frac{d_{\cal R}}{d_A}
\Tilde\tr\left(\Tilde U_{\bm z} \Tilde U_{\bm x}^\dagger \right)
\ ,
\\
  \label{eq:z=x,y-limit}
\lim\limits_{
      \bm z\to \bm y\ \text{or}\ \bm x}    
\big[\Tilde U_{\bm{z}}\big]^{a b}\
   \overset{{\scriptscriptstyle\cal R}}\tr(
\overset{{\scriptscriptstyle\cal R}} t^a 
\overset{{\scriptscriptstyle\cal R}} U_{\bm{x}}
\overset{{\scriptscriptstyle\cal R}} t^b 
\overset{{\scriptscriptstyle\cal R}}
  U^\dagger_{\bm{y}}) 
& = C_{\cal R}   \,
\overset{{\scriptscriptstyle\cal R}}\tr(
\overset{{\scriptscriptstyle\cal R}} U_{\bm{x}}
\overset{{\scriptscriptstyle\cal R}} U_{\bm{y}}^\dagger
) 
\ ,
\\
\lim\limits_{
      \bm z\to \bm y;\ \bm y\to \bm x}    
\big[\Tilde U_{\bm{z}}\big]^{a b}\
   \overset{{\scriptscriptstyle\cal R}}\tr(
\overset{{\scriptscriptstyle\cal R}} t^a 
\overset{{\scriptscriptstyle\cal R}} U_{\bm{x}}
\overset{{\scriptscriptstyle\cal R}} t^b 
\overset{{\scriptscriptstyle\cal R}}
  U^\dagger_{\bm{y}}) 
& = C_{\cal R} \, d_{\cal R}
\ ,
\end{align}
\end{subequations}
where $d_{\cal R}$ stands for the dimension of the representation
($d_f=N_c$ for the fundamental representation, $d_A=N_c^2-1$ for
adjoint, etc.) and $\Tilde\tr$ denotes the trace in the adjoint
representation.  While the third statement is merely a normalization
statement, the first two are remarkable: we read off that in the limit
of small parent dipole the three point operator on the left-hand side
reduces to a gluon dipole, no matter what representation ${\cal R}$
refers to, while in the limit $\bm z\to\bm x$ or $\bm y$ it reduces to
an ${\cal R}\Bar{\cal R}$-dipole.  The latter, (\ref{eq:z=x,y-limit}),
is crucial to ensure the real virtual cancellations
in~(\ref{eq:prefactUR}).

For {\em correlators}, the implications
of~\eqref{eq:3-point-generic-coincidence} go far beyond the isolated points
featuring in the limits shown. Since the correlation (saturation) length
$R_s=1/Q_s$ is the only dimensionful parameter, the only scale in the
problem,~\eqref{eq:3-point-generic-coincidence} determines the generic
behavior of $\langle \big[\Tilde U_{\bm{z}}\big]^{a b}\
\overset{{\scriptscriptstyle\cal R}}\tr( \overset{{\scriptscriptstyle\cal R}}
t^a \overset{{\scriptscriptstyle\cal R}} U_{\bm{x}}
\overset{{\scriptscriptstyle\cal R}} t^b \overset{{\scriptscriptstyle\cal R}}
U^\dagger_{\bm{y}}) \rangle(Y)$ in all of configuration space.

\begin{figure}[!t]
  \centering
  \includegraphics[height=6cm]{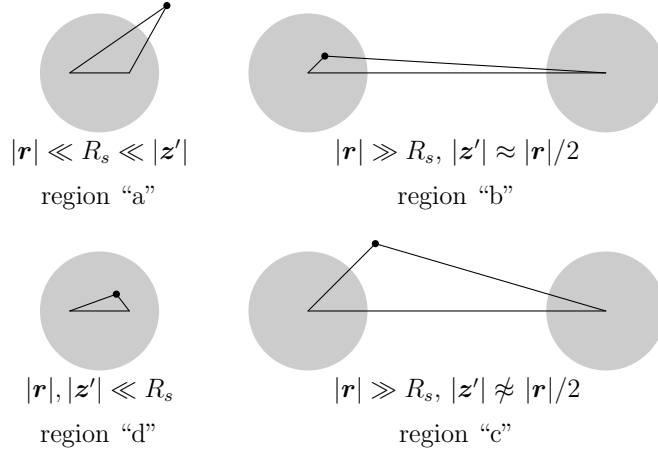}
  \caption{Schematic representation of physically distinct cases for
    correlator configurations. The shaded areas denote regions of size $R_s$,
    the particle coordinates $\bm x$, $\bm z'$, $\bm y$ are at the corners of
    the triangles with the gluon position $\bm z'$ marked by a dot.}
  \label{fig:scalecases}
\end{figure}

\begin{figure}[!thb]
  \centering
  \includegraphics[width=.49\textwidth]{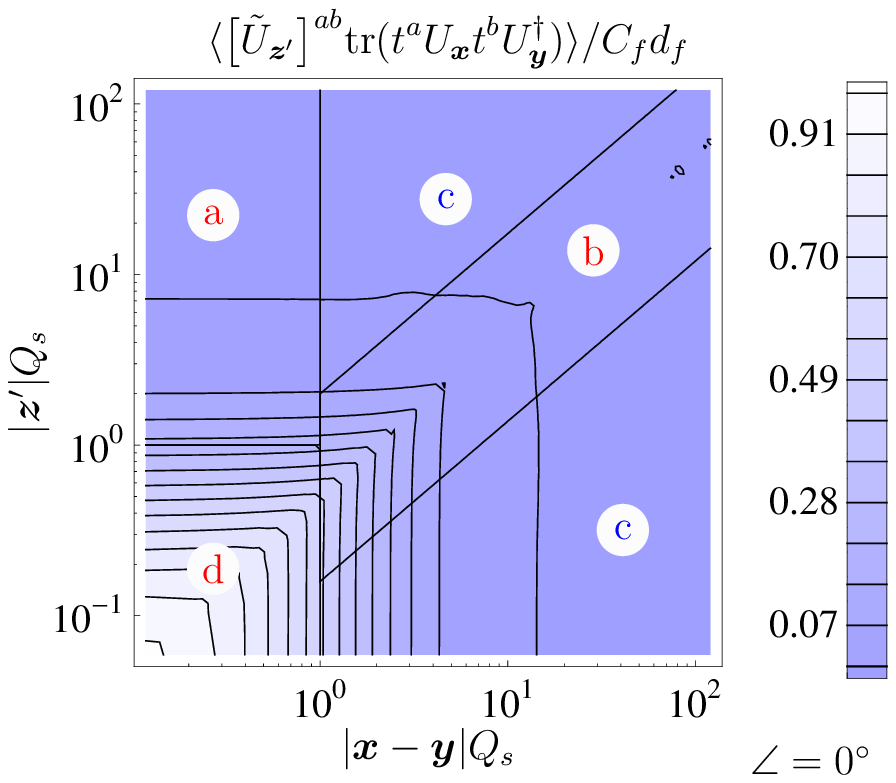}\hfill
  \includegraphics[width=.49\textwidth]{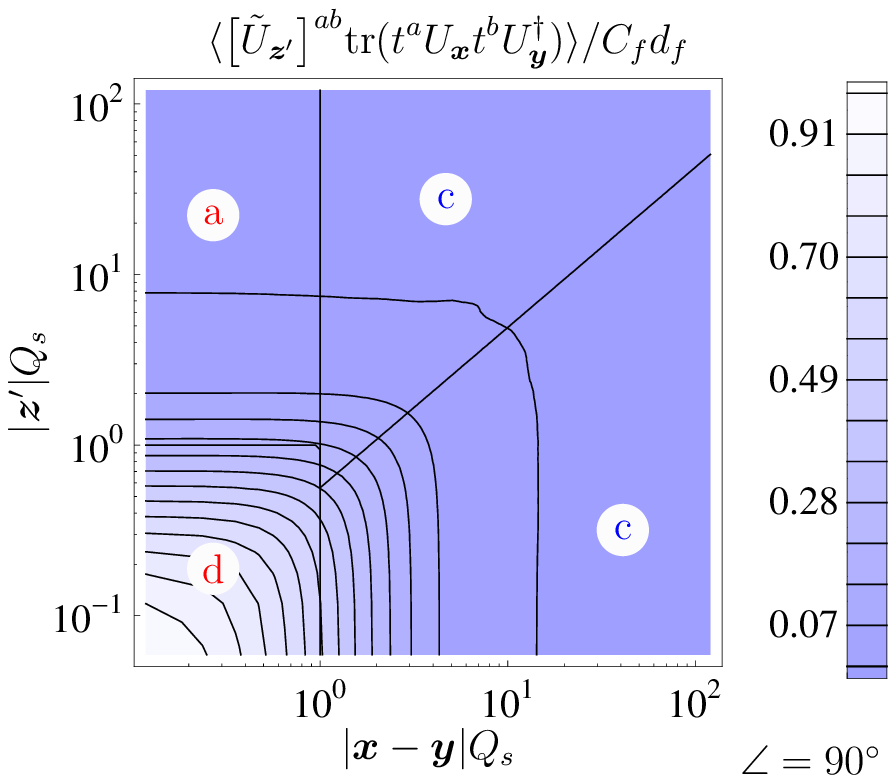}
  \caption{Behavior of three point correlator as discussed in
    Eq.~\eqref{eq:corr-regions} taken from our numerical solution of the
    JIMWLK equation at some intermediate $Y$.  $\bm z'$ is varied along rays
    of fixed angle with respect to $\bm r$ ($0^\circ$ and $90^\circ$), c.f.
    Fig.~\ref{fig:z-plane}. The regions are labeled in correspondence to
    Fig.~\ref{fig:scalecases}, the correlators display the generic behavior
    anticipated in Eq.~\eqref{eq:corr-regions}. Note that region ``b'' is only
    present near $0^\circ$ and completely disappears for $90^\circ$.}
  \label{fig:regions}
\end{figure}

Configuration space is first divided into two classes in which $\bm r$ is
either smaller or larger than $R_s$.  For each of these classes one has to
distinguish two cases according to whether the distance between the gluon and
the nearest quark is larger or smaller than $R_s$. The configurations are
shown in Fig.~\ref{fig:scalecases} and exhaust all physically different
situations (labels ``a'' through ``d'' in the figure are in correspondence to
the equation labels in (\ref{eq:corr-regions}) below).
One infers from~\eqref{eq:3-point-generic-coincidence} that in regions ``a''
and ``b'' the falloff is dipole-like
\begin{subequations}
  \label{eq:corr-regions}
\begin{align}
  \label{eq:gluon-dipole-range}
  & 
 |\bm r| \ll R_s \ll |\bm z'|  : 
& 
\hspace{-.5cm}\langle
  \big[\Tilde U_{\bm{z}}\big]^{a b}\ \overset{{\scriptscriptstyle\cal
      R}}\tr( \overset{{\scriptscriptstyle\cal R}} t^a
  \overset{{\scriptscriptstyle\cal R}} U_{\bm{x}}
  \overset{{\scriptscriptstyle\cal R}} t^b
  \overset{{\scriptscriptstyle\cal R}} U^\dagger_{\bm{y}}) \rangle(Y)
  \approx C_{\cal R} \frac{d_{\cal R}}{d_A} \langle
  \Tilde\tr\left(\Tilde U_{\bm z} \Tilde U_{\bm x}^\dagger \right)
  \rangle(Y) \ ,
  \\
  \label{eq:quark-dipole-range}
& 
|\bm r| \approx 2\,|\bm z'| \gg R_s :
&
\hspace{-.5cm} \langle
\big[\Tilde U_{\bm{z}}\big]^{a b}\
   \overset{{\scriptscriptstyle\cal R}}\tr(
\overset{{\scriptscriptstyle\cal R}} t^a 
\overset{{\scriptscriptstyle\cal R}} U_{\bm{x}}
\overset{{\scriptscriptstyle\cal R}} t^b 
\overset{{\scriptscriptstyle\cal R}}
  U^\dagger_{\bm{y}}) 
\rangle(Y)
\approx C_{\cal R}   
\langle
\overset{{\scriptscriptstyle\cal R}}\tr(
\overset{{\scriptscriptstyle\cal R}} U_{\bm{x}}
\overset{{\scriptscriptstyle\cal R}} U_{\bm{y}}^\dagger
) 
\rangle(Y)
\ ,
\intertext{i.e., $\langle
\big[\Tilde U_{\bm{z}}\big]^{a b}\
   \overset{{\scriptscriptstyle\cal R}}\tr(
\overset{{\scriptscriptstyle\cal R}} t^a 
\overset{{\scriptscriptstyle\cal R}} U_{\bm{x}}
\overset{{\scriptscriptstyle\cal R}} t^b 
\overset{{\scriptscriptstyle\cal R}}
  U^\dagger_{\bm{y}}) 
\rangle(Y)$ 
vanishes like a gluon or ${\cal R}\Bar{\cal R}$-dipole where $|\bm
  z'|\gg R_s \gg |\bm r|$ or $|\bm r| \approx |\bm z'|/2 \gg R_s$ (in 
  the latter region
  the gluon is near either the $q$ or $\Bar q$, implying either 
  $R_s\gg |\bm z-\bm y|\ \text{or}\ |\bm z-\bm x|$ respectively. This only
  occurs when the angle between $\bm r$ and $\bm z'$ is near $0^\circ$). 
  It also vanishes trivially in region ``c'', where $|\bm r|\gg R_s$, 
  {\em and} the
  gluon is far from both $q$ and $\Bar q$ ($|\bm z'|
  \approx\hspace*{-3.5mm}/\hspace*{2.0mm} |\bm r|/2$). In this region all
  three inter-particle distances are large and force exponential 
  suppression, although~\eqref{eq:3-point-generic-coincidence} gives no 
  additional information about the falloff, leaving us with
}
    \label{eq:three-scales-large}
&    
|\bm r|\gg R_s, |\bm z'|
  \approx\hspace*{-3.5mm}/\hspace*{2.0mm} |\bm r|/2:
&
\hspace{-.5cm}
\langle\big[\Tilde U_{\bm{z}}\big]^{a b}\
   \overset{{\scriptscriptstyle\cal R}}\tr(
\overset{{\scriptscriptstyle\cal R}} t^a 
\overset{{\scriptscriptstyle\cal R}} U_{\bm{x}}
\overset{{\scriptscriptstyle\cal R}} t^b 
\overset{{\scriptscriptstyle\cal R}}
  U^\dagger_{\bm{y}}) 
\rangle(Y) \to 0
\ .
\intertext{      This leaves only one region, labeled ``d'' in  Fig.~\ref{fig:scalecases}, 
  in which the contributions are not suppressed. In this remaining region, 
  all scales 
  are small simultaneously, as one would naively expect in a system with
  a finite correlation length $R_s$.  In region ``d'' we have}
\label{eq:corr-const-range}
& 
|\bm z'|, |\bm r| \lsim R_s : 
&
\hspace{-.5cm}
\langle\big[\Tilde U_{\bm{z}}\big]^{a b}\
   \overset{{\scriptscriptstyle\cal R}}\tr(
\overset{{\scriptscriptstyle\cal R}} t^a 
\overset{{\scriptscriptstyle\cal R}} U_{\bm{x}}
\overset{{\scriptscriptstyle\cal R}} t^b 
\overset{{\scriptscriptstyle\cal R}}
  U^\dagger_{\bm{y}}) 
\rangle(Y)
\lsim C_{\cal R} d_{\cal R}
\ .
\end{align}
\end{subequations}  
Fig.~\ref{fig:regions} illustrates this theoretical discussion with contour
plots of the three point function as obtained from actual JIMWLK
simulations. The plots show dependence on $q\Bar q$ separation $|\bm r|$ and
the distance of the gluon location with respect to the $q\Bar q$ midpoint
$|\bm z'|$, with $\bm z'$ perpendicular to and parallel to $\bm r$, i.e.,
along two of the lines indicated in Fig.~\ref{fig:z-plane}. One may notice
that the contributions on the axes, $|\bm r|=0$ and $|\bm z'|=0$ have no
angular dependence: the first corresponds to zero size parent dipoles in which
case $\bm r$ does not single out any direction to refer to, the second keeps
$\bm z$ firmly in the middle of the $q\Bar q$ pair while varying its size so
that again the angle does not play a role.

Eqs.~(\ref{eq:3-point-generic-coincidence}) and~\eqref{eq:corr-regions}
represent but one example of a much larger set of group constraints for more
complicated correlators that are all inherently true in the full JIMWLK
setting, but broken by the BK factorization.  The BK truncation is geared
towards quark dipoles (where ${\cal R}$ is the fundamental representation),
where it approximates the Fierz identity~(\ref{eq:Fierz}) by dropping the
$1/N_c$ term
\begin{equation}
  \label{eq:Fierz-approx}
  \big[\Tilde U_{\bm{z}}\big]^{a b}
  2 \tr(t^a U_{\bm{x}}t^b U^\dagger_{\bm{y}}) \approx
  \tr( U_{\bm{x}}U^\dagger_{\bm{z}})
           \tr( U_{\bm{z}}U^\dagger_{\bm{y}})
           +{\cal O}(1/N_c)
\end{equation}
This distorts the coincidence limit of the quark dipole version of
Eqs.~(\ref{eq:3-point-generic-coincidence}) into their large $N_c$
approximations
\begin{subequations}
  \label{eq:BK-limits}
\begin{align}
  \label{eq:BK-x=y-limit}
\lim\limits_{y\to x}   
\tr( U_{\bm{x}}U^\dagger_{\bm{z}})
           \tr( U_{\bm{z}}U^\dagger_{\bm{y}})
& =  \left|\tr( U_{\bm{x}}U^\dagger_{\bm{z}}) \right|^2
\\
\lim\limits_{\bm z\to\bm y\ \text{or}\ \bm x}   
\tr( U_{\bm{x}}U^\dagger_{\bm{z}})
           \tr( U_{\bm{z}}U^\dagger_{\bm{y}})
& =\  \tr( U_{\bm{x}}U^\dagger_{\bm{y}}) N_c
\\ 
 \lim\limits_{\bm z\to\bm x;\ \bm y\to\bm x} \tr( U_{\bm{x}}U^\dagger_{\bm{z}})
           \tr( U_{\bm{z}}U^\dagger_{\bm{y}})
& = N_c^2 ,
\end{align}
\end{subequations}
i.e., it approximates the gluon dipole operator on the right-hand side
of Eq.~(\ref{eq:generic-x=y-limit}) by the square of the quark dipole
operator, and replaces the constants in the remaining equations by
their large $N_c$ counterparts.

For correlators, the implications of~\eqref{eq:BK-limits} mirror the
conclusions drawn in Eq.~\eqref{eq:corr-regions} up to corrections of order
$1/N_c^2$, hence one naively expects the factorization violations to be of
precisely that order, unless there is a stronger cancellation at work in the
coefficient of that $1/N_c^2$ term. Eq.~\eqref{eq:corr-regions} contains all
that is needed to assess this issue if one
uses~\eqref{eq:Fierz} to recast $\Delta$ in terms of two and three point
correlators only (the first two terms represent the unfactored correlator):
\begin{align}
  \label{eq:Delta-defierzed}
  \Delta_{\bm{x z y}}(Y)  =\frac1{N_c^2}\Big[   
 \langle\big[\Tilde U_{\bm z}\big]^{a b}\, 
2\,\tr(  t^a
 U_{\bm{x}}
 t^b 
U^\dagger_{\bm{y}})\rangle
+  \frac{\langle\tr(U_{\bm x} U_{\bm y}^\dagger)\rangle}{N_c}
-  \langle\tr(U_{\bm x} U_{\bm z}^\dagger)\rangle
\langle\tr(U_{\bm z} U_{\bm y}^\dagger)\rangle
 \Big](Y)
\ .
\end{align} 
The four distinct regions of Fig.~\ref{fig:scalecases} and
Eq.~\eqref{eq:corr-regions} can then be addressed in turn (all individual
correlators are real and positive):
\begin{itemize}
\item Region ``a'', $|\bm z'|\gg R_s \gg |\bm r|$: Both the first and the last
  term inside the brackets of~\eqref{eq:Delta-defierzed} are exponentially
  small, but the second term approaches $1$. In the extreme case $|\bm
  z'|\to\infty$, $|\bm r|\to 0$ one finds $\Delta_{\bm{x z y}}(Y)\to 1/N_c^2$.
  Contrary to $\langle\big[\Tilde U_{\bm z}\big]^{a b}\, 2\,\tr( t^a
  U_{\bm{x}} t^b U^\dagger_{\bm{y}})\rangle$, $\langle\tr(U_{\bm x} U_{\bm
    z}^\dagger)\tr(U_{\bm z} U_{\bm y}^\dagger)\rangle$ contains a $\bm
  z$-independent additive term that survives this limit. If region ``a'' were
  to contribute to evolution at all, this would destroy infrared safety of
  JIMWLK evolution (see below).
\item Region ``b'', $|\bm r| \approx 2\,|\bm z'| \gg R_s$ (gluon near $q$ or
  $\Bar q$):
        Since there are always {\em two} large distances involved, {\em all three}
  of the terms in~\eqref{eq:Delta-defierzed} are exponentially suppressed and
  the contribution is naturally much smaller than $1/N_c^2$.
\item Region ``c'', $|\bm r|\gg R_s, |\bm z'|
  \approx\hspace*{-3.5mm}/\hspace*{2.0mm} |\bm r|/2$: With all three
  inter-particle distances large, all terms are exponentially
  suppressed individually, rendering their sum much smaller than $1/N_c^2$.
\item Region ``d'', $|\bm z'|, |\bm r| \lsim R_s$: The terms inside the
  brackets are order $N_c^2-1$, $1$, and $N_c^2$ respectively.  Moreover, in
  the strict coincidence limit $\bm x=\bm y=\bm z$, they cancel exactly! This
  guarantees a very strong (albeit not exponential) reduction of the
  coefficient in front of $1/N_c^2$. The cancellation is slightly
  less pronounced farther from exact coincidence, for scales $|\bm z'|, |\bm
  r|$ of order $R_s$, before  large distance damping at the boundary
  to the previous regions sets in.
\end{itemize}
One concludes that $\Delta$ is strictly bounded from above by $1/N_c^2$, but
there is only one region left in which this bound is actually reached --
region ``a''. In {\em all} other regions strong cancellations reduce the
contributions to values significantly below this bound.

So far we have used general arguments based on coincidence limits
(\ref{eq:3-point-generic-coincidence}) and on the effects of
saturation on the dipole scattering amplitude (two-Wilson line
correlators) to argue that $\Delta$ is in fact much smaller than
$1/N_c^2$ for much of its phase space. To understand the impact of
$\Delta$ on the evolution let us rewrite \eqref{eq:preBKS} using
\eqref{eq:Delta-old}
\begin{align}
  \label{eq:BK+D}
  \frac{d}{d Y} \langle \Hat S_{\bm{x y}} \rangle(Y) =\frac{\alpha_s
    N_c}{2\pi^2}\int d^2z\, \ {\cal K}_{\bm{x z y}} \ \left[ \langle
    \Hat S_{\bm{x z}} \rangle (Y) \ \langle \Hat S_{\bm{z y}} \rangle
    (Y) - \langle \Hat S_{\bm{x y}} \rangle(Y) + \Delta_{\bm{x z
        y}}(Y) \right].
\end{align}
Eq.~(\ref{eq:BK+D}) shows how $\Delta$ enters the full, untruncated
evolution equations.

As we saw above, somewhat surprisingly, region ``a'' where the maximal
possible factorization violation occurs is characterized by small
parent dipole size $|\bm r|Q_s\to 0$ but with the gluon produced far
away, with $|\bm z'|Q_s\gg 1$.  This region, however, has no impact on
evolution at all: it is completely power--suppressed by the evolution
kernel (\ref{K}) in \eqref{eq:BK+D} which goes to zero as the sixth power
of distances involved:
\begin{equation}
  \label{eq:kernel-power-supp}
  \frac{(\bm x-\bm y)^2}{(\bm x-\bm z)^2(\bm z-\bm y)^2}\approx
  \frac{\bm r^2}{(\bm z')^4}\to 0 
\ .
\end{equation}
Were it not for this kernel suppression, JIMWLK evolution (represented via the
$q\Bar q$ Balitsky hierarchy) would receive large distance contributions from
this region -- infrared safety would be lost.  Notably, region ``a'' is the
only large distance region that requires suppression from the kernel. The
remaining large distance regions (``b'' and ``c'') are exponentially
suppressed on the correlator level and automatically decouple from
evolution. That leaves the last region (region ``d'') with its strong
cancellation of contributions as dictated by the properties of the coincidence
limits: it remains as the sole channel through which the non-factorized
contributions affect the energy dependence of the $q\Bar q$-dipole. It is this
region which was quoted in~\cite{Rummukainen:2003ns} to contribute the
``typical'' factorization violations without connecting this to the
coincidence limits~\eqref{eq:3-point-generic-coincidence}.

While the argument given in this section does not give a parametric
estimate for the size of the factorization violations, it does explain
why the contributions are naturally {\em much} smaller than $1/N_c^2$.
We see that factorization violation $\Delta_{\bm{x z y}}$ is bounded
by $1/N_c^2$ from above. However this $1/N_c^2$ value is reached only
in a small subset of $({\bm x}, {\bm y},{\bm z})$ configuration space
(in region ``a''), which is suppressed by the evolution kernel. The
relative suppression of the integral of the factorization violation
over all $\bm z$'s in \eqref{eq:BK+D} compared to the first term on
the right-hand side of the equation is therefore much stronger than
the $1/N_c^2$ one would naively expect. Note that the generic
arguments given here also do not allow to determine the sign of the
contribution and thus do not allow to infer if one should expect
JIMWLK evolution to be slower or faster than the factorized BK
truncation.

We might now just push ahead and map out configuration space of the JIMWLK
3-point correlators using numerical results from our simulations, to
systematically supplement the numerical results of\cite{Rummukainen:2003ns}
and Fig.~\ref{fig:fact-viols-fixed-xmy} with contributions from the regions
not shown there (numerical results will be shown in Figs.~\ref{fig:DeltaC}
and~\ref{fig:KDeltaC}).  Let us instead first give a simple generalization of
BK factorization that treats the set of equations~(\ref{eq:prefactUR})
consistently and respects the coincidence
limits~(\ref{eq:3-point-generic-coincidence}). This generalization will
restore at least part of the true factorization violation and respect the
configuration space pattern deduced from~\eqref{eq:corr-regions}
and~\eqref{eq:Delta-defierzed}. This will likely improve agreement with JIMWLK
evolution and give some insight into the question in which direction evolution
speed is changed by the factorization violations.

\section{Gaussian truncation of JIMWLK}
\label{sect:gauss}

\subsection{A step beyond BK}

\label{sec:rest-coinc-limits}

Our argument for the suppression of $1/N_c^2$ corrections in the previous
section were based on saturation effects and coincidence limits. We observed
that the BK equation, while incorporating the saturation effects, violates the
coincidence limits at the subleading $N_c$ level.

Approaches that both incorporate saturation physics and respect the
coincidence limits of the general argument given in Sec.~\ref{sec:orig-fact}
are well established in the literature. They take the form of variants of the
McLerran-Venugopalan model~\cite{McLerran:1994ni, McLerran:1994ka,
  McLerran:1994vd} and the closely related Glauber-Mueller
approximation to high energy scattering~\cite{Mueller:1986wy,
  Mueller:1994rr, Mueller:1994jq}, which can be rigorously
established by summing QCD diagrams {\em without} taking into account small
$x$ evolution.  All these descriptions fall into a class of approximations of
the JIMWLK average $\langle \ldots \rangle(Y)$ over Wilson lines $U_{\bm x}$,
that is characterized by a longitudinally local Gaussian averaging procedure
that can be cast as
\begin{align}
  \label{eq:func-Gaussian-average}
  \langle \ldots \rangle(Y) = \exp\biggl\{ 
    -\frac12 \int\limits^Y dY' \int\! d^2x\, d^2y\ 
    G_{Y',\bm{x y}}\ \frac{\delta}{\delta A^{a +}_{\bm x,Y'}}
    \frac{\delta}{\delta A^{a +}_{\bm y,Y'}}
  \biggr\} \ldots
\ .
\end{align}
We refer the reader to~\cite{Weigert:2005us} for a discussion of how various
well known models can be recovered from the generic form shown in
Eq.~(\ref{eq:func-Gaussian-average}) by choosing specific expressions for
$G_{Y',\bm{x y}}$. In the quasi-classical limit $G$ encodes a two gluon
exchange with the target in the $t$-channel. That this same generic approach
automatically satisfies the coincidence limits has also been demonstrated
in~\cite{Weigert:2005us} for the case of ${\cal R}$ being the fundamental
representation.

Stepping beyond the quasi-classical limit in~\cite{Kovner:2001vi}, Kovner and
Wiedemann have suggested an all $N_c$ evolution equation that merges BK
principles with the Gaussian treatment of correlators incorporated in
Eq.~(\ref{eq:func-Gaussian-average}) by what amounts to applying the averaging
prescription to quark and gluon dipoles.

In fact, Eq.~(\ref{eq:func-Gaussian-average}) allows us to extend the
treatment of~\cite{Kovner:2001vi} beyond the specific case of quark
(fundamental) and gluon dipole evolution. This results in a self-consistent
treatment of the evolution of all generic dipole operators in which one
replaces the $q$ and $\Bar q$ by colored objects in arbitrary representations
${\cal R}$ and $\Bar{\cal R}$.  Doing so, one finds completely generic
expressions for the previously discussed correlators (see
Appendix~\ref{sec:gaussian-averages} for calculational details and also
\cite{Jalilian-Marian:1997xn, Kopeliovich:1998nw, Kopeliovich:1999am,
  Kovchegov:2001ni, Kovner:2001vi} for similar calculations):
\begin{subequations}
  \label{eq:simpleGcorr}
\begin{align}
  \label{eq:UUdaggersol}
  \langle \overset{{\scriptscriptstyle\cal R}}\tr(
\overset{{\scriptscriptstyle\cal R}} U_{\bm{x}}
\overset{{\scriptscriptstyle\cal R}}
  U^\dagger_{\bm{y}})
\rangle(Y) = \ d_{\cal R}\ e^{-C_{\cal R}{\cal G}_{Y,\bm{x y}}} 
\end{align}
\begin{align}
  \label{eq:UtrtUtUdagger}
\langle 
\big[\Tilde U_{\bm{z}}\big]^{a b} 
\overset{{\scriptscriptstyle\cal R}}\tr(
\overset{{\scriptscriptstyle\cal R}} t^a 
\overset{{\scriptscriptstyle\cal R}} U_{\bm{x}}
\overset{{\scriptscriptstyle\cal R}} t^b 
\overset{{\scriptscriptstyle\cal R}}
  U^\dagger_{\bm{y}})
\rangle(Y) = C_{\cal R} d_{\cal R}\ e^{
  -\frac{N_c}2\left(
  {\cal G}_{Y,\bm{x z}} + {\cal G}_{Y,\bm{z y}}- {\cal G}_{Y,\bm{x y}}  
\right)
-C_{\cal R} {\cal G}_{Y,\bm{x y}} 
}  
\end{align}
\end{subequations}
For convenience we have introduced
\begin{align}
  \label{eq:calGdef}
  {\cal G}_{Y,\bm{x y}} := \int\limits^Y\! dY'\biggr(G_{Y',\bm{x y}}
    -\frac12\bigl(G_{Y',\bm{x x}}+G_{Y',\bm{y y}}\bigr)\biggr)
\end{align}
to denote the combination in which the t-channel gluons enter these
expressions. Note that ${\cal G}_{Y,\bm{x x}}\equiv 0$ as required by
consistency in~(\ref{eq:UUdaggersol}). Quick inspection reveals
that~(\ref{eq:UtrtUtUdagger}) indeed complies
with~(\ref{eq:3-point-generic-coincidence}) as advertised.  In fact, this
property is not specific to this particular set of correlators. Any correlator
calculated using~(\ref{eq:func-Gaussian-average}) (or any generalization
thereof) will automatically satisfy all necessary group constraints by
construction.

This procedure then is a candidate to generalize the BK factorization in
which one simply trades an evolution equation for $\langle \Hat S_{\bm{x
    y}}\rangle(Y)$ for an evolution equation for ${\cal G}_{Y,\bm{x y}}$. The
procedure at least qualitatively repairs the flaw that is the source of
factorization violation in the BK equation. We will find below that it
provides quite good qualitative insights on factorization violation but
is not sufficient to obtain quantitatively correct results.

The equation for ${\cal G}$ has already been derived in~\cite{Weigert:2005us}
(and in a somewhat different form earlier in~\cite{Kovner:2001vi}\footnote{To
  connect with the form given in~\cite{Kovner:2001vi}, Eq.~(5.3), one
  should reconstruct the evolution equation for the $q\Bar q$-dipole operator
  by multiplying~(\ref{eq:tilde-G-evo-short}) with $\exp(-C_f {\cal
    G}_{Y,\bm{x y}})$ and note that our ${\cal G}_{Y,\bm{x y}}$ corresponds to
  $v(\bm x,\bm y)$ in~\cite{Kovner:2001vi}.  }), starting from the $q\Bar
q$-dipole evolution equation~(\ref{eq:preBKU}).  We reiterate that this
average treats {\em all} dipole equations consistently:
Inserting~(\ref{eq:simpleGcorr}) into the generic dipole evolution
equation~(\ref{eq:prefactUR}) yields one and the same equation for ${\cal G}$,
\begin{equation}
  \label{eq:tilde-G-evo-short}
 \frac{d}{d Y} {\cal G}_{Y,\bm{x y}}  =  \frac{\alpha_s}{\pi^2} \int\!\!d^2z\  
 {\cal K}_{\bm{x z y}} \biggl(
 1-  e^{-\frac{ N_c}{2} \bigl(
 {\cal G}_{Y,{\bm{x z}}} +{\cal G}_{Y,{\bm{y z}}}
 - {\cal G}_{Y,{\bm{x y}}}\bigr)}
\biggr)
\ ,
\end{equation}
{\em irrespective} of the representation ${\cal R}$. Note 
that~\eqref{eq:tilde-G-evo-short} 
is similar to (though not exactly the same as) the 
Ayala-Gay Ducati-Levin (AGL) evolution 
equation~\cite{Ayala:1996em, Ayala:1996ed, AyalaFilho:1997du}.

Contrary to BK evolution which systematically discards all $1/N_c^2$
suppressed terms contained in JIMWLK, the Gaussian truncation has no
expansion parameter justifying the approximation. Nevertheless we
expect it to lead to a good approximation of JIMWLK evolution since
\begin{itemize}
\item the equation incorporates a subset of these $1/N_c^2$ corrections that
  is sufficient to restore the coincidence limits;
\item the low density limit of Eq.~(\ref{eq:tilde-G-evo-short}) (viewed as its
  small $\cal G$ limit) reduces to the BFKL equation;
\item it has a large $N_c$ limit that is compatible with the BK equation as
  will be seen below.
\end{itemize}

Despite \eqref{eq:tilde-G-evo-short} being surprisingly more generic than
the BK equation, in the sense that the procedure addresses arbitrary
dipoles irrespective of representation, one remains with an
approximation to the true JIMWLK evolution, and does not obtain an
exact solution of the JIMWLK equation: the evolution equation for $\langle
\big[\Tilde U_{\bm{z}}\big]^{a b} \overset{{\scriptscriptstyle\cal
    R}}\tr( \overset{{\scriptscriptstyle\cal R}} t^a
\overset{{\scriptscriptstyle\cal R}} U_{\bm{x}}
\overset{{\scriptscriptstyle\cal R}} t^b
\overset{{\scriptscriptstyle\cal R}} U^\dagger_{\bm{y}}) \rangle(Y) $
resulting from JIMWLK would impose additional conflicting conditions
on ${\cal G}$, and can only be satisfied by introducing degrees of
freedom beyond ${\cal G}$. The Gaussian truncation still deviates from
JIMWLK evolution at the level of evolution equations for three point
functions.

It is worth noting two particular features that the
average~(\ref{eq:func-Gaussian-average}) entails. First,
Eq.~(\ref{eq:UUdaggersol}) implies Casimir scaling for dipole
correlators:\footnote{Approximate Casimir scaling  has been observed for
  Wilson line correlators in the
  context of heavy quark potentials in~\cite{Bali:1999hx,Bali:2000un}.} 
Given two representations ${\cal R}_{1,2}$ the normalized dipole correlators
are related by a simple power law
\begin{align}
  \label{eq:casimir-scaling-app}
  \frac{1}{d_{{\cal R}_1}}\langle \overset{{\scriptscriptstyle\cal R}_1}\tr(
\overset{{\scriptscriptstyle\cal R}_1} U_{\bm{x}}
\overset{{\scriptscriptstyle\cal R}_1}
  U^\dagger_{\bm{y}})
\rangle(Y)
= \left(
 \frac{1}{d_{{\cal R}_2}}\langle \overset{{\scriptscriptstyle\cal R}_2}\tr(
\overset{{\scriptscriptstyle\cal R}_2} U_{\bm{x}}
\overset{{\scriptscriptstyle\cal R}_2}
  U^\dagger_{\bm{y}})
\rangle(Y)\right)^{C_{{\cal R}_1}/C_{{\cal R}_2}}
\ .
\end{align}
Second, somewhat surprisingly, Eq.~(\ref{eq:tilde-G-evo-short}) can be mapped
back onto the BK equation. This implies that the dynamical content of the
Gaussian truncation is the same as that of the BK equation. The main
improvement is how this information is mapped onto the correlators. As we
shall see, this leads to a slightly better approximation of JIMWLK results.
On the practical side, this turns into a time saver: one can recycle the
numerical tools written to solve the BK equation, provided one relates
correlators and initial conditions accordingly.

One way to see that the dynamical content is the same is based on a
simple {\em re-parametrization} of the BK $S$-matrix in as close an
analogy to~(\ref{eq:UUdaggersol}) as possible. We write
\begin{align}
  \label{eq:S-BK-reparam}
  1-N_{Y,\bm{x y}}^{\text{BK}} = S_{Y,\bm{x y}}^{\text{BK}} = 
  \langle \tr( U_{\bm{x}} U^\dagger_{\bm{y}}) \rangle(Y)/N_c 
   = e^{-\frac{N_c}2  \Tilde{\cal G}_{Y,\bm{x y}}}  
\ .
\end{align}
$\Tilde{\cal G}_{Y,\bm{x y}}$ should be thought of as simply being {\em
  defined} by the solutions of the BK equation
through~\eqref{eq:S-BK-reparam}. The $N_c$ dependent constant is a convention
chosen in keeping with the $N_c$ lore of the Mueller dipole model and the BK
equation. Next one inserts this into the BK equation for $S$,
\begin{align}
  \label{eq:BK-S}
   \frac{d}{d Y}  S_{Y,\bm{x y}}^{\text{BK}}
   = &
\frac{\alpha_s N_c}{2 \pi^2} \int\!\!d^2z\  
{\cal K}_{\bm{x z y}} \
\Big(S_{Y,\bm{x z}}^{\text{BK}}S_{Y,\bm{z y}}^{\text{BK}}
-S_{Y,\bm{x y}}^{\text{BK}} \Big)
\ ,
\end{align}
and obtains
\begin{equation}
  \label{eq:tilde-G-evo-short-BK}
 \frac{d}{d Y} \Tilde{\cal G}_{Y,\bm{x y}}  =  \frac{\alpha_s }{  \pi^2} \int\!\!d^2z\  
 {\cal K}_{\bm{x z y}} \ \Big(
 1-  e^{-\frac{N_c}2 \big(
 \Tilde{\cal G}_{Y,{\bm{x z}}} +\Tilde{\cal G}_{Y,{\bm{y z}}}
 - \Tilde{\cal G}_{Y,{\bm{x y}}}\big)}
\Big)
\end{equation}
which is {\em identical} to~\eqref{eq:tilde-G-evo-short}, thus establishing
our claim of identical dynamical content, despite the different treatment of
correlators.

The procedure to obtain solutions for the Gaussian truncation that would serve
to determine, say, the $q\Bar q$--proton cross section for DIS at HERA would
then be to choose an initial condition for $S$, read off ${\cal G}$
via~(\ref{eq:UUdaggersol}), insert it in place of $\Tilde{\cal G}$
in~(\ref{eq:S-BK-reparam}) to determine the initial condition on
$S^{\text{BK}}$ to be used in the BK equation~(\ref{eq:BK-S}). After solving
Eq.~(\ref{eq:BK-S}) to obtain $S^{\text{BK}}$ at all rapidities one reverses
the procedure to recover ${\cal G}$ via~(\ref{eq:S-BK-reparam})
and~(\ref{eq:UUdaggersol}) at each rapidity $Y$. This ${\cal G}$ then
determines all correlators of the Gaussian truncation
through~\eqref{eq:func-Gaussian-average} and the special cases shown in
Eq.~(\ref{eq:simpleGcorr}).

As an immediate consequence we conclude that the asymptotic scaling shape for
the dipole correlators in the Gaussian truncation can be obtained from those
of the BK equation using a simple power law relationship
(\ref{eq:casimir-scaling-app}). As an immediate corollary also evolution
speeds of the Gaussian truncation and BK evolution must coincide in that
region. This link does not extend to the pre-asymptotic regime and we will see
that GT tends to be slower that BK in that range in
Sec.~\ref{sec:quant-cons-fact}.

Two further points are worth noting: (\i) one may recover BK
factorization (wherever it can be meaningfully applied) as the leading
$N_c$ contribution of the new procedure. (This can be verified for the
contributions entering the BK equation by taking the large $N_c$
contributions in the exponents in (\ref{eq:simpleGcorr}).) (\i\i) In
the small density limit, i.e., the limit of weak target fields where
${\cal G}$ is small its evolution equation,
Eq.~(\ref{eq:tilde-G-evo-short}), consistently reduces to the BFKL
equation for ${\cal G}$, in keeping with the underlying
interpretation.

In summary, one might think of the Gaussian truncation as a minimal
extension of the BK factorization to consistently incorporate group
constraints with an associated set of ``minimal'' subleading $1/N_c$
corrections without changing the dynamical content of the associated
evolution equation. Below we will frequently abbreviate Gaussian truncation
as GT.

\subsection{Factorization violations in the Gaussian truncation}

\label{sec:qual-struct-fact}

Now that, with JIMWLK, BK and GT, we have accumulated three different ways to
simulate small $x$ evolution for all of which the averaging procedure $\langle
\ldots \rangle(Y)$ leads to different results we need to refine our notations
to avoid confusion by distinguishing $\langle \ldots \rangle_J(Y)$, $\langle
\ldots \rangle_B(Y)$ and $\langle \ldots \rangle_G(Y)$ respectively. We also
are faced with different factorization violations and define
\begin{align}
  \label{eq:Delta-J}
  \Delta^J_{\bm{x z y}}(Y) := \bigl\langle
    \bigl( 
    \Hat S_{\bm{x z}} -\langle \Hat S_{\bm{x z}} \rangle_J(Y)
    \bigr)
    \bigl(
      \Hat S_{\bm{z y}} -\langle \Hat S_{\bm{z y}} \rangle_J(Y)
    \bigr)
    \bigr\rangle_J(Y)
\end{align}
to distinguish it from the analogous (non-vanishing) quantity taken in the
Gaussian truncation which we will denote $\Delta^G_{\bm{x z y}}(Y)$. Only in
BK this is set to zero by fiat: $\Delta^B_{\bm{x z y}}(Y) \equiv 0$.

$\Delta^J_{\bm{x z y}}(Y)$ is the only channel through which higher
order correlations of JIMWLK feed into the dipole equation of the
Balitsky hierarchy as can be made explicit by splitting \eqref{eq:preBKS}
into two parts as
\begin{align}
  \label{eq:dipole-Bal}
  \frac{d}{d Y} \langle \Hat S_{\bm{x y}} \rangle_J(Y)
  =&\ \frac{\alpha_s N_c}{2\pi^2}\int d^2z\,
  {\cal K}_{\bm{x z y}} \left(
 \langle \Hat S_{\bm{x z}} \rangle_J \langle \Hat S_{\bm{z y}} \rangle_J -\langle \Hat S_{\bm{x y}} \rangle_J      \right)(Y)
\notag \\ & \
+ \frac{\alpha_s N_c}{2\pi^2}\int d^2z\,
  {\cal K}_{\bm{x z y}}\  \Delta^J_{\bm{x z y}}(Y)
\ .
\end{align}
Without the $\Delta ^J$ term on the right-hand side this would reduce to the BK
equation and decouple from the rest of its Balitsky hierarchy. As such it is
also the only source of possible differences in evolution speed and
correlator shape between JIMWLK, BK and the Gaussian approximation. The
evolution equation of the latter can also be rendered in the
form~(\ref{eq:dipole-Bal}) with all $J$ replaced by $G$.

The Gaussian truncation provides explicit expressions for its associated
factorization violation $\Delta^G_{\bm{x z y}}(Y)$ which manifestly respects
the regional patterns outlined in Sec.~\ref{sec:orig-fact} on general grounds.

Using~(\ref{eq:simpleGcorr}) and the Fierz identity~(\ref{eq:Fierz}) one finds
\begin{align}
  \label{eq:DeltaG}
  \Delta^G_{\bm{x z y}}(Y) = \frac{ N_c^2\bigl[ 1- e^{ -C_f {\cal
        B}_{Y,\bm{x z y}} } \bigr] -(N_c^2-1)\bigl[1-
    e^{-\frac{N_c}{2} {\cal B}_{Y,\bm{x z y}} } \bigr] }{N_c^2}\ 
  e^{-C_f {\cal G}_{Y,\bm{x y}}} \ ,
\end{align}
where ${\cal B}_{Y,\bm{x z y}}$ is a shorthand notation for the
correlator combination already known from the evolution
equation~(\ref{eq:tilde-G-evo-short}), the expression for the three
point function~(\ref{eq:UtrtUtUdagger}) or the right-hand side of the
BFKL equation
\begin{align}
  \label{eq:calBdef}
  {\cal B}_{Y,\bm{x z y}} := 
  {\cal G}_{Y,\bm{x z}} + {\cal G}_{Y,\bm{z y}}- {\cal G}_{Y,\bm{x y}}  
\ .
\end{align}
$\Delta^G$ is positive semidefinite and strictly vanishes only where
${\cal B}$ vanishes. ${\cal B} =0$ is the hyper-surface on which the
integrand of~\eqref{eq:tilde-G-evo-short} changes sign, in analogy to
what was shown for the BK equation in Fig.~\ref{fig:z-plane}.  At
large positive ${\cal B}$, the fractional expression
in~\eqref{eq:DeltaG} approaches $1/N_c^2$, and it grows exponentially
at negative ${\cal B}$.  The region of large, positive ${\cal B}$
corresponds to $|\bm z'| \gg |\bm r|$ (region ``a'' in the notation of
Sect. \ref{sec:orig-fact}), the region power suppressed by the
evolution kernel. On the other hand, the region of maximally negative
${\cal B}$ corresponds to large parent dipoles $|\bm r| \gg R_s$
(regions ``b'' and ``c''  in Fig.~\ref{fig:scalecases}) and is is
strongly suppressed by the overall $e^{-C_f {\cal G}_{Y,\bm{x y}}}$.
This leaves the region near ${\cal B}=0$ to contribute.  This region
coincides with the region ``d'' singled out in
Sect.~\ref{sec:orig-fact} to yield the main contributions. Hence
$\Delta^G$ in \eqref{eq:DeltaG} illustrates the main feature of the
factorization violations derived in Sect. \ref{sec:orig-fact}).

The factorization violation in the Gaussian truncation
Eq.~\eqref{eq:DeltaG} also explicitly vanishes in the low density
(BFKL) limit: Contributions to Eq.~\eqref{eq:DeltaG} in fact start off
at order ${\cal B}^2$, i.e., are manifestly a nonlinear effect.

Since the right-hand side of the BK equation for $\langle\Hat
S\rangle$ (\ref{eq:BK}) is negative, one immediately concludes that
the positive contribution of the factorization violations included in
the Gaussian truncation slow down evolution for any parent dipole size
$\bm r$. One would expect this to carry over to JIMWLK evolution, i.e.
one would expect JIMWLK evolution to be slower than BK.

In the following we will show plots that map out factorization violations and
their influence on evolution. Since we want to show both the dependence on
parent dipole size $|\bm x-\bm y|$ and the position of the produced gluon $\bm
z$ we can not simultaneously plot $\Delta$ against all $\bm z$ degrees of
freedom.  We rely on the $\bm z'$-plane symmetries shown in
Fig.~\ref{fig:z-plane} and restrict ourselves to half rays at $0^\circ$ and
$90^\circ$ as done earlier and remind the reader that other half rays will
smoothly interpolate these extreme cases.

Contour plots of the factorization violations in the Gaussian truncation and
the full JIMWLK simulation in the continuum limit (see
Sect.~\ref{sec:quant-cons-fact} below) at some fixed rapidity are shown in
Fig.~\ref{fig:DeltaC}.
\begin{figure}[!thp]
  \centering
  \includegraphics[width=.49\textwidth]{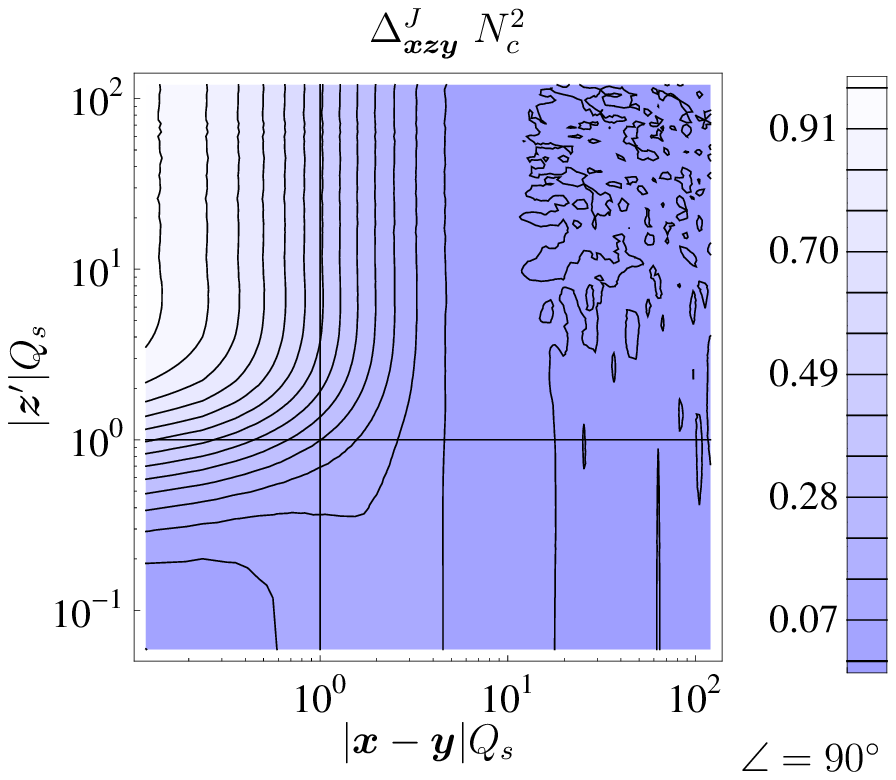}\hfill
  \includegraphics[width=.49\textwidth]{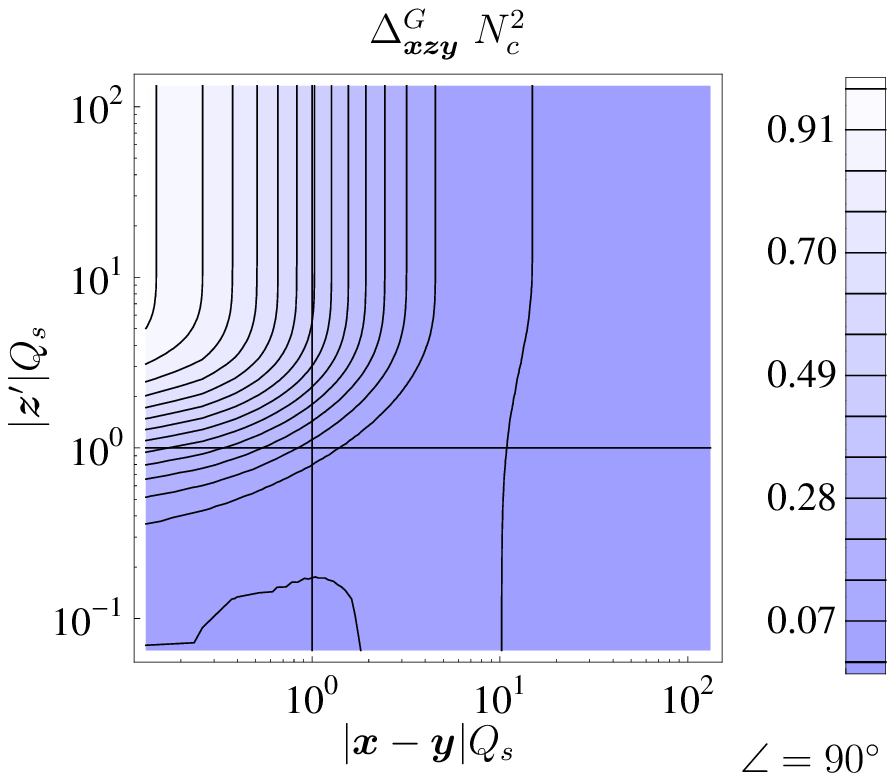} 
\\[.5cm]
  \includegraphics[width=.49\textwidth]{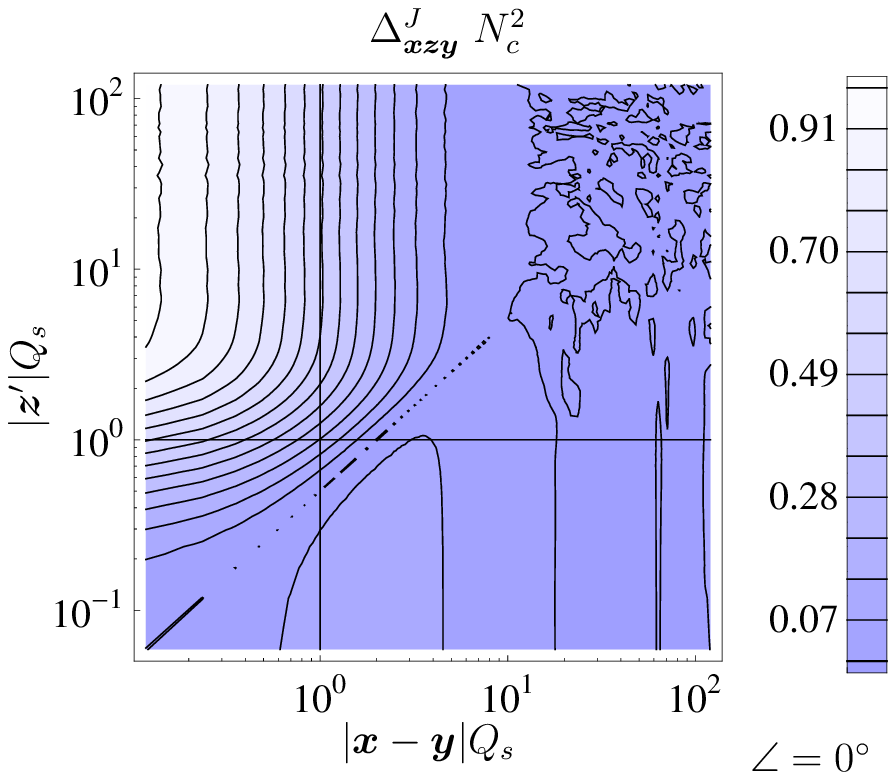}\hfill
  \includegraphics[width=.49\textwidth]{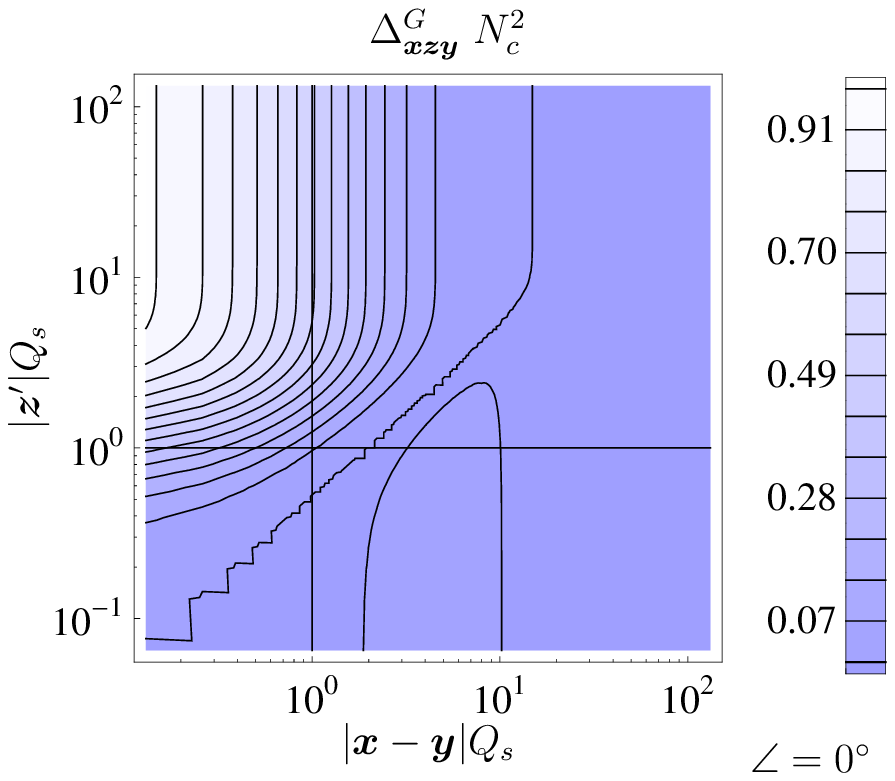}
  \caption{Contour plots of the factorization violation $\Delta$ scaled up 
    by $N_c^2$ at a fixed rapidity in JIMWLK (left) and in the
    Gaussian truncation (right).  The plots scan parent dipole size
    $|\bm x-\bm y|$ and distance $|\bm z'|$ from the midpoint $(\bm
    x+\bm y)/2$ at $90^\circ$ (top row) and $0^\circ$ (bottom row)
    with respect to $\bm x-\bm y$. The $0^\circ$ case is special since
    it contains contributions where $\Delta$ strictly vanishes due to
    the coincidence limit $\bm z=\bm x$ or $\bm y$ which appears here
    as a line with $|\bm z'|=|\bm x-\bm y|/2$. The bulk of the
    contributions is similar to the $90^\circ$ case. As was the case
    for the three-point-functions of Fig.~\ref{fig:regions}, the
    contributions along the axes $|\bm r|=0$ and $|\bm z'|=0$ are
    identical for all angles.  Fig.~\ref{fig:fact-viols-fixed-xmy}
    shows cuts along horizontal lines near the bottom of the two
    JIMWLK plots.  }
  \label{fig:DeltaC}
\end{figure}
They confirm that the regional pattern deduced in Sect.~\ref{sec:orig-fact} is
indeed seen in both JIMWLK evolution and the Gaussian truncation. Despite this
qualitative agreement, it becomes evident that the Gaussian truncation
underestimates the magnitude of the contributions significantly.
Fig.~\ref{fig:KDeltaC}
\begin{figure}[!thp]
  \centering
   \includegraphics[width=.49\textwidth]{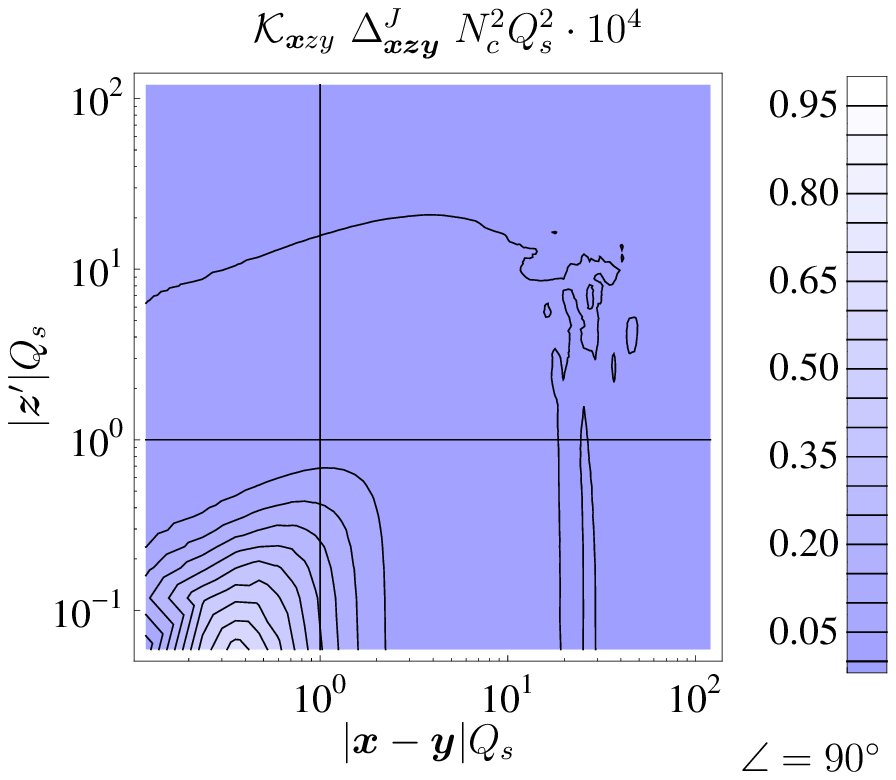}\hfill
  \includegraphics[width=.49\textwidth]{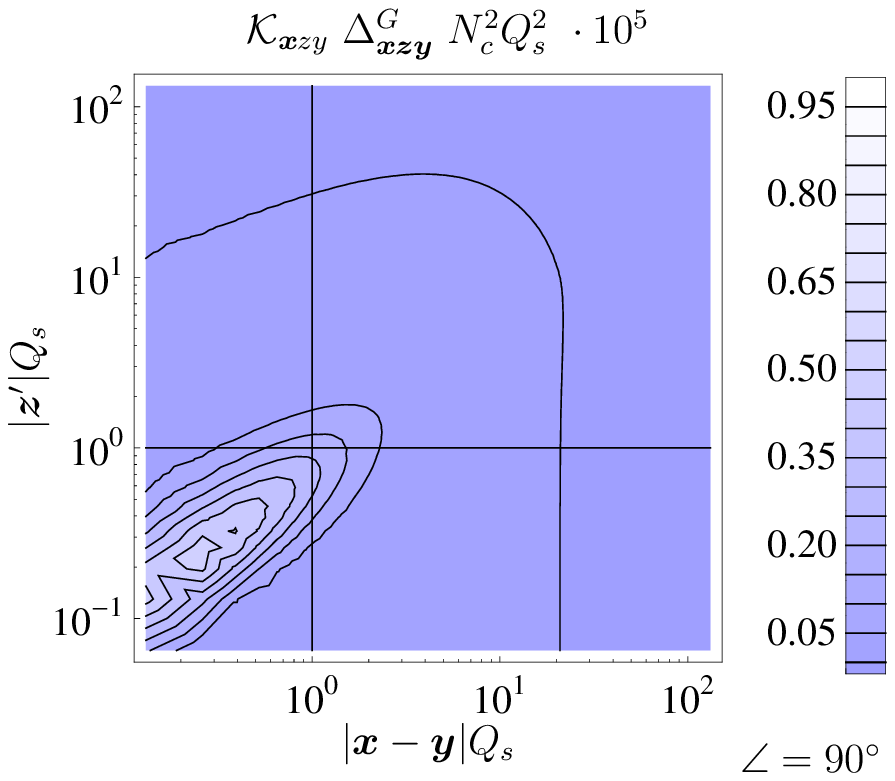}
\\[.5cm]
  \includegraphics[width=.49\textwidth]{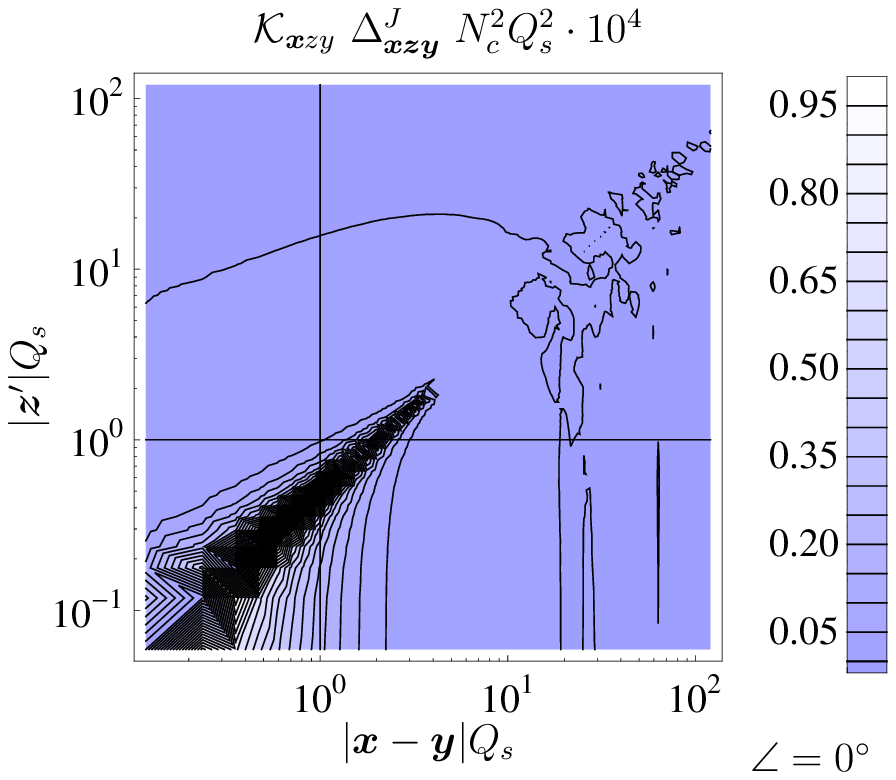}\hfill
  \includegraphics[width=.49\textwidth]{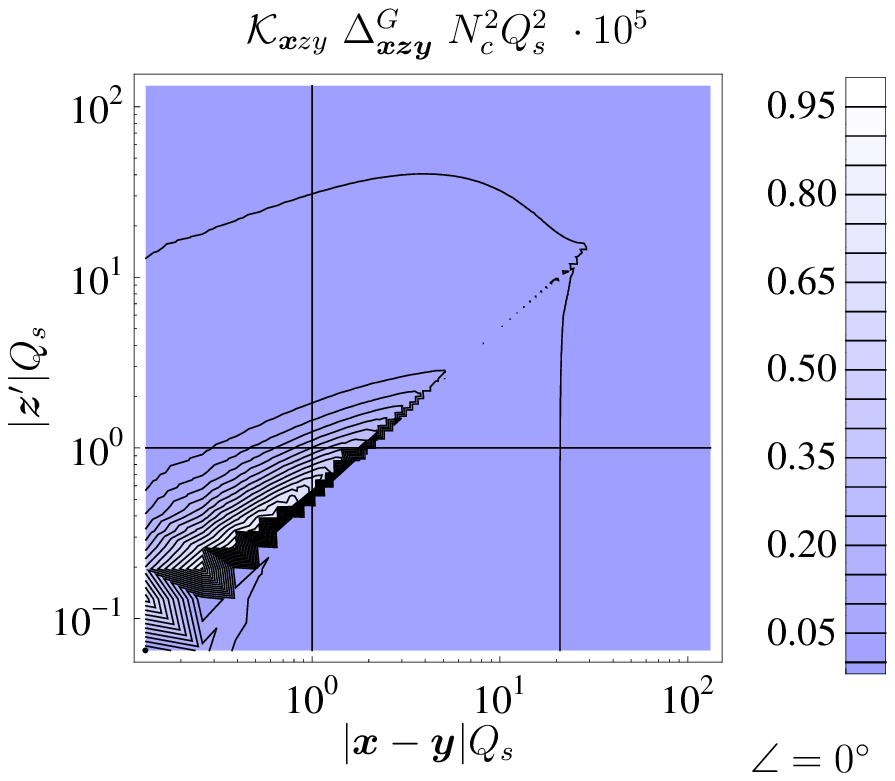}
\caption{Factorization violations shown in Fig.~\ref{fig:DeltaC} multiplied by
  the BK kernel (i.e. the integrand of the $\Delta ^J$ term in the
  evolution equation~(\ref{eq:dipole-Bal}) that determines how much
  they contribute to evolution), by $Q_s^2 \, N_c^2$, and by a scaling
  factor with varies from panel to panel. The notation and geometry
  are the same as explained in the caption of Fig.~\ref{fig:DeltaC}. In
  both cases the region where $\Delta$ reaches $1/N_c^2$ is completely
  suppressed by the kernel. Note the marked difference in size of the
  rescaling factor: the contribution of factorization violations in
  JIMWLK (left, with $10^4$) is an order of magnitude larger than in
  the Gaussian truncation (right, with $10^5$). }
  \label{fig:KDeltaC}
\end{figure}
shows the corresponding contributions to the integrand of the evolution
equation (\ref{eq:dipole-Bal}), i.e. after the kernel has been multiplied in.
The kernel both power {\em suppresses} the large $|\bm z'|$ region and {\em
  enhances} the contributions of small $|\bm z'|$ and, as already anticipated,
the results become quite sensitive to the short range behavior as ${\cal B}$
goes to zero in the lower left hand corners of the plots.

\section{Quantitative consequences: slowdown of evolution}
\label{sec:quant-cons-fact}

The notion that the presence (or absence) of a factorization violation
term in Eq.~(\ref{eq:dipole-Bal}) would affect evolution speed can be
made more precise.  By extension of an argument in~\cite{Iancu:2002tr}
we define evolution speed\footnote{Evolution speed as defined
  in\cite{Iancu:2002tr} refers to the scaling regime with a uniquely
  defined $Q_s(Y)$ and is defined as $\lambda(Y):=\tfrac{d}{d Y} \ln
  Q_s^2(Y)$. Outside the scaling region the starting point $Q_s(Y)$ is
  no longer uniquely defined and one may take this as one of many
  possible definitions for a well behaved measure of evolution speed.}
as a $\int d^2r/{\bm r^2}$-integral of the right-hand side of the
evolution equation~(\ref{eq:preBKS})
\begin{align}
  \label{eq:lambda-int-def}
  \lambda(Y) := - \frac{\alpha_s N_c}{2\pi^3}
  \int\! \frac{d^2r}{\bm r^2} \int\! d^2z\,
  \ {\cal K}_{\bm{x z y}} \
 \langle \Hat S_{\bm{x z}} 
\Hat S_{\bm{z y}} 
 -
\Hat S_{\bm{x y}} \rangle(Y)
\end{align}
and then proceed to split the contributions according
to~(\ref{eq:dipole-Bal}). (Again ${\bm r} = {\bm x} - {\bm y}$.)

To obtain a quantitative comparison of JIMWLK and BK
equations, one should take note that due to non-vanishing $\Delta^J$
\begin{align}
  \label{eq:JIMWLK-BK-S-diff}
  \langle \Hat S_{\bm{x y}} \rangle_J(Y) \neq 
  \langle \Hat S_{\bm{x y}} \rangle_B(Y)
\intertext{even if one chooses them to be equal at the initial rapidity $Y_0$:}
  \label{eq:JIMWLK-BK-S-init} 
 \langle \Hat S_{\bm{x y}} \rangle_J(Y_0) = 
  \langle \Hat S_{\bm{x y}} \rangle_B(Y_0)
\end{align}
as will be the case in all the numerical comparisons shown.

To calculate the difference of evolution speeds at some finite $Y-Y_0$
one has to compare Eq.~\eqref{eq:dipole-Bal} with the right-hand side
of
\begin{align}
  \label{eq:BK-S-proper}
\frac{d}{d Y} \langle \Hat S_{\bm{x y}} \rangle_B(Y)
  =\frac{\alpha_s N_c}{2\pi^2}\int d^2z\,
  {\cal K}_{\bm{x z y}} \left(
 \langle \Hat S_{\bm{x z}} \rangle_B \langle \Hat S_{\bm{z y}} \rangle_B -\langle \Hat S_{\bm{x y}} \rangle_B      \right)(Y).  
\end{align}
The difference between the JIMWLK and BK evolution speeds from Eqs.
(\ref{eq:dipole-Bal}) and (\ref{eq:BK-S-proper}) is due to the
difference in the JIMWLK and BK averagings of the $S$-matrices and to
the presence of the $\Delta^J$-term in \eqref{eq:dipole-Bal}.  Indeed
these two causes of difference are interconnected. Introducing the
shape dependent correlator difference
\begin{align}
  \label{eq:delta_s}
  \Delta^{JB}_{\bm{x z y}}(Y):=
 \left(
 \langle \Hat S_{\bm{x z}} \rangle_J \langle \Hat S_{\bm{z y}} \rangle_J-\langle \Hat S_{\bm{x z}} \rangle_B  \langle \Hat S_{\bm{z y}} \rangle_B -\left(\langle \Hat S_{\bm{x y}} \rangle_J-\langle \Hat S_{\bm{x y}} \rangle_B\right)
      \right)(Y)  
\end{align}
the difference in evolution speed arises from two non-vanishing
contributions according to
\begin{align}
  \label{eq:deltalambda}
  \Delta\lambda^{JB}(Y) :=&\ \lambda_J(Y)-\lambda_B(Y) 
\\
  =&\ - \frac{\alpha_s N_c}{2\pi^3}
  \int\! \frac{d^2r}{\bm r^2} \int\! d^2z\,
  {\cal K}_{\bm{x z y}}\  \Delta^J_{\bm{x z y}}(Y)
  -\frac{\alpha_s N_c}{2\pi^3}\int\! \frac{d^2r}{\bm r^2}\int\! d^2z\,
  {\cal K}_{\bm{x z y}}\  \Delta^{JB}_{\bm{x z y}}(Y)
\ .
\notag
\end{align}
Given identical initial conditions,~(\ref{eq:JIMWLK-BK-S-init}), the
$\Delta^J$-term is the sole reason for a non-vanishing $\Delta\lambda^{JB}(Y)$
to be generated at all, but the second term may in some cases be
quantitatively not less important once that has happened.

It is worth noting that in the definition of $\lambda$,
Eq.~(\ref{eq:lambda-int-def}), the contributions at small $\bm r^2$ are
enhanced by the conformal measure $d^2r/\bm r^2$. This carries over to the
differences of evolution speed between factorized and unfactorized evolution,
Eq.~(\ref{eq:deltalambda}). We have already shown in Fig.~\ref{fig:KDeltaC}
that this is where the main contributions to the product of kernel $\cal K$
times $\Delta$ lie, so that one might expect that $\Delta\lambda$ receives an
enhanced contribution from any non-vanishing $\Delta$ observed in our
simulations.

Since it is the product of kernel and $\Delta$ that enters one would expect
that $\lambda$ should also be sensitive to any modifications to the kernel as
we step beyond leading order by, for example, inclusion of running coupling
effects. We would expect running coupling effects to generically suppress
contributions at small $|\bm r|$ and thus reduce some of the speed difference
visible at leading order. At the same time one should be aware that
contributions beyond running coupling would also affect evolution speed and
the way $1/N_c$ corrections couple into dipole evolution. Those might be of
the same size as the running coupling contributions, but it would be very
peculiar if they were to affect the general pattern of the leading oder
behavior observed in the simulations shown below.

We have performed a numerical solution of the JIMWLK evolution equation at one
loop accuracy with fixed coupling along the lines of and using the techniques
developed in \cite{Rummukainen:2003ns}. The method is based on the fact that
the JIMWLK equation takes the form of a functional Fokker-Planck equation
which has an equivalent Langevin formulation~\cite{Weigert:2000gi,
  Blaizot:2002xy} in which the averages $\langle\ldots\rangle_J(Y)$ are
expressed as ensemble averages of fields $U_{Y,\bm x}$, with the
$Y$-dependence implemented by a Langevin equation for the ensemble members. To
implement such a description numerically one is forced to discretize
transverse space to represent the ensemble member fields $U_{Y,\bm x}$ in
terms of a finite number of degrees of freedom. This automatically introduces
a UV regulator in the guise of the lattice spacing $a$ and a IR regulator, the
lattice size $L$.  The reader interested in further details of the numerical
solution and in the implementation of the numerical procedure is referred
to~\cite{Rummukainen:2003ns}. In order to be able to directly compare our
JIMWLK simulations with BK and GT results at fixed $a$ and $L$, we have chosen
to run our BK and GT simulation on the same type of regular lattice with
identical $a$ and $L$ values although that is decidedly {\em not} the most
efficient way to perform standalone BK or GT simulations.

\begin{figure}[!thb]
  \centering
\includegraphics[width=.32\linewidth]{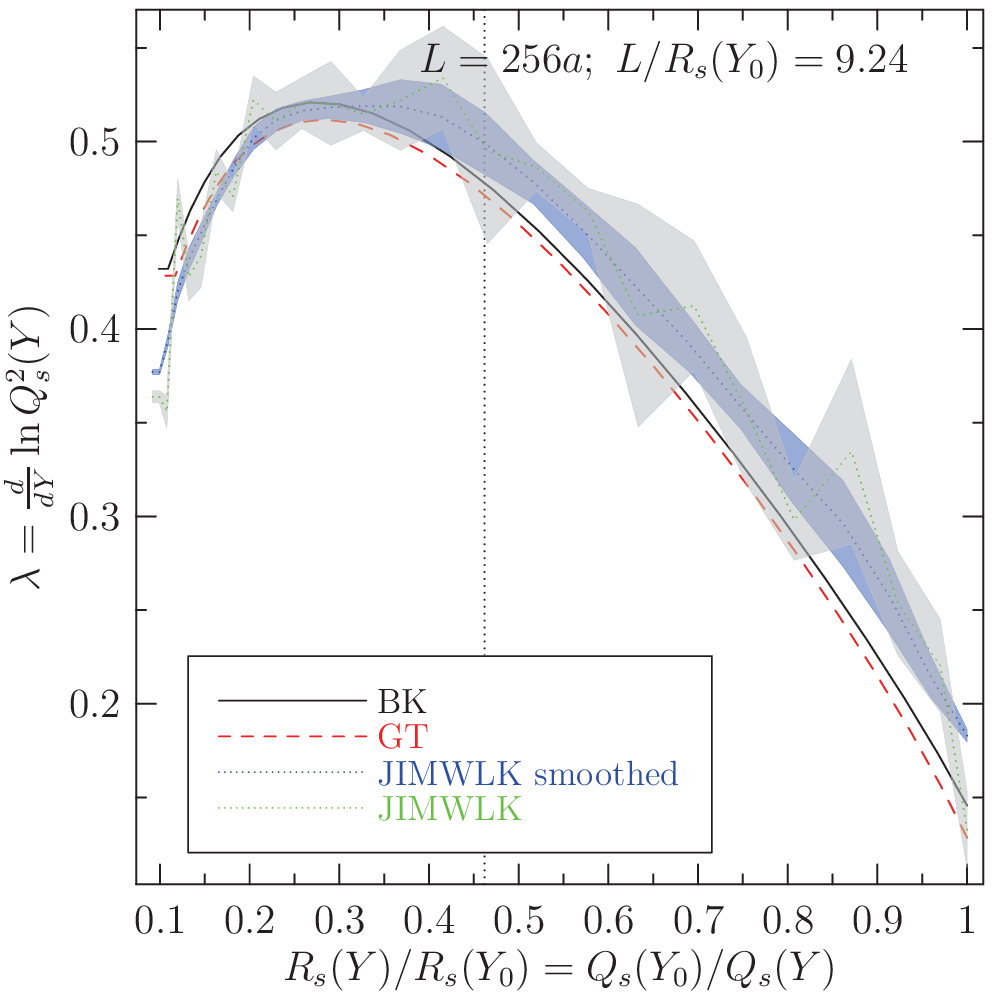}\hfill
  \includegraphics[width=.32\linewidth]{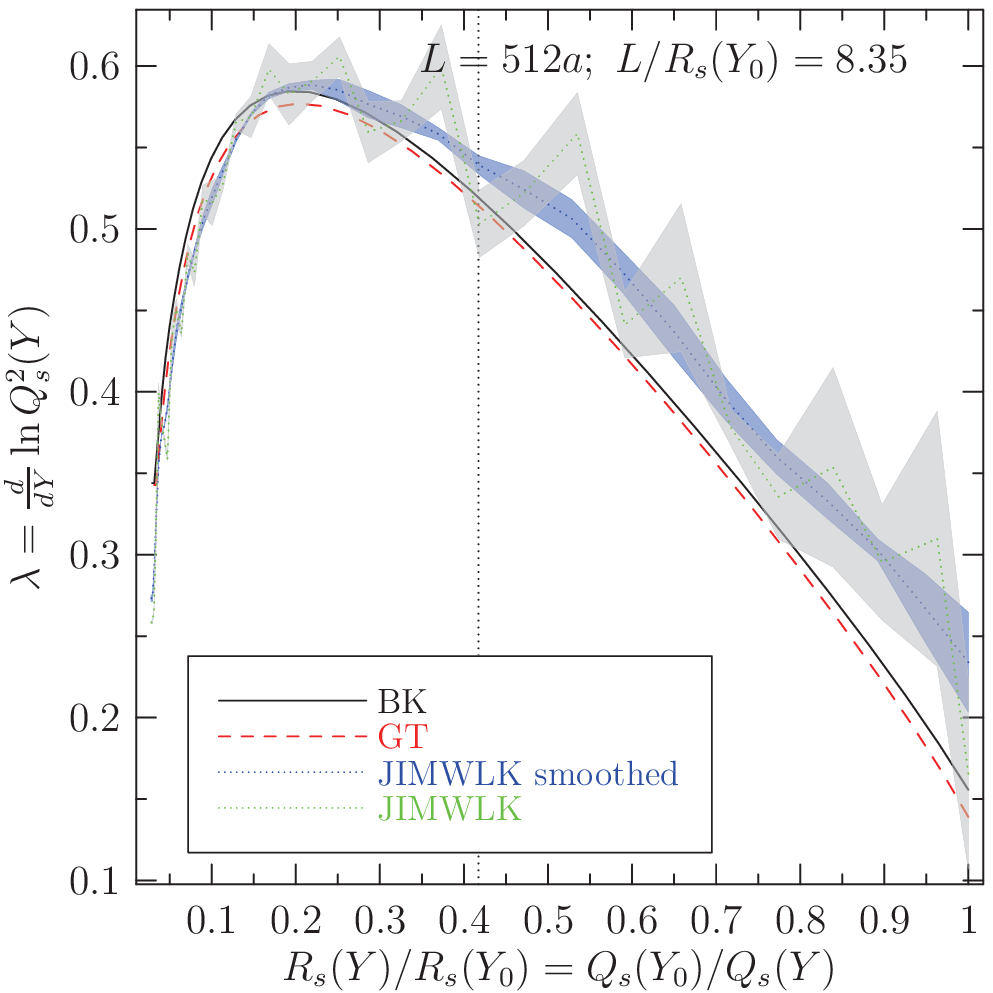}\hfill
  \includegraphics[width=.32\linewidth]{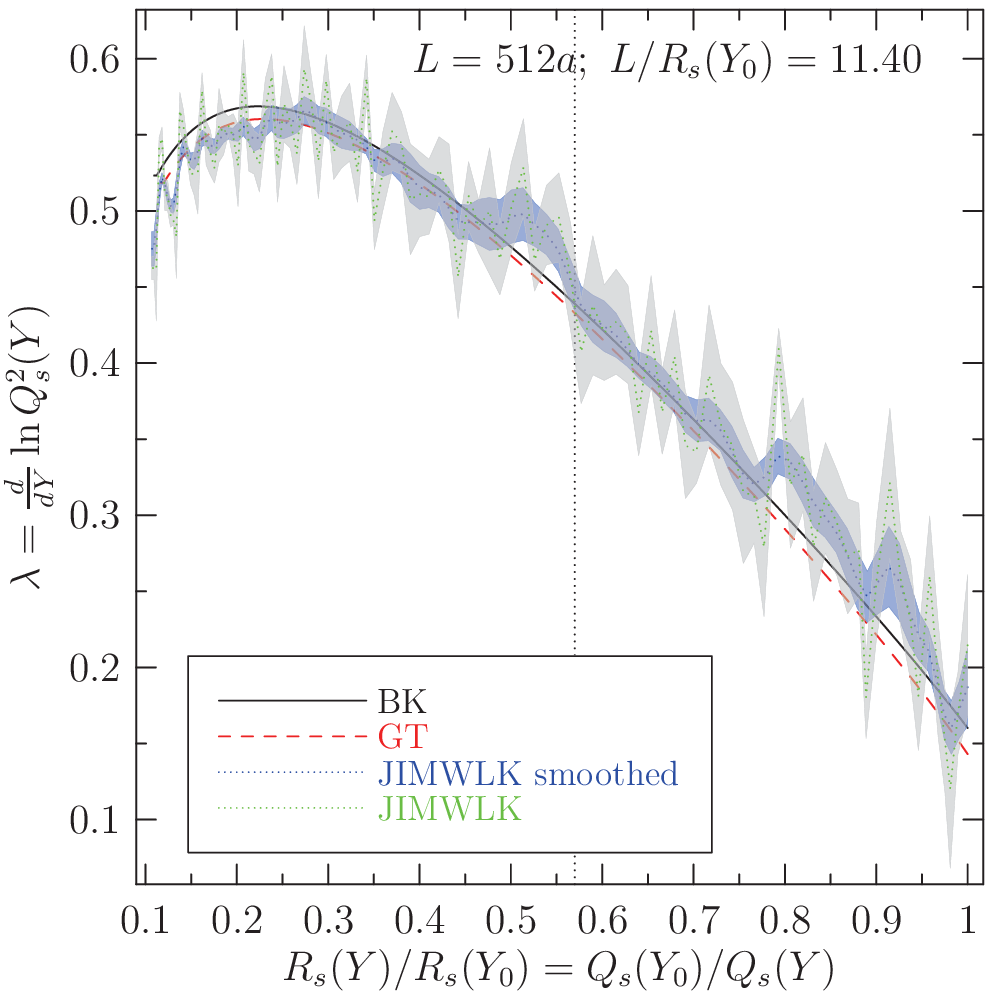}\\[.5cm]
  \includegraphics[width=.32\linewidth]{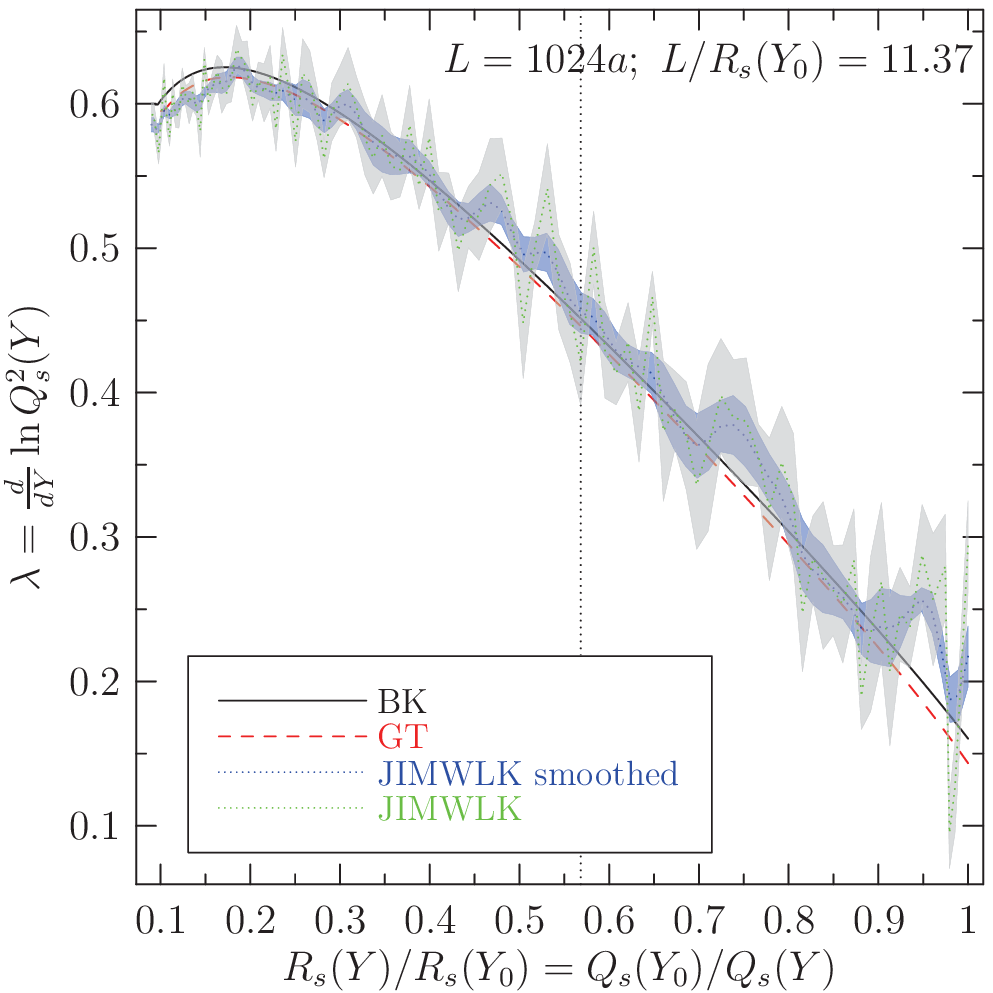}\hfill
\includegraphics[width=.32\linewidth]{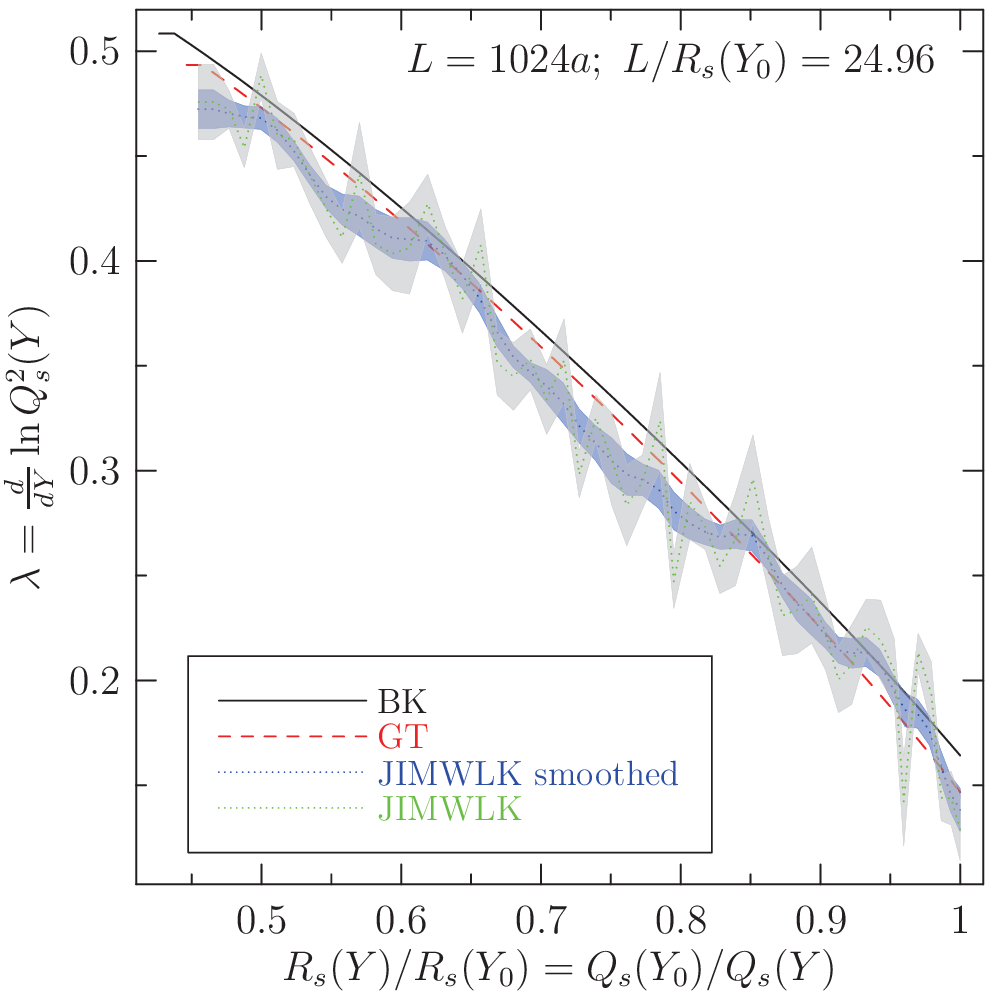}\hfill
\includegraphics[width=.32\linewidth]{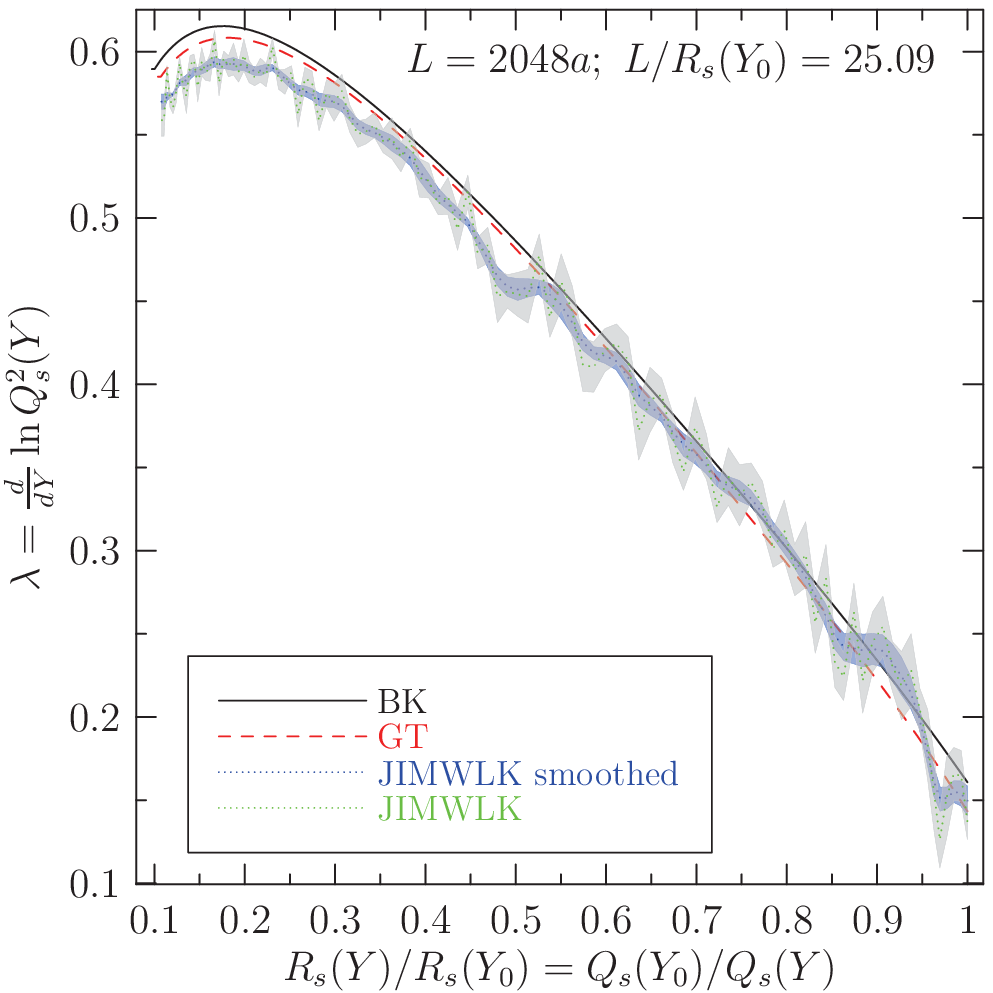}\\[.5cm]
\includegraphics[width=.32\linewidth]{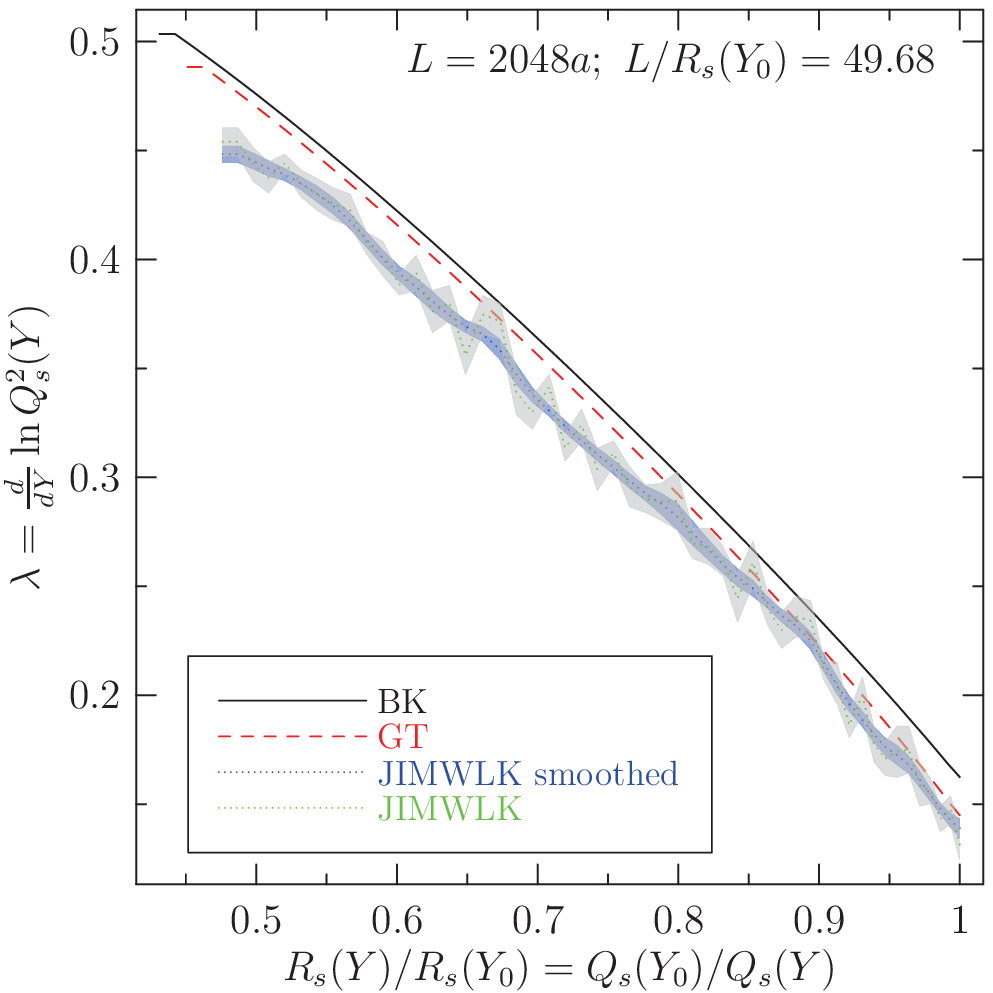}\hfill
\includegraphics[width=.32\linewidth]{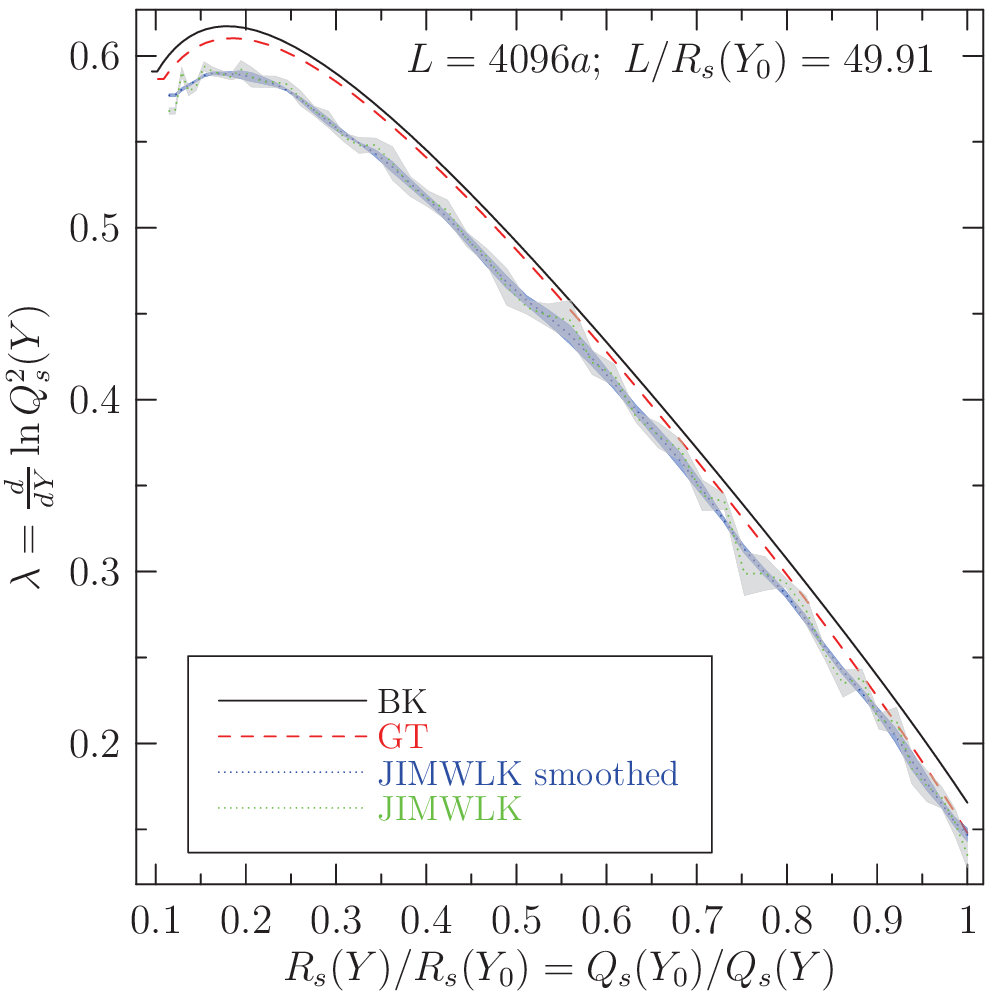}\hfill
\includegraphics[width=.32\linewidth]{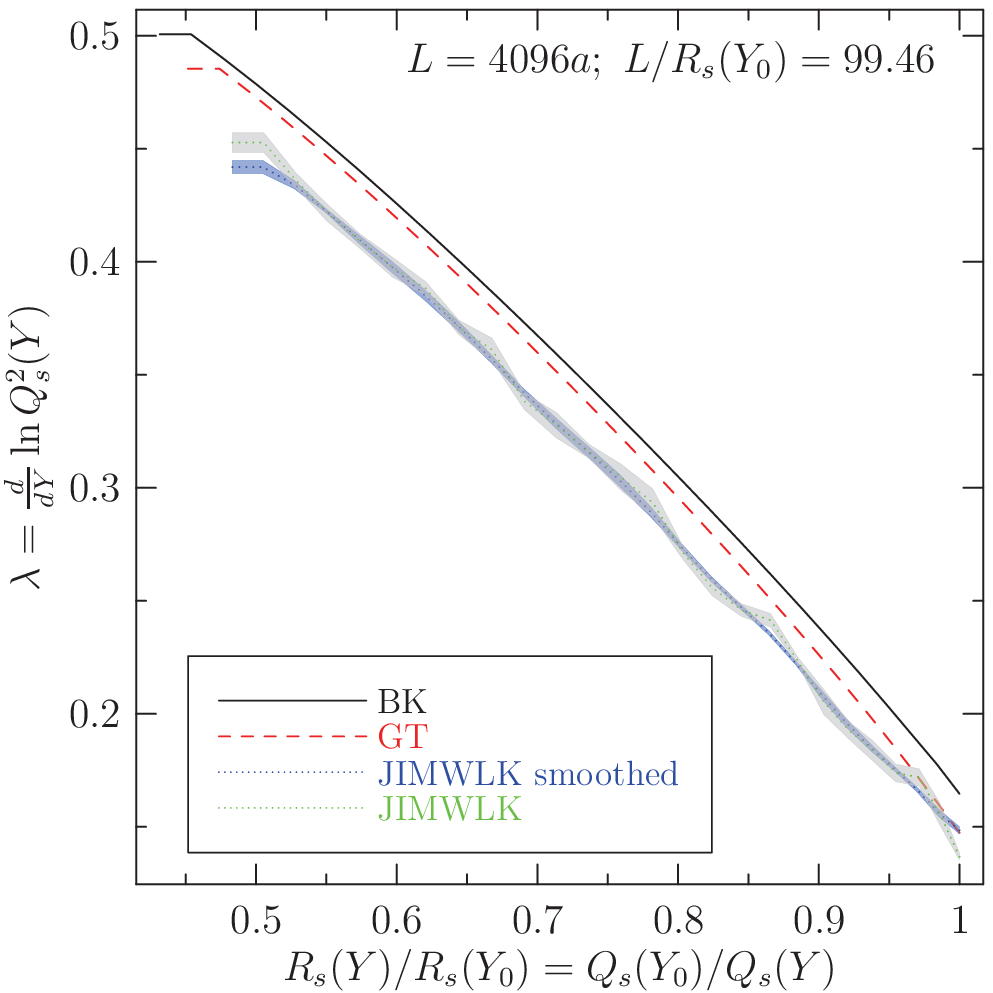}
\caption{Comparing evolution speed in JIMWLK, BK and GT on different size
  lattices, starting from initial conditions with identical dipole correlators
  $ \langle \Hat S_{\bm{x y}} \rangle_J(Y_0)= \langle \Hat S_{\bm{x y}}
  \rangle_B(Y_0)= \langle \Hat S_{\bm{x y}} \rangle_G(Y_0) $. The vertical line
  in the first four plots marks $L/R_s(Y)=20$. The bands show statistical
  (Jackknife) errors.
  }
  \label{fig:lambda-data}
\end{figure}
Fig.~\ref{fig:lambda-data} shows numerical results for evolution speeds
$\lambda(Y)$ as a function of $R_s(Y)=1/Q_s(Y)$, for lattices sizes varying
from $256^2$ to $4096^2$. The JIMWLK results of the first two plots are
compiled from the simulations presented in~\cite{Rummukainen:2003ns}, this
were the largest lattice sizes taken into account then. All the others are
based on new simulations, with measurements taken at smaller $Y$ intervals
(which explains the more fine-grained raggedness of the JIMWLK results --
adding additional intermediate steps would further enhance the phenomenon).

For the new runs, initial conditions were chosen to carefully explore the
convergence to both the continuum limit and the infinite volume limit, i.e.,
to scan for UV- and IR-cutoff artifacts.  IR phase space available in the
simulation at the initial condition at $Y_0$ is varied by increasing lattice
size $L$ compared to initial correlation length $R_s(Y_0)$, UV phase space is
varied by increasing lattice size at (approximately) fixed $L/R_s(Y_0)$.
During evolution active phase space, which is centered around $1/R_s(Y)$,
moves towards the UV, so that available IR phase space grows with $L/R_s(Y)$
while the UV shrinks with $R_s(Y)/a$.

The most striking feature of the JIMWLK simulations are the large fluctuations
of evolution speed on all but the largest lattices. The fluctuations turn out
to be IR dominated, they average out as we increase the number of points in
the IR. UV cutoff effects manifest themselves only for the longest runs, as a
relatively sharp turn downwards as one follows the curves from right to left
as $R_s$ shrinks while $Y$ grows. This downturn (where present) indicates that
the simulation is running out of UV phase space with $R_s(Y)/a \lsim
10$. While this behavior is not physical, it affects all our simulations in
the same way and one does not prevent us from comparing the behavior of the
simulations with each other.

This can be read off from Fig.~\ref{fig:lambda-data} by tracing the following
systematic features: With the saturation scale safely more than an order of
magnitude smaller than the inverse lattice spacing, only varying IR phase space
affects $\lambda$. As $L/R_s(0)$ increases from $8.35$ to $99.46$ JIMWLK
evolution becomes less and less affected by IR fluctuations. Evolution speeds
(as compared to BK and GT, which both are not affected by fluctuations) slow
down until they settle at their infinite volume limit at around
$L/R_s(0)\approx 50$.  

Note that in the first four panels (with smallest $L/R_s(0)$) JIMWLK is
initially faster than both BK and the Gaussian truncation, before this becomes
less and less pronounced as $L/R_s(Y)$ grows with evolution.  This is a direct
consequence of the impact of fluctuations becoming less pronounced as
shrinking $R_s(Y)$ cuts off contributions from the infrared.  It turns out
that for very small $L/R_s(Y)$ where IR fluctuations contribute most, the
$\Delta^{JB}$ contribution completely overwhelms the $\Delta^J$ contribution
which in all cases gives a contribution that slows evolution down. The
relative size of this contributions shrinks strongly when $L/R_s(Y) \gsim 20$
and is no longer able to overwhelm $\Delta^J$ in the runs with $L/R_s(Y_0)
\gsim 25$.

\begin{figure}[!thb]
  \centering
  \includegraphics[width=.32\textwidth]{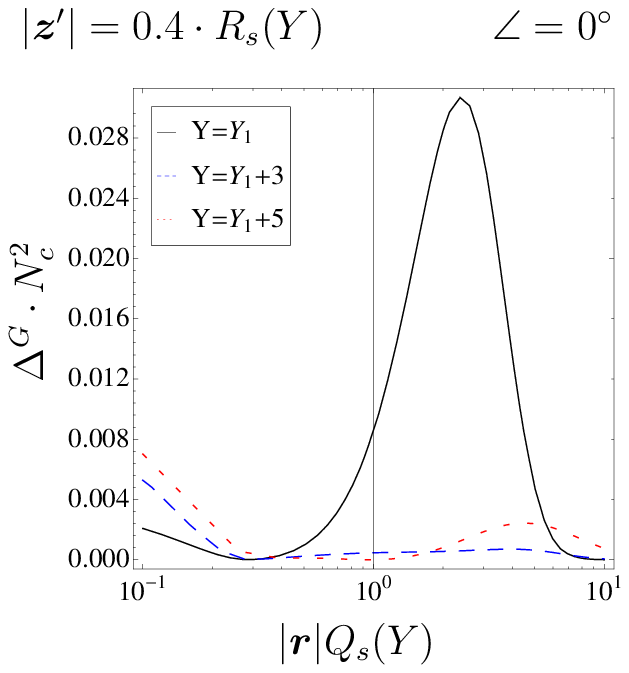}\hfill
  \includegraphics[width=.32\textwidth]{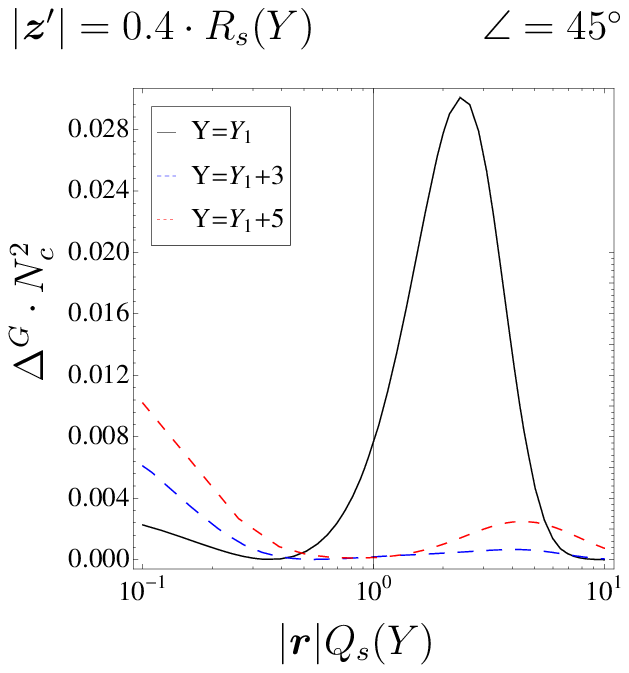}\hfill
  \includegraphics[width=.32\textwidth]{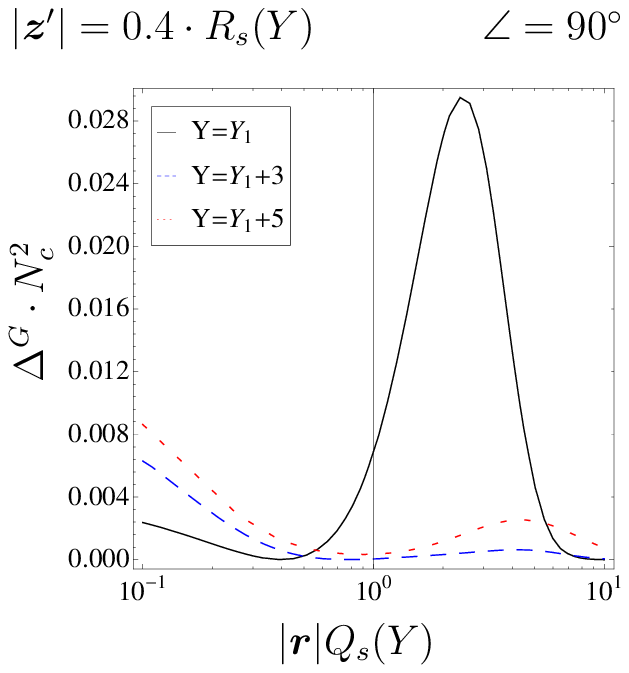}
  \caption{Factorization violations in the Gaussian truncation (scaled up by
    $N_c^2$) against varying parent dipole size at fixed $|\bm z'| = 0.4\cdot
    R_s(Y)$ depicted here for comparison with JIMWLK results shown in
    Fig.~\ref{fig:fact-viols-fixed-xmy}. The larger factorization violations
    in JIMWLK lead to slower evolution.  The strong change form $Y_1$ to
    higher rapidities is mirrored by a convergence of evolution speed between
    GT and BK with evolution towards asymptotic regime shown in
    Fig.~\ref{fig:lambda-ratios}.  }
  \label{fig:fact-viols-Y-dep-GT}
\end{figure}

\begin{figure}[!thb]
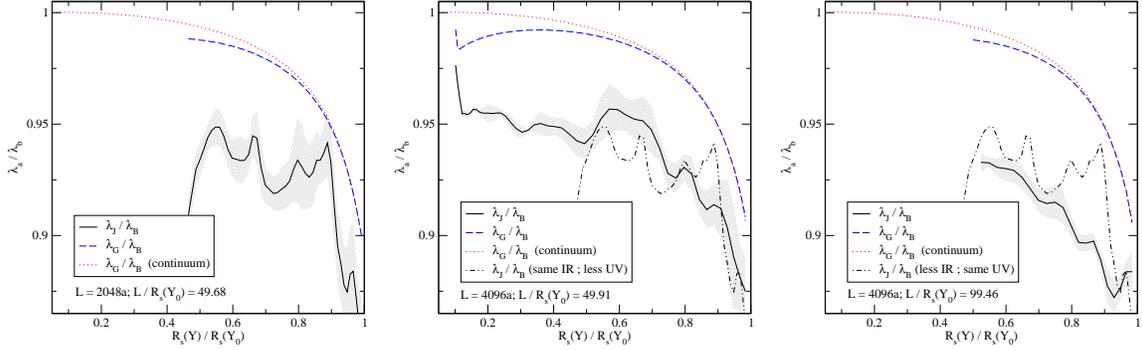

  \centering
    \includegraphics[width=.32\textwidth]{ratio_2048_64}
   \hfill
    \includegraphics[width=.32\textwidth]{med_ir}
\hfill
    \includegraphics[width=.32\textwidth]{high_ir}
    \caption{Ratios of evolution speed with finite lattice spacing and in the
      continuum limit. The plots correspond one to one to the bottom row in
      Fig.~\ref{fig:lambda-data}. [The sharp up- or downturn of the curves
      with finite $a$ towards the left indicate complete breakdown of the
      simulations in the UV.]  Shown are in all plots, from bottom to top
      (excepting the dash-dotted lines): $\lambda_{J}(Y)/\lambda_{B}(Y)$ and
      $\lambda_{G}(Y)/\lambda_{B}(Y)$ (both at finite $a$) and
      $\lambda_{G}(Y)/\lambda_{B}(Y)$ (in the continuum limit). A comparison
      of the two upper curves indicates the size of the UV cutoff effects
      which remain fairly small compared to the error on the JIMWLK results.
      Evolution in JIMWLK is slower than in both BK and GT in all plots. The
      left- and rightmost plots serve to assess lattice artifacts in JIMWLK
      evolution. Compared to the middle plot, the left plot has the same IR
      but less UV phase space, the right plot less IR but the same UV phase
      space.  Both of these ``fall through the lattice'' already at
      $R_s(Y)/R_s(Y_0)\approx .5$, they agree with the middle plot within
      errors. (To facilitate this comparison the solid line from the left
      panel is repeated as a dash-dotted line in the other panels.) Using the
      middle plot where the JIMWLK simulations reach closest to the asymptotic
      regime (where $\lambda_{G}(Y)/\lambda_{B}(Y)\to 1$ in the continuum
      limit, top curve), we read off roughly a 3-5\% slowdown of JIMWLK-
      relative to BK-evolution due to factorization violations. This is
      expected to be strongly reduced by running coupling effects.}
  \label{fig:lambda-ratios}
\end{figure}

Similarly, for all other runs, where $L/R_s(Y) \gsim 25$ from the outset, we
observe (on average) a clear hierarchy of evolution speeds with BK the
fastest, the Gaussian truncation in the middle and JIMWLK the slowest.  In
this range this is simply a reflection of the relative size of factorization
violations: the larger these are in the regions enhanced by the kernel, the
slower the evolution becomes. This hierarchy is already visible in the
rescaling factors of Figs.~\ref{fig:DeltaC} and~\ref{fig:KDeltaC}: scaling
violations in GT are consistently an order of magnitude smaller than in JIMWLK
across the $Y$ range explored.  This becomes evident again, if we contrast the
JIMWLK results of Fig.~\ref{fig:fact-viols-fixed-xmy} with their counterpart
in the Gaussian truncation, Fig.~\ref{fig:fact-viols-Y-dep-GT}.
Were we to extend our comparison of BK and GT evolutions into the asymptotic
range, however, evolution speeds would necessarily become identical as
discussed in Sec.~\ref{sec:rest-coinc-limits}. We can use this to assess how
closely the simulations shown in Fig.~\ref{fig:lambda-data} approach the
asymptotic scaling region. With the regular grid necessary to compare with
JIMWLK this is not practical, but a simulation that only compares BK and GT
can make be implemented more efficiently and in fact reach the asymptotic
limit. Using this freedom we find that this occurs just beyond the region
where we loose UV accuracy in the longest JIMWLK simulations such as that in
the bottom middle plot of Fig.~\ref{fig:lambda-data}. The result of this
comparison is shown in Fig.~\ref{fig:lambda-ratios}.  Comparing the ratios of
evolutions speeds in Fig.~\ref{fig:lambda-ratios} we conclude that in the
asymptotic scaling region (at fixed coupling!) one should expect a
factorization violation induced 3-5\% slowdown in evolution speed in JIMWLK
evolution compared to BK evolution at one loop accuracy.  There is no sign
from the simulation and no theoretical reason to argue that factorization
violations in JIMWLK should disappear in the asymptotic region. As already
noted we would expect this relative slowdown effect to become less pronounced
at NLO.

\section{JIMWLK beyond the Gaussian truncation: higher order correlators}
\label{sec:jimwlk-beyond-gauss}

The simulation results shown in the previous section show a persistently small
but measurable improvement of the Gaussian truncation over the BK
approximation. Still, JIMWLK evolution is much more general than either
truncation. Both approximations restrict the information retained in evolution
to two point functions that, in the low density limit matches up with double
reggeon exchange as incorporated in BFKL evolution. JIMWLK evolution, on the
other hand, allows for multi--reggeon exchange in evolution and even the
limited set of correlators discussed in the above is sensitive to their
contributions. An example of this is the simplistic form in which dipole
correlators of higher representations are mapped back onto the quark dipole
correlator. This is a direct consequence of the fact that the Gaussian
truncation only iterates two reggeon exchange into Glauber exponents.

One might attempt to include multi--reggeon exchanges by
generalizing~(\ref{eq:func-Gaussian-average}) to include higher order
terms; naive inclusion of three--reggeon terms would
modify~(\ref{eq:func-Gaussian-average}) to
\begin{align}
  \label{eq:func-2n3-average-cartesian}
  \langle \ldots \rangle(Y) &=  \exp\left\{ 
    -\frac12 \int\limits^Y\! dY' \int\! d^2x\, d^2y\ 
    G_{Y',\bm{x y}} \frac{\delta}{\delta A^{a +}_{\bm x,Y'}}
    \frac{\delta}{\delta A^{a +}_{\bm y,Y'}}
\right. \\ & \left.
    -\frac1{3!}
  \int\limits^Y\! dY' \int\! d^2x\, d^2y\, d^2z\ 
    \left(G_{Y'\bm{x y z}}^f\ f_{a b c} 
      +
      G_{Y'\bm{x y z}}^d\ d_{a b c} 
      \right)
    \frac{\delta}{\delta A^{a +}_{\bm x,Y'}}
    \frac{\delta}{\delta A^{b +}_{\bm y,Y'}}
    \frac{\delta}{\delta A^{c +}_{\bm z,Y'}}
 \right\} \ldots 
\notag
\end{align}
and include odderon contributions \cite{Lukaszuk:1973nt}.  However,
starting with three reggeon terms, locality in $Y$ is an assumption
that might prove to be too restrictive and will generally not lead to
a consistent treatment of the Balitsky hierarchies. Moreover, one has
to be careful in simply exponentiating the 3-reggeon terms, as is done
in \eqref{eq:func-2n3-average-cartesian}. Here the best way to keep
the calculations under parametric control is to employ the power
counting developed for the classical gluon fields in
\cite{Kovchegov:1997pc,Kovchegov:1996ty}. In the quasi-classical limit
the leading term in the exponent of
\eqref{eq:func-2n3-average-cartesian} corresponds to a two-gluon
exchange with a nucleon in the nucleus, such that $G_{Y,\bm{x y}} \sim
\as^2 \, A^{1/3}$ with $A$ the atomic number of the nucleus. For $\as
\ll 1$ and $A \gg 1$ there exists a regime where $\as^2 \, A^{1/3}
\sim 1$ and the GT approximation of Sect. \ref{sect:gauss} resums all
powers of $\as^2 \, A^{1/3}$. From the standpoint of this
quasi-classical power counting the second term in the exponent of
\eqref{eq:func-2n3-average-cartesian} corresponds to a 3-gluon
$t$-channel exchange with a single nucleon, such that $G_{Y \bm{x y
    z}}^f \sim G_{Y \bm{x y z}}^d \sim \as^3 \, A^{1/3}$, i.e., it is
suppressed by one power of the coupling $\as$ compared to the leading
term. Iteration of such term more than once would be beyond the
precision of the approximation: two 3-gluon exchanges are of the same
order in $\as$ and $A$ as a 2-gluon exchange combined with a 4-gluon
exchange. From this perspective the contributions of
Eq.~\eqref{eq:func-2n3-average-cartesian} are only under parametric
control up to linear order in $G^f$ and $G^d$:
\begin{align}
  \label{eq:func-2n3-average-cartesian-2}
   \langle \ldots & \rangle(Y) =  \exp\left\{ 
    -\frac12 \int\limits^Y\! dY' \int\! d^2x\, d^2y\ 
    G_{Y',\bm{x y}} \frac{\delta}{\delta A^{a +}_{\bm x,Y'}}
    \frac{\delta}{\delta A^{a +}_{\bm y,Y'}}
 \right\}  \\ & \times \left[1
    -\frac1{3!}
  \int\limits^Y\! dY' \int\! d^2x\, d^2y\, d^2z\ 
    \left(G_{Y'\bm{x y z}}^f\ f_{a b c} 
      +
      G_{Y'\bm{x y z}}^d\ d_{a b c} 
      \right)
    \frac{\delta}{\delta A^{a +}_{\bm x,Y'}}
    \frac{\delta}{\delta A^{b +}_{\bm y,Y'}}
    \frac{\delta}{\delta A^{c +}_{\bm z,Y'}} \right] .
\notag
\end{align}

A discussion of truncations that systematically include multi--reggeon
contributions goes beyond the scope of this paper, but it is not hard
to play with~(\ref{eq:func-2n3-average-cartesian}) as an ansatz to
explore the consequences of the inclusion of an odderon term in this
manner (see Appendix~\ref{sec:odder-contr-lead}): it becomes quite
manifest that such multi--reggeon contributions naturally break Casimir
scaling. The reason for this is that in general, higher representation
contributions in the $t$-channel start to pick up on the more
complicated decomposition of a general $s$-channel ${\cal R}\Bar{\cal
  R}$-dipole into irreducible representations. (Higher representations
were also considered in \cite{Popov:2008ck}.) For the odderon
contribution they account for the fact that the $q\Bar q$ dipole
acquires an imaginary part (with a specific $N_c$ dependence) while a
$g g$-dipole remains real since the adjoint representation is real by
definition.

\begin{figure}[htb]
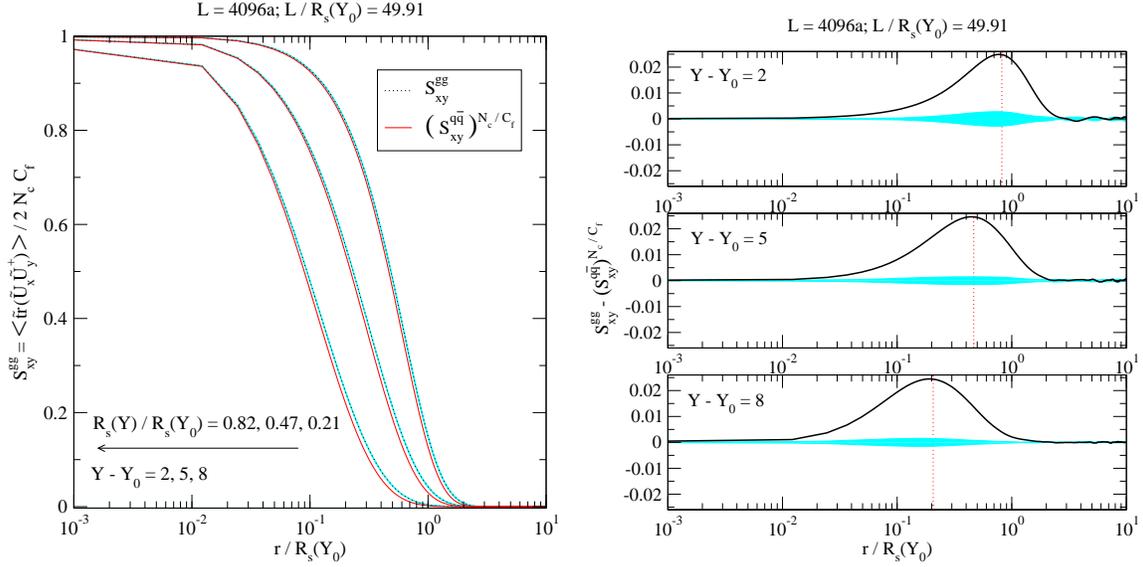

  \centering
  \includegraphics[width=.49\linewidth]{adj_comp_4096_128}
\hfill
  \includegraphics[width=.49\linewidth]{adj_diff_4096_128}
  \caption{Violation of Casimir scaling for fundamental and adjoint two point
    correlators for three different rapidities plotted as a function
    of dipole size $r$ in units of correlation length $R_s(Y)$.  Left:
    adjoint correlators compared to the Casimir-rescaled fundamental
    correlators. Error bands indicating numerical uncertainty are too
    narrow to be clearly visible. Right: differences of the two.
    (These are more stable than ratios which tend to become
    uncontrollable at large distances.) The bands indicate the size of
    the errors.  The violation follows correlation length $R_s(Y)$,
    indicated by vertical lines. It would appear to scale with $Q_s$
    well within errors. Note that this is in a regime where the dipole
    cross section has not yet reached its scaling regime and decidedly
    does \emph{not} scale yet. }
  \label{fig:casimir-scaling-viol}
\end{figure}

With the numerical results from JIMWLK at hand it is straightforward
look for the actual presence of Casimir scaling-violating effects in
JIMWLK evolution. This is explicitly shown in
Fig.~\ref{fig:casimir-scaling-viol}.
We note in particular that the violations of Casimir scaling do not
grow with energy: they seem to qualitatively scale with $Q_s$ and
might, if anything even be slowly erased, but any firm conclusion to
that effect is beyond the present numerical accuracy, in particular
because of the short lever arm available before the simulation starts
to ``fall through the lattice'' (i.e. runs out of UV phase space).
Note that this qualitative $Q_s$-scaling behavior occurs in a region
far outside the $Q_s$-scaling region of the dipole correlator itself.
Presently we have no systematic explanation for this observation. (The
$Q_s$-scaling region of the dipole correlator is not reachable in the
current simulations of JIMWLK equation due to the limited UV phase
space on the lattices used.)

\begin{figure}[!thb]
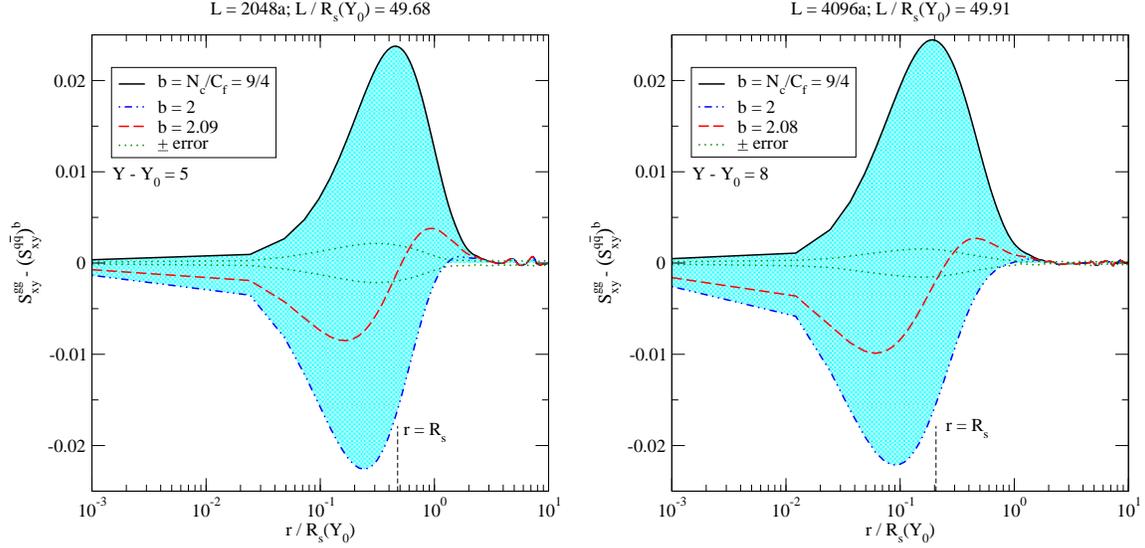

  \centering
    \includegraphics[width=.49\textwidth]{exponent_2048_fix}\hfill
    \includegraphics[width=.49\textwidth]{exponent_4096_fix}
    \caption{The violation of Casimir scaling in JIMWLK evolution for two
      lattice sizes with the same infrared phase space as indicated by
      the $L/R_s (Y_0)$ value.  The plots explore the correlator
      difference $S^{gg}_{\bm x\bm y}-\bigl(S^{q\Bar q}_{\bm x\bm
        y}\bigr)^b$ as a function of dipole size $r$ for values of
      $b=2$ (the BK result) up to the GT value of
      $N_c/C_{\text{f}}=9/4$ at $N_c=3$. Intermediate values of $b$
      fall into the shaded areas. The result for ``a best fit'' is
      indicated as a dashed (red) line. Clearly a simple modification
      of the power alone is not sufficient to remove the mismatch. To
      remedy the situation one would either need a more complex
      functional relationship between dipole correlators or, more
      likely, new degrees of freedom to alter the $(\bm x-\bm y)^2$
      dependence in at least one of the correlators shown. }
  \label{fig:new-dof-needed}
\end{figure}

To illustrate that the violations of Casimir scaling are driven by nontrivial
coordinate dependence we show in Fig.~\ref{fig:new-dof-needed} that no power
law relationship modeled on Eq.~(\ref{eq:casimir-scaling-app}) 
can provide a good explanation for the differences observed in
Fig.~\ref{fig:casimir-scaling-viol}.
By itself, this does not exclude a more intricate functional relationship, but
that in turn would require its own explanation. Our earlier arguments would
lead us to believe that it is much more likely and natural that new degrees of
freedom (starting with four point contributions to the
average~(\ref{eq:func-2n3-average-cartesian})) are needed to explain this
difference.

\section{Conclusions}
\label{sec:conclusions}

We have explored the size and nature of $1/N_c$ corrections in the JIMWLK
equation and have found that quite natural cancellations lead to the much
stronger suppression than the naively expected $1/N_c^2 \approx 10\%$ (for
$N_c=3$) observed in~\cite{Rummukainen:2003ns}.  The argument is based first
on both group theoretical coincidence limits for singlet Wilson line
correlators and scaling with the saturation scale $Q_s(Y)$ to establish
suppression of contributions in most of configuration space. All remaining
contributions are then shown to be then decoupled from evolution after
suppression by the BK kernel has been taken into account.

We have shown that $1/N_c$ corrections enter through the factorization
violation $\Delta$ and have explored its properties. It is bounded by
$1/N_c^2$ from above, but is much smaller than $1/N_c^2$ for most of its phase
space, due to saturation effects controlled by coincidence limits leading to
extra suppression. The argument is generic and can be easily applied beyond
the leading $\ln(1/x)$ approximation used here.  Thus saturation effects
provide an extra suppression of $1/N_c^2$ corrections to the BK evolution that
reduces the difference between the JIMWLK and BK results for the dipole
scattering amplitude considerably at any accuracy. While NLO corrections will
have some quantitative impact, we do not expect them to grossly change the LO
result for $\Delta$ or correlator differences (when compared at the same
$R_s(Y)$ or $Q_s(Y)$). At one loop we find typical contributions to $\Delta$
of the order of $10^{-3}$ or $0.1\%$ of the individual correlator values for
correlator differences.

To complement the qualitative discussion for correlators in JIMWLK evolution
not only numerically but also analytically, we have made a step beyond the
leading $1/N_c$ BK truncation by exploring an alternative truncation of JIMWLK
evolution in the spirit of a Glauber-Mueller iterated two-reggeon exchange
truncation that we dubbed the Gaussian truncation. This truncation includes a
minimal set of subleading $1/N_c$ corrections necessary to restore the
coincidence limits for correlators that are at the core of our cancellation
argument for subleading contributions. Correspondingly, it includes a minimal
set of factorization violations. They turn out to have the right qualitative
structure but are numerically noticeably smaller than the factorization
violation of full JIMWLK evolution. The Gaussian truncation allows access to a
larger subset of the Balitsky hierarchies than BK truncation by treating
evolution equations for dipole operators in arbitrary representations
consistently. As a result the Gaussian truncation proves to be a somewhat
better approximation to full JIMWLK evolution. Accordingly, one of its main
consequences, Casimir scaling between dipole operators of different
representations, turns out not to be strongly violated in full JIMWLK.

We have firmly established that factorization violations slow down
evolution compared to the BK truncation, both in the minimalist from
introduced in the Gaussian truncation and in full JIMWLK. One of
main differences between the two is that factorization violation in
JIMWLK persist during evolution on a level comparable with the
factorization violation present in the GB-W--like initial condition
used, while they become notably smaller in the Gaussian truncation. 

Evolution speed is somewhat more sensitive to $1/N_c$ corrections than the
factorization violations in the correlators due to an enhancement of
contributions from small parent ``dipoles.'' At one loop accuracy this leads
to a relative slowdown of JIMWLK evolution that approaches 3-5\% near the
scaling region. Running coupling corrections are known to reduce evolution
speed by suppressing the relative importance of small parent dipoles. We have
argued that the same mechanism is likely to also reduce the {\em relative}
slowdown, i.e. the impact of $1/N_c$ corrections on evolution speed.

The subleading-$N_c$ terms present in JIMWLK but absent in BK can likely be
attributed to multi--reggeon $t$-channel exchanges and multi--reggeon
splitting vertices. The smallness of the difference between the dipole
scattering amplitude obtained from JIMWLK and BK appears to indicate that
multi--reggeon effects are not important for this observable. Further
investigation is needed to clarify if this is indeed the case.

Casimir scaling violations present in JIMWLK evolution provide a means to
assess multi--reggeon exchange contributions. We have illustrated this both
with a sketch model that includes odderon contributions and a numerical
comparison of $g g$ and $q\Bar q$ dipoles. While the odderon contributions
play no role for the observables considered here and are generically
suppressed by evolution, the multi--reggeon contributions present here are
small but contribute throughout evolution. We have numerical hints at
$Q_s(Y)$-scaling behavior of these contributions that sets in much earlier
than geometric scaling of dipole correlators.  At present we have no
systematic explanation for this observation other than ``naturalness.''

Our whole discussion was carried out at the leading log level, without any NLO
contributions taken into account, despite the fact that they are known to
strongly influence evolution speed. This is partly due to necessity: no
numerically practical way has been devised to include higher order effects, for
example the running coupling corrections obtained
in~\cite{Gardi:2006rp,Kovchegov:2006vj,Balitsky:2006wa}. However, since the
arguments given for the suppression of factorization violations are completely
generic, one expects the observations made here to persist to higher orders in
the perturbative expansion, even though quantitative modifications are
expected to affect the precise numerical result of the cancellations observed
at leading order.

\section*{Acknowledgments}

We would like to thank G\"{o}sta Gustafson for encouraging us to put our
insights on the smallness of factorization violations into writing. In the
process we found ourselves prompted to extend the numerical treatment
of\cite{Rummukainen:2003ns} to complement the analytical insight.

The research of Yu.K. is sponsored in part by the U.S. Department of Energy
under Grant No. DE-FG02-05ER41377. The research of K.R. is partially supported
by Academy of Finland grant 114371.  J.K. acknowledges support from the Jenny
and Antti Wihuri Foundation. H.W. is supported by a research grant of the
University of Oulu.

\appendix

\section{Generic coincidence limits}
\label{sec:gener-coinc-limits}
\renewcommand{\theequation}{A\arabic{equation}}
  \setcounter{equation}{0}

To understand the coincidence limits for the operator $   \big[\Tilde U_{\bm{z}}\big]^{a b}\
   \overset{{\scriptscriptstyle\cal R}}\tr(
\overset{{\scriptscriptstyle\cal R}} t^a 
\overset{{\scriptscriptstyle\cal R}} U_{\bm{x}}
\overset{{\scriptscriptstyle\cal R}} t^b 
\overset{{\scriptscriptstyle\cal R}}
  U^\dagger_{\bm{y}})
 $, generic properties of the adjoint representation are useful. They
 immediately give rise to the identities
\begin{align}
    \label{eq:three-adj-rewrites-app}
    \big[\Tilde U_{\bm{z}}\big]^{a b}\
   \overset{{\scriptscriptstyle\cal R}}\tr(
\overset{{\scriptscriptstyle\cal R}} t^a 
\overset{{\scriptscriptstyle\cal R}} U_{\bm{x}}
\overset{{\scriptscriptstyle\cal R}} t^b 
\overset{{\scriptscriptstyle\cal R}}
  U^\dagger_{\bm{y}})
 = \big[\Tilde U_{\bm{x}}^\dagger \Tilde U_{\bm{z}}\big]^{c b}\
\overset{{\scriptscriptstyle\cal R}}\tr(
\overset{{\scriptscriptstyle\cal R}} U_{\bm{x}}
\overset{{\scriptscriptstyle\cal R}} t^c 
\overset{{\scriptscriptstyle\cal R}} t^b 
\overset{{\scriptscriptstyle\cal R}} U^\dagger_{\bm{y}})
  = \big[\Tilde U_{\bm z} \Tilde U_{\bm y}^\dagger\big]^{a c}\
\overset{{\scriptscriptstyle\cal R}}\tr(
\overset{{\scriptscriptstyle\cal R}} t^a 
\overset{{\scriptscriptstyle\cal R}} U_{\bm{x}}
\overset{{\scriptscriptstyle\cal R}} U^\dagger_{\bm{y}}
\overset{{\scriptscriptstyle\cal R}} t^b 
)
\end{align}
by pulling adjoint factors out of the trace. 

In the limit $\bm x=\bm y$ this
is proportional to 
\begin{align}
\label{eq:alpha_R-def-app}
\overset{{\scriptscriptstyle\cal R}}\tr(
\overset{{\scriptscriptstyle\cal R}} t^a \overset{{\scriptscriptstyle\cal R}}
t^b ) = \alpha_{\cal R} \delta^{a b},
\end{align}
and the first task is to understand the constant. Clearly it has to be
proportional to the Casimir of the representation, but normalization
conventions do also play a role: With standard conventions the fundamental
representation has $\alpha_{\text{fund}}=\alpha_F=\frac12$, while in the
adjoint representation one obtains
$\alpha_{\text{adjoint}}=\alpha_A=N_{\text{c}}$. This can, in fact, be
expressed via the Casimir values and the dimension of the representation. The
usual definition of the quadratic Casimir,
\begin{align}
  \label{eq:Cas-const-app}
  \left[\overset{{\scriptscriptstyle\cal R}} t^a 
  \overset{{\scriptscriptstyle\cal R}} t^a 
  \right]_{i j} =C_{\cal R} \delta_{i j}
\ ,
\end{align}
implies
\begin{align}
  \label{eq:cas-dim-app}
  \overset{{\scriptscriptstyle\cal R}}\tr\left(
\overset{{\scriptscriptstyle\cal R}} t^a 
\overset{{\scriptscriptstyle\cal R}} t^a 
\right) = C_{\cal R} d_{\cal R}
\end{align}
and thus, together with the definition of $\alpha_{\cal R}$ above,
$\alpha_{\cal R} d_{A}=C_{\cal R} d_{\cal R}$ or
\begin{align}
  \label{eq:alpha_R-app}
  \alpha_{\cal R} = C_{\cal R} \frac{d_{\cal R}}{d_A}
\end{align}
with $d_A$ the dimension of the adjoint representation. This readily leads to
$\alpha_A = C_A = N_{\text{c}}$ and
$\alpha_F=C_f\frac{N_{\text{c}}}{2N_{\text{c}}C_f}=\frac12$ as obtained from
direct calculation.

This allows to write the $\bm x= \bm y$ limit as
\begin{subequations}
  \label{eq:3-point-generic-coincidence-app}
\begin{align}
  \label{eq:generic-x=y-limit-app}
\lim\limits_{y\to x}     \big[\Tilde U_{\bm{z}}\big]^{a b}\
  \overset{{\scriptscriptstyle\cal R}}\tr(
\overset{{\scriptscriptstyle\cal R}} t^a 
\overset{{\scriptscriptstyle\cal R}} U_{\bm{x}}
\overset{{\scriptscriptstyle\cal R}} t^b 
\overset{{\scriptscriptstyle\cal R}}
  U^\dagger_{\bm{y}})
= & \lim\limits_{y\to x} \big[\Tilde U_{\bm z} \Tilde U_{\bm y}^\dagger\big]^{a c}\
\overset{{\scriptscriptstyle\cal R}}\tr(
\overset{{\scriptscriptstyle\cal R}} t^a 
\overset{{\scriptscriptstyle\cal R}} U_{\bm{x}}
\overset{{\scriptscriptstyle\cal R}} U^\dagger_{\bm{y}}
\overset{{\scriptscriptstyle\cal R}} t^b 
)
=
\big[\Tilde U_{\bm z} \Tilde U_{\bm x}^\dagger\big]^{a c}\
\overset{{\scriptscriptstyle\cal R}}\tr(
\overset{{\scriptscriptstyle\cal R}} t^a 
\overset{{\scriptscriptstyle\cal R}} t^c 
)
\notag \\ 
= & \ \big[\Tilde U_{\bm z} \Tilde U_{\bm x}^\dagger\big]^{a c} \alpha_{\cal R}
\delta^{a c}
=  C_{\cal R} \frac{d_{\cal R}}{d_A}
\Tilde\tr\left(\Tilde U_{\bm z} \Tilde U_{\bm x}^\dagger \right) .
\end{align}
Note that this leads to a correlator in the adjoint representation,
irrespective of ${\cal R}$. The only reference to ${\cal R}$ is the
proportionality factor.

The limit $\bm z\to \bm y\ \text{or}\ \bm x$ is simpler:
 \begin{align}
 \lim\limits_{
      \bm z\to \bm y\ \text{or}\ \bm x}    
\big[\Tilde U_{\bm{z}}\big]^{a b}\
   \overset{{\scriptscriptstyle\cal R}}\tr(
\overset{{\scriptscriptstyle\cal R}} t^a 
\overset{{\scriptscriptstyle\cal R}} U_{\bm{x}}
\overset{{\scriptscriptstyle\cal R}} t^b 
\overset{{\scriptscriptstyle\cal R}}
  U^\dagger_{\bm{y}}) 
= &\ \lim\limits_{
      \bm z\to \bm y\ \text{or}\ \bm x}    
   \overset{{\scriptscriptstyle\cal R}}\tr(
\overset{{\scriptscriptstyle\cal R}} t^a 
\overset{{\scriptscriptstyle\cal R}} U_{\bm{x}}
\overset{{\scriptscriptstyle\cal R}} U_{\bm{z}}^\dagger
\overset{{\scriptscriptstyle\cal R}} t^a 
\overset{{\scriptscriptstyle\cal R}} U_{\bm{z}} 
\overset{{\scriptscriptstyle\cal R}} U^\dagger_{\bm{y}})
=  C_{\cal R} \overset{{\scriptscriptstyle\cal R}}\tr(
\overset{{\scriptscriptstyle\cal R}} U_{\bm x}
\overset{{\scriptscriptstyle\cal R}} U^\dagger_{\bm y}) 
\ .
\end{align}

The completely local limit can be obtained directly
  from~\eqref{eq:generic-x=y-limit-app} to be $C_{\cal R} d_{\cal R}$ or via
  \begin{align}
 \lim\limits_{
      \bm z\to \bm y;\ \bm y\to \bm x}    
\big[\Tilde U_{\bm{z}}\big]^{a b}\
   \overset{{\scriptscriptstyle\cal R}}\tr(
\overset{{\scriptscriptstyle\cal R}} t^a 
\overset{{\scriptscriptstyle\cal R}} U_{\bm{x}}
\overset{{\scriptscriptstyle\cal R}} t^b 
\overset{{\scriptscriptstyle\cal R}}
  U^\dagger_{\bm{y}}) 
=\lim\limits_{
      \bm z\to \bm y;\ \bm y\to \bm x}    
   \overset{{\scriptscriptstyle\cal R}}\tr(
\overset{{\scriptscriptstyle\cal R}} t^a 
\overset{{\scriptscriptstyle\cal R}} U_{\bm{x}}
\overset{{\scriptscriptstyle\cal R}} U_{\bm{z}}^\dagger
\overset{{\scriptscriptstyle\cal R}} t^a 
\overset{{\scriptscriptstyle\cal R}} U_{\bm{z}}
\overset{{\scriptscriptstyle\cal R}}
  U^\dagger_{\bm{y}}) 
=  \overset{{\scriptscriptstyle\cal R}}\tr\left(
\overset{{\scriptscriptstyle\cal R}} t^a 
\overset{{\scriptscriptstyle\cal R}} t^a 
\right) = C_{\cal R} d_{\cal R}
\ .
\end{align}
\end{subequations}
This establishes Eqs.~(\ref{eq:3-point-generic-coincidence}) of Sect.~\ref{sec:orig-fact}.

\section{Gaussian target averages}

\label{sec:gaussian-averages}
\renewcommand{\theequation}{B\arabic{equation}}
  \setcounter{equation}{0}

While direct calculation of specific correlators using the averaging
procedure~(\ref{eq:func-Gaussian-average}) is in many cases straightforward,
the calculation can be often simplified significantly by using differential
equations.  This relies on the observation that
\begin{align}
  \label{eq:gen-dY-eq-app}
  \frac{d}{dY}\langle \ldots \rangle(Y) =\ & -\frac12 \langle \int\!
  d^2u\, d^2v\ G_{Y,\bm{u v}} \frac{\delta}{\delta A^{a +}_{\bm u,y}}
  \frac{\delta}{\delta A^{a +}_{\bm v,y}} \ldots \rangle(Y) .
\end{align}
Applied to concrete examples this takes its simplest form if the right-hand
side is directly proportional to $\langle \ldots \rangle(Y)$ itself, but can
be useful even in more general cases. We will explicitly address the examples
needed in Sect.~\ref{sec:orig-fact}.

\paragraph{Two point projectile ${\cal R}$-$\bar {\cal R}$ correlators:}
Using the notation~(\ref{eq:calGdef}) and a prime to denote a $Y$
derivative, straightforward algebra leads to
\begin{align}
  \label{eq:trUUdagger-app}
   \frac{d}{dY}\langle \overset{{\scriptscriptstyle\cal R}}\tr(
\overset{{\scriptscriptstyle\cal R}} U_{\bm{x}}
\overset{{\scriptscriptstyle\cal R}}
  U^\dagger_{\bm{y}})
\rangle
= -{\cal G}'_{Y,\bm{x y}}\ 
\langle \overset{{\scriptscriptstyle\cal R}}\tr(
\overset{{\scriptscriptstyle\cal R}} t^a 
\overset{{\scriptscriptstyle\cal R}} U_{\bm{x}}
\overset{{\scriptscriptstyle\cal R}}
  U^\dagger_{\bm{y}}\overset{{\scriptscriptstyle\cal R}} t^a )
\rangle(Y)
= -C_{\cal R} {\cal G}'_{Y,\bm{x y}}\ 
\langle \overset{{\scriptscriptstyle\cal R}}\tr(
\overset{{\scriptscriptstyle\cal R}} U_{\bm{x}}
\overset{{\scriptscriptstyle\cal R}}
  U^\dagger_{\bm{y}}
\rangle(Y)
\end{align}
which is readily solved to obtain
\begin{align}
  \label{eq:UUdaggersol-app}
  \langle \overset{{\scriptscriptstyle\cal R}}\tr(
\overset{{\scriptscriptstyle\cal R}} U_{\bm{x}}
\overset{{\scriptscriptstyle\cal R}}
  U^\dagger_{\bm{y}})
\rangle(Y) = d_{\cal R} e^{-C_{\cal R}{\cal G}_{Y,\bm{x y}}}
\ . 
\end{align}
The freedom in the initial condition was used to accommodate the normalization
factor $d_{\cal R}$.

\paragraph{Three point projectile adjoint-${\cal R}$-$\bar {\cal R}$ correlators:}
These involve several distinct color structures.
\begin{align}
  \label{eq:UtrtUtUdagger-eq-app}
   \frac{d}{dY} & \langle 
\big[\Tilde U_{\bm{z}}\big]^{a b} 
\overset{{\scriptscriptstyle\cal R}}\tr(
\overset{{\scriptscriptstyle\cal R}} t^a 
\overset{{\scriptscriptstyle\cal R}} U_{\bm{x}}
\overset{{\scriptscriptstyle\cal R}} t^b 
\overset{{\scriptscriptstyle\cal R}}
  U^\dagger_{\bm{y}})
\rangle(Y)
=
-{\cal G}'_{Y,\bm{x z}} \langle 
\big[\Tilde t^i \Tilde U_{\bm{z}}\big]^{a b}\
\overset{{\scriptscriptstyle\cal R}}\tr(
\overset{{\scriptscriptstyle\cal R}} t^a 
\overset{{\scriptscriptstyle\cal R}} t^i 
\overset{{\scriptscriptstyle\cal R}} U_{\bm{x}}
\overset{{\scriptscriptstyle\cal R}} t^b 
\overset{{\scriptscriptstyle\cal R}}
  U^\dagger_{\bm{y}})
\rangle(Y)
\notag \\ & \
+
{\cal G}'_{Y,\bm{z y}} \langle 
\big[\Tilde t^i \Tilde U_{\bm{z}}\big]^{a b}\
\overset{{\scriptscriptstyle\cal R}}\tr(
\overset{{\scriptscriptstyle\cal R}} t^a 
\overset{{\scriptscriptstyle\cal R}} U_{\bm{x}}
\overset{{\scriptscriptstyle\cal R}} t^b 
\overset{{\scriptscriptstyle\cal R}}
  U^\dagger_{\bm{y}} \overset{{\scriptscriptstyle\cal R}} t^i 
)
\rangle(Y)
+
{\cal G}'_{Y,\bm{x y}}
\langle 
\big[\Tilde U_{\bm{z}}\big]^{a b}\
\overset{{\scriptscriptstyle\cal R}}\tr(
\overset{{\scriptscriptstyle\cal R}} t^a 
\overset{{\scriptscriptstyle\cal R}} t^i 
\overset{{\scriptscriptstyle\cal R}} U_{\bm{x}}
\overset{{\scriptscriptstyle\cal R}} t^b 
\overset{{\scriptscriptstyle\cal R}}
  U^\dagger_{\bm{y}}
\overset{{\scriptscriptstyle\cal R}} t^i 
)
\rangle(Y)
\notag \\
\notag \\
= \ &
-\left[\frac{N_c}2\left(
  {\cal G}'_{Y,\bm{x z}} + {\cal G}'_{Y,\bm{z y}} 
\right)
+ {\cal G}'_{Y,\bm{x y}} \left(C_{\cal R}-\frac{N_c}2\right)
\right]
\langle 
\big[\Tilde U_{\bm{z}}\big]^{a b}\
\overset{{\scriptscriptstyle\cal R}}\tr(
\overset{{\scriptscriptstyle\cal R}} t^a 
\overset{{\scriptscriptstyle\cal R}} U_{\bm{x}}
\overset{{\scriptscriptstyle\cal R}} t^b 
\overset{{\scriptscriptstyle\cal R}}
  U^\dagger_{\bm{y}})
\rangle(Y)
\end{align}
where we have used
\begin{align}
  \label{eq:t-resuffle-1-app}
\overset{{\scriptscriptstyle\cal R}} t^i 
\overset{{\scriptscriptstyle\cal R}} t^a 
\overset{{\scriptscriptstyle\cal R}} t^i & = 
\overset{{\scriptscriptstyle\cal R}} t^a 
\overset{{\scriptscriptstyle\cal R}} t^i 
\overset{{\scriptscriptstyle\cal R}} t^i 
+
[\overset{{\scriptscriptstyle\cal R}} t^i
, 
\overset{{\scriptscriptstyle\cal R}} t^a 
]
\overset{{\scriptscriptstyle\cal R}} t^i \notag \\
& = 
\overset{{\scriptscriptstyle\cal R}} t^a 
C_{\cal R} 
+ i f_{i a k}
\overset{{\scriptscriptstyle\cal R}} t^k 
\overset{{\scriptscriptstyle\cal R}} t^i
= \overset{{\scriptscriptstyle\cal R}} t^a 
C_{\cal R} 
+ i f_{i a k}
\frac12 i f_{k i j} \overset{{\scriptscriptstyle\cal R}} t^j
=  
\overset{{\scriptscriptstyle\cal R}} t^a \left(C_{\cal R}-\frac{N_c}2\right).
\end{align}
The nontrivial point is that this holds for any representation ${\cal R}$.
Integrating~(\ref{eq:UtrtUtUdagger-eq-app}) one finds
\begin{align}
  \label{eq:UtrtUtUdagger-app}
\langle 
\big[\Tilde U_{\bm{z}}\big]^{a b} &
\overset{{\scriptscriptstyle\cal R}}\tr(
\overset{{\scriptscriptstyle\cal R}} t^a 
\overset{{\scriptscriptstyle\cal R}} U_{\bm{x}}
\overset{{\scriptscriptstyle\cal R}} t^b 
\overset{{\scriptscriptstyle\cal R}}
  U^\dagger_{\bm{y}})
\rangle(Y) = C_{\cal R} d_{\cal R} e^{
  -\frac{N_c}2\left(
  {\cal G}_{Y,\bm{x z}} + {\cal G}_{Y,\bm{z y}}- {\cal G}_{Y,\bm{x y}}  
\right)
-C_{\cal R} {\cal G}_{Y,\bm{x y}} 
}  
\ ,
\end{align}
again with the free initial condition used to set the normalization properly.

\section{Odderon contributions lead to Casimir scaling violations}
\label{sec:odder-contr-lead}
\renewcommand{\theequation}{C\arabic{equation}} \setcounter{equation}{0}
  
Here we explore the results of
applying~\eqref{eq:func-2n3-average-cartesian} to the calculation of
dipole and 3-point correlators, as was done earlier with the Gaussian
truncation.

It turns out that starting from this level of three-point $t$-channel
correlators generic expressions for arbitrary representations ${\cal R}$ can
not be given. This is a consequence of the arbitrarily complicated
decomposition of a general $s$-channel ${\cal R}\Bar{\cal R}$-dipole into
irreducible representations. These start to mix in nontrivial ${\cal
  R}$-dependent ways beyond the Gaussian truncation.

We have therefore restricted ourselves to ${\cal R}$ being either the
fundamental or the adjoint representation. Here it turns out that $G^f$ in
\eqref{eq:func-2n3-average-cartesian} does not contribute at all to correlators
of Wilson lines, and that $G^d$, as expected, can be thought of as an
odderon contribution. This will allow us to compare the results we are about
to obtain to~\cite{Kovchegov:2003dm} by counting $G^d$ as ${\cal O}(\alpha_s)$
(using the quasi-classical counting) and correspondingly expanding the
equations we get to the lowest order in $G^d$, as was done in
\eqref{eq:func-2n3-average-cartesian-2}. Our results can also be compared
to~\cite{Hatta:2005as} if we keep the $G^d$ contributions to all orders.

For the correlators in question, this contribution generates imaginary parts
wherever the representation ${\cal R}$ is not intrinsically {\em real}, such
as the adjoint representation. For the limited set of correlators we are
looking at, only the $\bm x \leftrightarrow\bm y$ antisymmetric combination
$\int^Y\!dY'\bigl(
  G^d_{Y',\bm{y y x}}- G^d_{Y',\bm{y x x}}
  \bigr)
$
enters. For compactness, we will also absorb a constant in
the shorthand to be used below. We define
\begin{align}
  \label{eq:CalGO}
  {\cal G}_{Y,\bm{x y}}^{\cal O} :=\frac{C_d}4 \int\limits^Y\!dY'\bigl(
  G^d_{Y',\bm{y y x}}- G^d_{Y',\bm{y x x}}
  \bigr)
\ ,
\end{align}
with 
\begin{equation}
  \label{eq:C_d}
  C_d := \frac{N_c^2-4}{N_c}
\end{equation}
characterizing the symmetric ``octet'' (in SU(3) parlance) in the
decomposition of a $g g$-dipole into invariant multiplets
(see.~\cite{Cvitanovic-GroupTheory}, Sec 9.12 for a systematic treatment that
is much more practical than most). Note that ${\cal G}_{Y,\bm{x x}}^{\cal O}
=0$.

Let us begin with the three point $q {\bar q} g$ function. With ${\cal
  R}$ the fundamental representation one finds
\begin{align}
  \label{eq:UtrtUtUdagger3p}
 \langle 
& \big[\Tilde U_{\bm{z}}\big]^{a b} 
\tr(
 t^a 
 U_{\bm{x}}
 t^b 
  U^\dagger_{\bm{y}})
\rangle(Y)
=
N_c C_f e^{
-\left\{\bigr[\frac{N_c}2\left(
  {\cal G}_{\bm{x z}} + {\cal G}_{\bm{z y}} 
\right)
+ {\cal G}_{\bm{x y}} \left(C_f-\frac{N_c}2\right)
\bigr]
+\frac{i}2\bigr[
-\frac{
{\cal G}_{\bm{x y}}^{\cal O}
}{N_c}
+N_c \left(
{\cal G}_{\bm{x z}}^{\cal O}+{\cal G}_{\bm{z y}}^{\cal O}
\right)
\bigr]
\right\}(Y)
} \notag \\ 
& = N_c C_f e^{
-\bigr[\frac{N_c}2\left(
  {\cal G}_{\bm{x z}} + {\cal G}_{\bm{z y}} 
\right)
+ {\cal G}_{\bm{x y}} \left(C_f-\frac{N_c}2\right)
\bigr]} \ \left\{ 1
-\frac{i}{2} \left[ 
-\frac{
{\cal G}_{\bm{x y}}^{\cal O}
}{N_c}
+N_c \left(
{\cal G}_{\bm{x z}}^{\cal O}+{\cal G}_{\bm{z y}}^{\cal O}
\right)
\right]
 + o \left( {\cal G}^{{\cal O} \, 2} \right) \right\}(Y)
\ .
\end{align}
The expanded expression in the second line serves to recall that
higher orders in this expansion in powers of ${\cal G}^{\cal O}$
are beyond the control of our approximation.

The coincidence limits~(\ref{eq:3-point-generic-coincidence}), which
are at the heart of factorization violations provide relationships
with dipole correlators also here.  Eq. (\ref{eq:UtrtUtUdagger3p}) in
the limit $\bm x=\bm y$ provides the expression for the adjoint
correlator
\begin{align}
  \label{eq:adjoint-noodderon}
  \langle\Tilde\tr(\Tilde U_{\bm{z}}\Tilde U_{\bm{x}}^\dagger)\rangle(Y)
 =\ \langle\big[\Tilde U_{\bm{z}}\big]^{a b} 
\ 2\, \tr(
 t^a 
 U_{\bm{x}}
 t^b 
  U^\dagger_{\bm{x}})
\rangle(Y)
= \ 2 N_c C_f\ e^{-N_c\, {\cal G}_{Y,\bm{z x}}}
\end{align}
which turns out to be unmodified from the Gaussian truncation and remains
completely independent of the odderon term ${\cal G}^{\cal O}$ even without
expanding in the odderon contribution. Note that this is more stringent than
the group theoretical requirement that the $g g$-dipole has to be real, which
would have allowed even powers of ${\cal G}^{\cal O}$ to appear in an all
orders expression in terms of ${\cal G}^{\cal O}$.

The fundamental correlator, on the other hand, does acquire modifications both
to real and imaginary parts. In accordance with the limits $\bm z=\bm x$ and
$\bm z=\bm y$ of~(\ref{eq:UtrtUtUdagger3p}) one finds
\begin{align}
  \label{eq:qbarq-odd}
\langle \tr( U_{{\bm x}} U_{{\bm y}}^\dagger) \rangle(Y)
= & \frac1{C_f} \langle\big[\Tilde U_{\bm x}\big]^{a b} 
 \tr(
 t^a 
 U_{\bm{x}}
 t^b 
  U^\dagger_{\bm{y}})
\rangle(Y)
= \frac1{C_f} \langle\bigl(\big[\Tilde U_{\bm y}\big]^{a b} 
 \tr(
 t^a 
 U_{\bm{x}}
 t^b 
  U^\dagger_{\bm{y}})
\bigr)^\dagger
\rangle(Y)
\notag \\ = &\
 N_c\  e^{-C_f \bigl({\cal G}_{Y,\bm{x y}} + i\, 
{\cal G}_{Y,\bm{x y}}^{\cal O}\bigr)
} = N_c\  e^{-C_f {\cal G}_{Y,\bm{x y}}} \, \bigl[ 1 - i\, C_f \,
{\cal G}_{Y,\bm{x y}}^{\cal O} + o \left( {\cal G}^{{\cal O} \, 2} \right) \bigr].
\end{align}
Contrary to what happens in the adjoint representation, this result
would emerge from our earlier expression for the $q\Bar q$-dipole
using the substitution
\begin{align}
  \label{eq:GT-to-Odd-subst}
  {\cal G}_{Y,\bm{x y}} \to {\cal G}_{Y,\bm{x y}} + i\, 
{\cal G}_{Y,\bm{x y}}^{\cal O} 
\ .
\end{align}
Comparing Eqs. (\ref{eq:qbarq-odd}) and (\ref{eq:adjoint-noodderon})
we can see that the odderon contribution introduces violation of the
Casimir scaling of \eqref{eq:casimir-scaling-app}.  We therefore can
conjecture that Casimir scaling violating effects are due to multiple
reggeon exchanges.

The evolution equation for the $q\Bar q$ dipole (Eq.~(\ref{eq:prefactUR}) with
${\cal R}$ the fundamental representation, after insertion
of~(\ref{eq:qbarq-odd}) and~(\ref{eq:UtrtUtUdagger3p})) leads to
\begin{align}
  \label{eq:fund-odd-final}
  \frac{d}{dY} \bigl({\cal G}_{Y,\bm{x y}} + & i\, 
{\cal G}_{Y,\bm{x y}}^{\cal O} \bigr)
=  \ \frac{\alpha_s}{\pi^2} \int\! d^2z\ K_{\bm{x z y}}
\left(
1-e^{
 -
   \frac{N_c}2\left[
   \left({\cal G}_{\bm{x z}} +
i\,{\cal G}_{\bm{x z}}^{\cal O}
\right)
+
\left({\cal G}_{\bm{z y}} +
i\,{\cal G}_{\bm{z y}}^{\cal O}
\right)
-\left({\cal G}_{\bm{x y}} +
i\,{\cal G}_{\bm{x y}}^{\cal O}
\right)
 \right](Y)
}
\right)
\end{align}
which repeats the structure of~(\ref{eq:tilde-G-evo-short}), again with the
simple substitution~(\ref{eq:GT-to-Odd-subst}).

As we have control only over the terms linear in ${\cal G}^{\cal O}$
we expand \eqref{eq:fund-odd-final} and use \eqref{eq:tilde-G-evo-short} to
obtain
\begin{align}
  \label{eq:fund-odd-final-2}
  \frac{d}{dY} \, {\cal G}_{Y,\bm{x y}}^{\cal O} 
=  \ \frac{\alpha_s \, N_c}{2 \, \pi^2} \int\! d^2z\ K_{\bm{x z y}}
\ e^{
 -
   \frac{N_c}2\left[
   {\cal G}_{\bm{x z}} + {\cal G}_{\bm{z y}}
- {\cal G}_{\bm{x y}}
 \right](Y)
} \ \left[
   {\cal G}_{Y,\bm{x z}}^{\cal O}  + {\cal G}_{Y,\bm{z y}}^{\cal O} 
- {\cal G}_{Y,\bm{x y}}^{\cal O} 
 \right] .
\end{align}
Defining the real part of the $S$-matrix and the odderon exchange
amplitude $\cal O$ by
\begin{align}
  S_{Y, \bm{x y}} = e^{- C_f \, {\cal G}_{\bm{x y}}}, \ \ \ {\cal
    O}_{Y, \bm{x y}} = - i \, C_f \, {\cal G}_{Y, \bm{x y}}^{\cal O}
  \, e^{-C_f \, {\cal G}_{\bm{x y}}}
\end{align}
and using these definitions in \eqref{eq:fund-odd-final-2} in the
large-$N_c$ limit (which is needed here just like it was needed to
derive BK equation (\ref{eq:BK-S}) from the GT truncation
(\ref{eq:tilde-G-evo-short}) in Sect. \ref{sec:rest-coinc-limits}) one
derives the non-linear evolution equation for the odderon found
in~\cite{Kovchegov:2003dm} (see also \cite{Hatta:2005as})
\begin{align}\label{odd}
  \frac{d}{dY} \, {\cal O}_{Y, \bm{x y}} = \frac{\as \, N_c}{2 \,
    \pi^2} \, \int d^2 z \, {\cal K}_{\bm{x z y}} \, \left[ {\cal
      O}_{Y, \bm{x z}} \, S_{Y, \bm{z y}} + S_{Y, \bm{x z}} \, {\cal
      O}_{Y, \bm{z y}} - {\cal O}_{Y, \bm{x y}} \right].
\end{align}

In \cite{Kovchegov:2003dm, Hatta:2005as} the authors discuss how this
equation maps onto the BJKP hierarchy \cite{Bartels:1978fc,
  Bartels:1980pe, Jaroszewicz:1980mq, Kwiecinski:1980wb}, i.e., onto
the systematic inclusion of multi--reggeon exchanges in the t-channel.
Beyond the low density limit [where ${\cal G}^{\cal O}$, ${\cal G}$
(and all higher $n$-point t-channel insertions) are small] our
procedure provides a generalization consistent with JIMWLK evolution.
In the linear regime our solution for $\cal O$ stays constant with
energy in agreement with~\cite{Kovchegov:2003dm,Bartels:1999yt}.

One of the main properties of the resulting Eqs.
(\ref{eq:fund-odd-final-2}) and (\ref{odd}) is that ${\cal G}^{\cal
  O}\equiv 0$ (or, equivalently, ${\cal O}\equiv 0$) is a stable
solution of the equation, and that (as already discussed
in~\cite{Hatta:2005as, Kovchegov:2003dm}) non-vanishing odderon
contributions in the initial condition are erased very quickly due to
nonlinear effects. This may provide a glimpse of how the Casimir
scaling-violating multi--reggeon contributions may be erased by
non-linear evolution even if they are present in the initial
conditions.

Note that already the inclusion of ${\cal G}^{\cal O}$ breaks the Casimir
scaling relation~(\ref{eq:casimir-scaling-app}) between the dipole correlators
in the fundamental and adjoint representation (see Eqs. (\ref{eq:qbarq-odd})
and (\ref{eq:adjoint-noodderon})). It is only natural that higher t-channel
$n$-point exchanges will also contribute to this breaking of Casimir
scaling. The odderon contribution in our simulations, however, vanishes from
the outset: the initial conditions necessary to accommodate the total cross
section in DIS (as was the case for all our simulations) require ${\cal
  G}^{\cal O}\equiv 0$.

Still, this discussion does clarify the nature of what we expect to arise as
one includes corrections to the Gaussian truncation of JIMWLK evolution.

\providecommand{\href}[2]{#2}\begingroup\raggedright\endgroup

\end{document}